\documentclass[10pt]{article}
\usepackage{amsfonts}
\usepackage{amsmath,graphicx,array}
\usepackage{latexsym,amssymb,enumerate,ulem,pifont,calc,cite}
\usepackage{rotating,afterpage}
\usepackage{xcolor}
\usepackage{multirow}

  \numberwithin{equation}{section}
\begin{document}

\begin{titlepage}

\vskip 2.0 cm
\begin{center}  {\huge{\bf Non-BPS Black Branes in M-theory over Calabi-Yau Threefolds}} \\\vskip 0.5 cm {\Large{\bf (Non-)Uniqueness and Recombination of Non-BPS Black Strings in Two-Moduli CICY and THCY Models}}

\vskip 2.5 cm

{\large{\bf Alessio Marrani$^1$}, {\bf Anshul Mishra$^2$}, and {\bf Prasanta K. Tripathy$^2$}}

\vskip 1.0 cm

$^1${\sl Instituto de F\'{i}sica Teorica, Dep.to de F\'{i}sica,\\Universidad de Murcia, Campus de Espinardo, E-30100, Spain\\
\texttt{jazzphyzz@gmail.com}}\\

\vskip 0.5
cm

$^2${\sl Department of Physics, Indian Institute of Technology Madras,\\
 Chennai 600 036, India\\
\texttt{anshulmishra2025@gmail.com},\texttt{prasanta@iitm.ac.in}}\\

\vskip 0.5 cm


\vskip 0.5
cm


 \end{center}

 \vskip 2.0 cm

\begin{abstract}

We study extremal solutions arising in M-theory compactifications on
Calabi-Yau threefolds, focussing on non-BPS attractors for their importance in relation to the Weak Gravity Conjecture (WGC); M2 branes wrapped on two-cycles give rise to black
holes, whereas M5 branes wrapped on four-cycles result in black strings. In
the low-energy/field theory limit one obtains minimal $N=2$, $D=5$
supergravity coupled to Abelian vector multiplets. By making use of the
effective black hole potential formalism with Lagrange multipliers and of
the Attractor Mechanism, we obtain the explicit expressions of the
attractor moduli for BPS and non-BPS solutions, and we compute the
Bekenstein-Hawking black hole entropy and the black string tension.
Furthermore, by focussing on two-moduli complete intersection (CICY) or toric
hypersurface (THCY) Calabi-Yau threefolds, we investigate the
possible non-uniqueness of the attractor solutions, as well as the stability of
non-BPS black holes and black strings (restricting to doubly-extremal
solutions, for simplicity's sake). In all models taken into consideration,
we find that both BPS and non-BPS extremal black hole attractors are always
unique for a given, supporting electric charge configuration; moreover,
non-BPS black holes are always unstable, and thus they decay into
constituent BPS/anti-BPS pairs : this confirms the WCG, for which macroscopic non-supersymmetric solutions are bound to decay. For what concerns extremal black strings, it is well known they are unique in the BPS case; we confirm uniqueness also for non-BPS strings in two-moduli CICY
models. On the other hand, we discover multiple non-BPS extremal black string attractors
(with different tensions) in most of the two-moduli THCY models, and we
determine the corresponding magnetic configurations supporting them; this
indicates the existence of volume-minimizing representatives in the same
homology class having different values of their local minimal volume.
Moreover, we find that non-BPS (doubly-)extremal black strings, both for
single and multiple solutions, are stable and thus enjoy recombination of
their constituent BPS/anti-BPS pairs; in Calabi-Yau geometry, this means
that the volume of the representative corresponding to the black string is
less than the volume of the minimal piecewise-holomorphic representative,
predicting recombination for those homology classes and thus leading to
stable, non-BPS string solutions, which for the WGC are microscopic with small charges.

 \end{abstract}
\vspace{24pt} \end{titlepage}


\newpage \tableofcontents \newpage

\section{Introduction}

In recent years, knowledge at the borders between Physics and Mathematics
has witnessed great advances, stemmed from an intensive investigation of
compactifications of string theory on special holonomy manifolds, which
preserve some amount of Supersymmetry, and whose most notable class is
provided by Calabi-Yau manifolds. So far, this is the unique framework in
which stable solutions of Quantum Gravity (in particular concerning
holography in string theory) have been discovered, related to BPS objects,
with intriguing relations to the mathematical theory of invariants. In a
mathematical perspective, calibrated cycles in special holonomy have been
the subject of many studies, though it is now well known that many cycle
classes do not admit calibrated representatives, thus begging for the
investigation of minimal volume cycles, which is a topic hard to deal with,
in which very few results are currently established.

Supersymmetry, if any, is broken at low energies in our Universe. Thus,
string theory, as a candidate for a theory of Quantum Gravity, has to deal
with non-supersymmetric solutions, which have however been hitherto plagued
by the lack of stability. A way out to such a conundrum may be provided by
the study of non-supersymmetric solutions in supersymmetric string theory :
thus, the theory is itself stable, and all instability can be ascribed to
the decay of non-supersymmetric objects in an otherwise stable background.
Along the years, a remarkable success in string theory has provided a
deep-rooted understanding of the microscopic origin of black hole entropy
\cite{Strominger:1996sh}. More recently, black holes have turned out to be
important ingredients in investigating the intensely studied conjecture
which goes under the name of Weak Gravity Conjecture (WGC) \cite%
{Arkani-Hamed:2006emk}, stating that gravity is always the weakest force,
and therefore that all macroscopic, non-supersymmetric (i.e., non-BPS)
objects are finally bound to decay into microscopic objects with small
charge, which may result in saturating or not the BPS bound. Moreover, the
WGC has proved to be instrumental \cite{Ooguri:2006in} in order to formulate
the so-called \textquotedblleft Swampland program\textquotedblright\ \cite%
{Vafa:2005ui}. Therefore, in this framework, the study of non-BPS black
holes and black strings and their evolution toward a final state may play a
crucial role in confirming or disproving the WGC.

Compactifications of eleven-dimensional M-theory over Calabi-Yau threefolds
provides an important playground to study Quantum Gravity, and their
low-energy limit is given by minimal, $N=2$ supergravity theories in 5
space-time dimensions. More specifically, extremal (electric) black holes
(with $AdS_{2}\times S^{3}$ near-horizon geometry) are realized by wrapping
M2 branes on 2-cycles (in particular, on non-holomorphic curves) of the
Calabi-Yau manifold, whereas (magnetic) black strings (with $AdS_{3}\times
S^{2}$ near-horizon geometry) are given by M5 branes wrapped on 4-cycles
(namely, on non-holomorphic divisors) of such a space. A wrapped cycle is
conjectured to be a connected, locally volume-minimizing representative of
its homology class. Hence, non-BPS black holes and black strings may provide
key clues in investigating the existence, stability, and asymptotic count
for minimal-volume, 2- and 4- cycles in Calabi-Yau threefolds.

Extremal black holes and black strings are characterized by the so-called
Attractor Mechanism \cite%
{Ferrara:1995ih,Ferrara:1996dd,Ferrara:1996um,Ferrara:1997tw}, in which the
moduli spaces of scalar fields may admit multiple basin of attractors \cite%
{Kallosh:1999mz}. As one extrapolates the scalar fields from spatial
infinity, where they can take arbitrary initial values, to the black hole
horizon, they run into one of the attractor points. Within the same basin of
attraction, the values of these scalar fields at the attractor point is
determined in terms of the black hole charges only. Thus, the macroscopic
properties of a black hole, such as its Bekenstein-Hawking thermodynamical
entropy, depend only on the conserved charges associated to the underlying
gauge invariance. In this respect, it is here worth remarking that the
attractor machanism is applicable to both BPS as well as non-BPS solutions,
as long as they remain extremal \cite{Ferrara:1997tw,Goldstein:2005hq}. One
of the most fruitful approaches to the Attractor Mechanism is the use of the
so-called black hole / black string effective potential, whose critical
points determine the attractor values of the moduli (scalar) fields at the
event horizon of the black object under consideration.

This has been recently exploited by Long, Sheshmani, Vafa and Yau in \cite%
{LSVY}, in which the procedure of minimization of the effective potential
has been shown to fix the moduli inside the K\"{a}hler cone, yielding to the
determination of the black hole entropy or of the black string tension, and
hinting to a conjectural formula for the volume of the non-holomorphic
cycles wrapped by the M2 or M5 branes. In a given homology class, non-BPS,
doubly-extremal configurations have been used to obtain non-calibrated
cycles of minimal volume, as well as to compute the asymptotic volumes of
the representative cycles which minimize the volume. By explicitly
considering Calabi-Yau threefold compactifications with a few moduli, the
authors of \cite{LSVY} have found that non-BPS, extremal black holes
correspond to \textit{local}, but not global, volume minimizers of the
corresponding curve classes, as there is always a disconnected,
piecewise-calibrated representative (union of holomorphic and
anti-holomorphic curves) which corresponds to the BPS/anti-BPS black hole
constituents, and whose smaller volume implies that the aforementioned WGC
is satisfied, yielding to the decay of non-BPS black holes into
widely-separated BPS and anti-BPS particles.

On the other hand, within some of the same few moduli models, in \cite{LSVY}
it was discovered that non-BPS extremal black strings correspond to \textit{%
global} minima of the corresponding effective magnetic potential, and thus
the existence of a phenomenon called \textit{\textquotedblleft
recombination\textquotedblright } was established : \ holomorphic and
anti-holomorphic constituents of the same homology class fuse together to
make a smaller cycle, and by the WGC this yields to the prediction that
there should be \textit{stable}, non-BPS black strings (with small charge)
in the spectrum of the resulting supergravity theory. However, it should be
here remarked that, for a given supporting electric or magnetic charge
configuration, in the whole treatment of \cite{LSVY} all extremal black hole
resp. black string solutions have been found to be \textit{unique}.

In this context, the investigation of the possible \textit{non-uniqueness}
of attractors with different entropy or tension is of considerable
importance, as it may provide evidence for the existence of
volume-minimizing representatives in the same homology class having
different values of their local minimal volume; this fact can actually be
traced back to the homological structure and topological data of the
Calabi-Yau threefolds under consideration. In the present paper, developing
and extending the analysis of \cite{LSVY}, we plan to carry out an in-depth
investigation of extremal (non-BPS) black hole and black string attractors
in five-dimensional, minimal supergravity theory arising from Calabi-Yau
compactification of M-theory, especially for what concerns their possible
non-uniqueness, as well as the stability against the decay into constituent
elements with small charges.

Our main result will be two-fold : on one hand, we will confirm the
existence of stable, non-BPS, extremal black strings (which, when combined
with the WGC, hints for the fact that such strings should be microscopic
with small charges); on the other hand, for a given supporting magnetic
charge configuration, we will also find evidence for the non-uniqueness of
such (non-BPS) stable extremal black string attractors, thus hinting for the
aforementioned existence of volume-minimizing representatives in the same
homology class. For our purposes, two-moduli Calabi-Yau manifolds provide
the simplest class of models in which these issues can be explored;
classification of two-moduli Calabi-Yau manifolds as complete intersections
of product of projective spaces (CICY) as well as hypersurfaces in toric
varieties (THCY) has been carried out in various studies \cite%
{Candelas:1987kf,Kreuzer:2000xy}; for instance, the relevant cohomology data
along with the respective K\"{a}hler cones have been reported for these
manifolds in \cite{Ruehle}, hence providing the needed ingredients for our
investigation.\bigskip

The plan of the present paper is as follows.

We will start and recall basic facts on extremal (electric) black hole
attractors in five dimensional, minimal supergravity in Sec. \ref{5D-BHs},
then focussing on two-moduli Calabi-Yau threefolds in Sec. \ref{2-moduli BHs}%
. After a general treatment in Sec. \ref{general}, in Sec. \ref%
{Uniqueness-BHs} we analyze the uniqueness of black hole attractor solutions
in a variety of two-moduli Calabi-Yau threefolds : in models with $c=d=0$
and $a=b=0$ in Secs. \ref{cdzero} and \ref{cdzero2}, respectively, and in
complete intersection Calabi-Yau (CICY) and in Calabi-Yau manifolds arising
as a hypersurface in a toric variety (THCY), respectively in Secs. \ref{CICY}
resp. \ref{thcy}. We will then compute the recombination factor for non-BPS
extremal black hole attractors in two-moduli THCY models in Sec. \ref{BH-rec}%
, obtaining instability of such solutions in all cases. Then, we proceed and
recall basic facts on extremal (magnetic) black string attractors in five
dimensional, minimal supergravity in Sec. \ref{5D-BSs}, then focussing again
on two-moduli Calabi-Yau threefolds in Sec. \ref{2mbs}. Then, in Sec. \ref%
{Uniqueness-BSs} we analyze the uniqueness of black string attractor
solutions in a variety of two-moduli Calabi-Yau threefolds : in models with $%
c=d=0$ and $a=b=0$ in Secs. \ref{cdzero-2} and \ref{cdzero-3}, respectively,
and in CICY two-moduli models in Sec. \ref{mcicy}. Then, in Sec. \ref{thcym}
we present evidence for multiple non-BPS extremal black strings in
two-moduli THCY models. After that, we will compute the recombination factor
for non-BPS extremal black strings in the same class of models in Sec. \ref%
{BS-rec}, obtaining stability of such solutions in most of the models under
consideration. We make some conclusive remarks and comments on further
possible developments in the final Sec. \ref{Conclusions}. Various
appendices conclude and complete the paper. In Apps. \ref{App-1} and \ref%
{App-1-magn} we respectively recall the electric attractor equations and
present the magnetic attractor equations in minimal $D=5$ supergravity.
Finally, in Apps. \ref{cicymodel}-\ref{thcys} and \ref{cicys2}-\ref{thcy2}
we report (in various Tables) explicit results on extremal black hole resp.
black string attractors in the two-moduli CICY and THCY models.

\section{\label{5D-BHs}5D Black Hole Attractors}

In this paper we will focus on extremal solutions arising from the
compactification of M-theory on a Calabi-Yau manifold \cite%
{Cadavid:1995bk,Ferrara:1996hh}, whose low-energy, field theory limit
results into minimal, $N=2$ supergravity theory in five space-time
dimensions \cite{Gunaydin:1984ak}. The K\"{a}hler moduli of the Calabi-Yau
manifold gives rise to vector multiplets in the resulting five dimensional
theory. The moduli space exhibits the structure of a very special geometry
\cite{deWit:1992cr}. Critical points in this theory has been studied
extensively \cite{Chou:1997ba}. In the following we will first outline the
basic formalism in order to obtain the stabilization equation as well as to
set up the notations and conventions. Here we will mostly use the
conventions of \cite{Chou:1997ba}. In the case of coupling to $n$ vector
multiplets, the black hole effective potential in five dimensions is given
by ($I,J=1,...,n+1$, and $i,j=1,...,n$)
\begin{equation}
V=G^{IJ}q_{I}q_{J}=Z^{2}+\frac{3}{2}g^{ij}\partial _{i}Z\partial _{j}Z,
\label{V}
\end{equation}%
where the central charge reads as\footnote{%
In the following treatment, the subscript \textquotedblleft $e$" (for
\textquotedblleft \textit{electric}") will be understood for brevity.}%
\begin{equation}
Z_{e}=t^{I}q_{I},  \label{Ze}
\end{equation}%
in terms of the electric charge $q_{I}$ and the (pull-back of the) K\"{a}%
hler moduli $t^{I}$. The metric $g_{ij}$ of the real special moduli space $%
\mathcal{M}$ (namely, the scalar manifold of the corresponding supergravity
theory, with dim$_{\mathbb{R}}\mathcal{M}=n$) is given as%
\begin{equation}
g_{ij}=\frac{3}{2}\partial _{i}t^{I}\partial
_{j}t^{J}G_{IJ},~~g^{ij}g_{jk}=\delta _{k}^{i},
\end{equation}%
where $G_{IJ}$ is its pull-back onto the $\left( n+1\right) $-dimensional
\textquotedblleft ambient space\textquotedblright , which is the canonical
metric associated to the cubic form $C_{IJK}t^{I}t^{J}t^{K}$,
\begin{equation}
G_{IJ}:=-\frac{1}{3}\left. \frac{\partial ^{2}\log C_{LMN}t^{L}t^{M}t^{N}}{%
\partial t^{I}\partial t^{J}}\right\vert _{\ast }=\left(
3C_{ILM}C_{JNP}t^{L}t^{M}t^{N}t^{P}-2C_{IJM}t^{M}\right) _{\ast },
\end{equation}%
where \textquotedblleft $\ast $" denotes the evaluation at%
\begin{equation}
C_{IJK}t^{I}t^{J}t^{K}=1.  \label{constraint}
\end{equation}%
Here $C_{IJK}$ are the triple intersection numbers associated with the
Calabi-Yau manifold, which ultimately fix the whole bosonic sector of the
Lagrangian density of $N=2$, $D=5$ Maxwell-Einstein supergravity \cite{GST}:%
\begin{equation}
\frac{\mathcal{L}}{\sqrt{-\mathbf{g}}}=-\frac{1}{2}R-\frac{1}{4}G_{IJ}F_{\mu
\nu }^{I}F^{J|\mu \nu }-\frac{1}{2}g_{ij}\partial _{\mu }\varphi
^{i}\partial ^{\mu }\varphi ^{j}+\frac{1}{6^{3/2}\sqrt{-\mathbf{g}}}%
C_{IJK}\varepsilon ^{\lambda \mu \nu \rho \sigma }F_{\lambda \mu }^{I}F_{\nu
\rho }^{J}A_{\sigma }^{K},
\end{equation}%
where $\mathbf{g}:=$det$\mathbf{g}_{\mu \nu }$, and $\mathbf{g}_{\mu \nu }$
is the space-time metric Introducing
\begin{equation}
\Pi ^{IJ}:=g^{ij}\partial _{i}t^{I}\partial _{j}t^{J},  \label{Pij}
\end{equation}%
we can write the effective potential as
\begin{equation}
V=Z^{2}+\frac{3}{2}\Pi ^{IJ}q_{I}q_{J}\ .  \label{V2}
\end{equation}%
It has been shown in \cite{Chamseddine:1996pi} that
\begin{equation}
\Pi ^{IJ}=-\frac{1}{3}\big(C^{IJ}-t^{I}t^{J}\big),  \label{Pij-2}
\end{equation}%
where $C^{IJ}$ is the inverse of the matrix\footnote{%
In homogeneous scalar manifolds, since $C^{IJK}C_{J(LM|}C_{K|N)P}=\frac{1}{%
108}\left( \delta _{P}^{K}C_{LMN}+3\delta _{(L}^{K}C_{MN)P}\right) $, it
holds that $C^{IJ}=108\left( C^{IJK}C_{KMN}t^{M}t^{N}-\frac{1}{36}%
t^{I}t^{J}\right) $.} $C_{IJ}=C_{IJK}t^{K}$. Substituting this, we find
\begin{equation}
V=\frac{3}{2}Z^{2}-\frac{1}{2}C^{IJ}q_{I}q_{J}  \label{V3}
\end{equation}%
We will use the method of Lagrange multiplier to extremize this potential
subjected to the constraint $C_{IJK}t^{I}t^{J}t^{K}=1$. Extremizing
\begin{equation}
\tilde{V}=\frac{3}{2}Z^{2}-\frac{1}{2}C^{IJ}q_{I}q_{J}+\lambda
(C_{IJK}t^{I}t^{J}t^{K}-1)
\end{equation}%
with respect to $t^{K}$ and $\lambda $ we find
\begin{eqnarray}
0 &=&C_{IJK}t^{I}t^{J}t^{K}-1; \\
0 &=&3Zq_{K}-\frac{1}{2}\partial _{K}C^{IJ}q_{I}q_{J}+3\lambda
C_{KIJ}t^{I}t^{J}.
\end{eqnarray}%
Multiplying by $t^{K}$ in the second of the above and using\footnote{%
In homogeneous scalar manifolds, one can compute that $\partial
_{K}C^{IJ}=216C^{IJR}C_{RKM}t^{M}-3\left( t^{I}\delta _{K}^{J}+t^{J}\delta
_{K}^{I}\right) $.}
\begin{equation}
\partial _{K}C^{IJ}=-C^{IL}C^{JM}C_{KLM},
\end{equation}%
we find
\begin{equation}
3\lambda =-3Z^{2}-\frac{1}{2}C^{IJ}q_{I}q_{J}\ .
\end{equation}%
Using this value of $\lambda $ we find the equations of motion
\begin{equation}
3Z\big(q_{K}-ZC_{KIJ}t^{I}t^{J}\big)+\frac{1}{2}\big(%
C^{IL}C^{JM}C_{KLM}-C^{IJ}C_{KLM}t^{L}t^{M}\big)q_{I}q_{J}=0  \label{5deom}
\end{equation}%
along with the constraint (\ref{constraint}). The equation of motion can be
rewritten in a compact form as
\begin{equation}
C_{KLM}\big(C^{IL}C^{JM}-C^{IJ}t^{L}t^{M}\big)\big(q_{J}+6ZC_{JN}t^{N}\big)%
q_{I}=0\ .  \label{eomr}
\end{equation}%
The supersymmetric critical points correspond to\footnote{%
Note that (\ref{bpseq}) perfectly matches the relation (3.14) of \cite%
{CFMZ1-D=5}, obtained in the so-called \textquotedblleft new attractor"
approach to the 5D attractor equations (cf. Sec. 3.1.4 of \cite{CFMZ1-D=5}).}
\begin{equation}
q_{K}-ZC_{KIJ}t^{I}t^{J}=0\ .  \label{bpseq}
\end{equation}%
%
%
%
%
%
%
%
%
%
%
%
%
%
%
%
%
%
%
%
%
%
%
%
%
%
%
%
%
%
%
%
%
%
%
%
%
%
%
%
%
%
%
%
The non-BPS black holes can be found upon solving \eqref{5deom} such that $%
\big(q_{K}-ZC_{KIJ}t^{I}t^{J}\big)\neq 0$. It is worth exploring whether we
can obtain an equation analogous to \eqref{bpseq} for the non-BPS critical
points. A na\"{\i}ve analysis of \eqref{eomr} might suggest the non-BPS
solutions to the equation of motion given in terms of $q_{J}+6ZC_{JN}t^{N}=0$%
. However, for such a solution $C_{IJK}t^{I}t^{J}t^{K}=-1/6$ and hence
it is not consistent with the constraint \eqref{constraint}.

In order to obtain the algebraic equation corresponding to non-BPS critical
points, we set
\begin{equation}
X_{I}:=q_{I}-ZC_{IJK}t^{J}t^{K}  \label{X}
\end{equation}%
Note that the constraint \eqref{constraint} implies
\begin{equation}
t^{I}X_{I}=0  \label{tX=0}
\end{equation}%
Substituting $q_{I}=X_{I}+ZC_{IJK}t^{J}t^{K}$ in \eqref{5deom} we find
\begin{equation}
8ZX_{K}+C_{KLM}\big(C^{IL}C^{JM}-C^{IJ}t^{L}t^{M}\big)X_{I}X_{J}=0
\label{eomxi}
\end{equation}%
Solving the above along with the constraint $X_{I}t^{I}=0$ we can obtain the
expression for $X_{I}$ as a function of $t^{I}$ and $q_{I}$. It might be
easier to solve $X_{I}t^{I}=0$ first. This will give a possible solution for
$X_{I}$ up to an overall multiplicative factor. We can determine it as
follows. Multiply both sides of \eqref{eomxi} with $C^{KN}X_{N}$ to obtain
\begin{equation}
8ZC^{IJ}X_{I}X_{J}+C_{KLM}C^{KI}C^{LJ}C^{MN}X_{I}X_{J}X_{N}=0\ .
\label{norm}
\end{equation}%
This will determine the overall multiplicative factor in $X_{I}$. We can
substitute the resulting expression back in \eqref{eomxi} to verify whether
it holds. The trivial solution $X_{I}=0$ corresponding to BPS critical
points whereas any non-zero solution for $X_{I}$ will correspond to non-BPS
black holes.

Note that, from the \textquotedblleft new attractor" approach to 5D
attractors treated in \cite{CFMZ1-D=5} (see App. \ref{App-1}), (\ref{X}) can
equivalently be rewritten as%
\begin{equation}
X_{I}=\frac{3^{3/2}}{2^{5/2}}\frac{1}{Z}T^{ijk}\partial _{j}Z\partial
_{k}Z\partial _{i}t_{I},  \label{X-2}
\end{equation}%
where $t_{I}=C_{IJK}t^{J}t^{K}$ \cite{CFMZ1-D=5, FG-D=5}, and
\begin{eqnarray}
T^{ijk} &=&g^{il}g^{jm}g^{kn}T_{lmn}; \\
T_{ijk} &=&\partial _{i}t^{I}\partial _{j}t^{J}\partial _{k}t^{K}C_{IJK}.
\end{eqnarray}%
The condition (\ref{tX=0}) is consistent with (\ref{X-2}), because%
\begin{equation}
t^{I}\partial _{i}t_{I}=0,~\forall i,
\end{equation}%
as a consequence of the normalization condition 
\begin{equation}
t^{I}t_{I}=1.
\end{equation}

\section{\label{2-moduli BHs}Two-moduli Models}

\subsection{\label{general}General treatment}

We will now focus our attention to two-moduli models, by setting $n=1$;
thus, $I,J=1,2$. In order to avoid cluttering of indices, we will use the
notation{\footnote{%
This differs from the notations of \cite{Kallosh:1999mz} where they
introduced $\bar{t}^{I}=Z^{1/2}t^{I}$ and $x=\bar{t}^{1},y=\bar{t}^{2}$.}} $%
t^{1}=x,t^{2}=y$. By defining $a:=C_{111}$, $b:=C_{112}$, $c:=C_{122}$, $%
d:=C_{222}$, we find
\begin{equation}
C_{IJ}=%
\begin{pmatrix}
ax+by & bx+cy\cr bx+cy & cx+dy%
\end{pmatrix}%
\ ,
\end{equation}%
and its inverse
\begin{equation}
C^{IJ}=\frac{1}{Lxy-Nx^{2}-My^{2}}%
\begin{pmatrix}
cx+dy & -(bx+cy)\cr-(bx+cy) & ax+by%
\end{pmatrix}%
,
\end{equation}%
where $L:=ad-bc,~M:=c^{2}-bd,~N:=b^{2}-ac$. Further\footnote{$%
C_{IJ}t^{J}=C_{IJK}t^{I}t^{K}$ is the (un-normalized) `Jordan dual' \cite{JD}
of $t^{I}$.}
\begin{equation}
C_{IJ}t^{J}=%
\begin{pmatrix}
ax^{2}+2bxy+cy^{2}\cr bx^{2}+2cxy+dy^{2}%
\end{pmatrix}
\label{cijtj}
\end{equation}%
and
\begin{equation}
C^{IJ}q_{J}=\frac{1}{Lxy-Nx^{2}-My^{2}}%
\begin{pmatrix}
q_{1}(cx+dy)-q_{2}(bx+cy)\cr-q_{1}(bx+cy)+q_{2}(ax+by)%
\end{pmatrix}%
\ .
\end{equation}

The equations of motion (\ref{5deom}) have the following lengthy
expressions:
\begin{eqnarray}
6Z(q_{1}-ZA_{1})+\frac{\big(aQ_{1}^{2}+2bQ_{1}Q_{2}+cQ_{2}^{2}\big)}{%
(Lxy-Nx^{2}-My^{2})^{2}}-\frac{A_{1}\big(q_{1}Q_{1}+q_{2}Q_{2}\big)}{%
(Lxy-Nx^{2}-My^{2})} &=&0;  \label{eom1} \\
6Z(q_{2}-ZA_{2})+\frac{\big(bQ_{1}^{2}+2cQ_{1}Q_{2}+dQ_{2}^{2}\big)}{%
(Lxy-Nx^{2}-My^{2})^{2}}-\frac{A_{2}\big(q_{1}Q_{1}+q_{2}Q_{2}\big)}{%
(Lxy-Nx^{2}-My^{2})} &=&0,  \label{eom2}
\end{eqnarray}%
where, for easy reading, we have introduced the notations
\begin{equation}
A_{1}:=ax^{2}+2bxy+cy^{2},\ A_{2}:=bx^{2}+2cxy+dy^{2}  \label{a1a2}
\end{equation}%
and\footnote{%
Note that $Q^{I}=(Lxy-Nx^{2}-My^{2})C^{IJ}q_{J}$ should have a contravariant $%
I$-index; we choose covariant indices, but this is irrelevant in our
treatment.}
\begin{equation}
Q_{1}:=q_{1}(cx+dy)-q_{2}(bx+cy),\ Q_{2}:=-q_{1}(bx+cy)+q_{2}(ax+by)\ .
\end{equation}%
These give rise to two coupled degree seven equations, which in general
cannot be solved to obtain exact analytic expression for the moduli $t^{I}$
in terms of the \textquotedblleft charges\textquotedblright\ $Q_{I}$. The
supersymmetric critical points, corresponding to BPS black hole attractors,
are obtained from (\ref{bpseq}),
\begin{eqnarray}
q_{1}-Z(ax^{2}+2bxy+cy^{2}) &=&0; \\
q_{2}-Z(bx^{2}+2cxy+dy^{2}) &=&0\ .  \label{2-eom}
\end{eqnarray}%
The general solution for these equations have been obtained in \cite%
{Kallosh:1999mz}. The analogous equations for the non-BPS critical points,
corresponding to non-BPS black hole attractors, can be obtained by solving %
\eqref{eomxi} for $X_{I}$ along with $X_{I}t^{I}=0$. The later condition can
easily be solved to find $X_{I}=\hat{X}\tilde{X}_{I}$ with $\tilde{X}_{1}=-y,%
\tilde{X}_{2}=x$. The overall multiplicative factor $\hat{X}$ can be
obtained from \eqref{norm} upon substituting the above form of $X_{I}$ in
it. We find
\begin{equation}
\hat{X}=-\frac{8ZC^{IJ}\tilde{X}_{I}\tilde{X}_{J}}{C_{KLM}C^{KP}C^{LQ}C^{MN}%
\tilde{X}_{P}\tilde{X}_{Q}\tilde{X}_{N}}.
\end{equation}%
This can be further simplified using
\begin{equation}
C^{IJ}\tilde{X}_{J}=\frac{1}{\mathrm{det}C}%
\begin{pmatrix}
-A_{2}\cr A_{1}%
\end{pmatrix}%
\ \mathrm{and}\ C^{IJ}\tilde{X}_{I}\tilde{X}_{J}=\frac{1}{\mathrm{det}C},
\end{equation}%
where $C:=\left( C_{IJ}\right) $. This gives rise to
\begin{equation}
\hat{X}=-\frac{8Z\ {\mathrm{det}}^{2}{C}}{C_{IJK}A^{I}A^{J}A^{K}},
\label{res}
\end{equation}%
where $A^{1}=-A_{2},A^{2}=A_{1}$. Upon using $C_{IJK}t^{I}t^{J}t^{K}=1$, the
degree six polynomial ${C_{IJK}A^{I}A^{J}A^{K}}$ can be shown to reduce to
the cubic
\begin{equation*}
(2b^{3}-3abc+a^{2}d)x^{3}+3(b^{2}c-2ac^{2}+abd)x^{2}y-3(bc^{2}-2b^{2}d+acd)xy^{2}-(2c^{2}-3bcd+ad^{2})y^{3}
\end{equation*}

To summarise, the equation of motion corresponding to the non-BPS critical
point for an arbitrary two-moduli model is obtained by rewriting (\ref{X})
with the positions written below (\ref{2-eom}) and using the result (\ref%
{res}) :
\begin{equation}
q_{I}-ZC_{IJK}t^{J}t^{K}+\frac{8Z\ {\mathrm{det}}^{2}{C}}{%
C_{IJK}A^{I}A^{J}A^{K}}\tilde{X}_{I}=0\ .  \label{nbps1b}
\end{equation}%
For a given value of $C_{IJK}$ and for a given set of charges, this equation
can be solved numerically to obtain the values of the moduli $t^{I}$
corresponding to a non-BPS critical point.

The effective black hole potential (\ref{V}) (or, equivalently, (\ref{V2})
or (\ref{V3})) has the expression
\begin{equation}
V=\frac{1}{2}\left[ 3(q_{1}x+q_{2}y)^{2}-\frac{%
q_{1}^{2}(cx+dy)-2q_{1}q_{2}(bx+cy)+q_{2}^{2}(ax+by)}{Lxy-Nx^{2}-My^{2}}%
\right] \ .  \label{vcrit}
\end{equation}%
By adopting the normalization of \cite{LSVY}, the (Bekenstein-Hawking) black
hole entropy $S$ can be determined from the critical value of the effective
black hole potential as\footnote{%
Actually, the effective black hole potential discussed in \cite{LSVY}
differs from ours by a factor of $2/3$ and hence we have $\left( V/9\right)
^{3/4}$ instead of $\left( V/6\right) ^{3/4}$.}%
\begin{equation}
S=2\pi \left( \frac{V}{9}\right) ^{3/4}.
\end{equation}%
For BPS solution :
\begin{equation}
V=Z^{2}=(q_{1}x+q_{2}y)^{2}\ \Rightarrow S\ =\frac{2}{3\sqrt{3}}\pi
\left\vert Z\right\vert ^{3/2}=\frac{2}{3\sqrt{3}}\pi \left\vert
q_{1}x+q_{2}y\right\vert ^{3/2}.  \label{vbps}
\end{equation}

We anticipate here that in \textit{all} $36+48=84$ two-moduli models (of
CICY and THCY type) we have considered in the present paper, we will find
that \textit{all} (BPS and non-BPS) black hole (electric) attractors are
\textit{unique}, confirming and generalizing the results of \cite{LSVY}.
Moreover, by analising the so-called \textit{recombination factor}, we will
also find that \textit{all} non-BPS black holes are actually unstable
against the decay into their BPS/anti-BPS constituent pairs, again
confirming and generalizing the results of \cite{LSVY}.

\subsection{\label{Uniqueness-BHs}Uniqueness of Attractors}

\subsubsection{\label{cdzero}$c=d=0$}

Before analysing the class of two-moduli models in detail, we observe that
the corresponding equations of motion (\ref{eom1})-(\ref{eom2}) take a
particularly simple form if we set $c=d=0$. A number of Calabi-Yau models do
possess intersection numbers satisfying either of these conditions; for
instance, the K3 fibration considered in \cite{LSVY} is one such example.
Upon setting $c=0=d$ and introducing $t=x/y$ and $q=q_{1}/q_{2}$, the
constraint \eqref{constraint} reduces to
\begin{equation}
(at^{3}+3bt^{2})y^{3}=1\ .
\end{equation}

Using the above, the BPS equation reduces to the simple form
\begin{equation}
t(a-bq)+2b=0\ ,
\end{equation}%
which gives rise to the \textit{unique} solution
\begin{equation}
t=\frac{2b}{bq-a}\ .
\end{equation}%
The critical values for the moduli $x,y$ are given by
\begin{equation}
x=\frac{2^{1/3}}{(3bq-a)^{1/3}},~~~y=\frac{bq-a}{2^{2/3}b(3bq-a)^{1/3}}.
\end{equation}%
The solution will lie inside the K\"{a}hler cone (or, in other words, the
conditions of positivity of critical values of the moduli $x$ and $y$ are
satisfied) provided $q>a/b$ and $3bq>a$. The effective potential for the
above critical point is
\begin{equation}
V=\frac{q_{2}^{2}\left( 3bq-a\right) ^{4/3}}{2^{4/3}b^{2}}\ ,
\label{effvbps}
\end{equation}%
and hence the entropy
\begin{equation}
S=2\pi \left( \frac{V}{9}\right) ^{3/4}=\frac{\pi (3bq-a)}{3\sqrt{3}}%
\left\vert \frac{q_{2}}{b}\right\vert ^{3/2}\ .
\end{equation}

A similar analysis for the non-BPS case leads to the linear equation
\begin{equation}
6b+t(a+3bq)=0\ ,
\end{equation}%
resulting the \textit{unique} solution
\begin{equation}
t=-\frac{6b}{a+3bq}\ .
\end{equation}%
The expression for the moduli are given as
\begin{equation}
x=\frac{2^{1/3}}{(a-3bq)^{1/3}}\ ,\ y=\frac{-(a+3bq)}{3\
2^{2/3}b(a-3bq)^{1/3}}\ .
\end{equation}%
The solution lies within the K\"{a}hler cone, provided $a/b+3q<0$ and $a>3bq$%
. The entropy of the black hole is given by
\begin{equation}
S=\frac{\pi (a-3bq)}{3\sqrt{3}}\left\vert \frac{q_{2}}{b}\right\vert ^{3/2}\
.
\end{equation}

\subsubsection{\label{cdzero2}$a=b=0$}

A similar analysis can be done for the case $a=b=0$. For the BPS solution,
we find
\begin{equation}
t=\frac{c-dq}{2cq}\ .
\end{equation}%
This gives rise to
\begin{equation}
x=\frac{c-dq}{c(2q)^{2/3}(3c-dq)^{1/3}}\ ,\ y=\left( \frac{2q}{3c-dq}\right)
^{1/3}\ .
\end{equation}%
The solution lies within the K\"{a}hler cone for $q<c/d$ and $d<3c/q$. The
entropy of the corresponding solution is
\begin{equation}
S=\frac{\pi (3c-dq)}{3\sqrt{3}q}\left\vert \frac{qq_{2}}{c}\right\vert
^{3/2}\ .
\end{equation}

For the non-BPS case we find
\begin{equation}
t = - \frac{3 c + d q}{6cq} \ .
\end{equation}
Thus, we have
\begin{equation}
x = \frac{3 c + d q}{3 c (2q)^{2/3}(3 c - d q)^{1/3}} \ , \ y = \left(\frac{%
2q}{d q - 3 c}\right)^{1/3} \ ,
\end{equation}
with entropy
\begin{equation}
S = \frac{\pi(d q - 3 c)}{3\sqrt{3}q}\left\vert\frac{q q_2}{c}%
\right\vert^{3/2} \ .
\end{equation}
The K\"ahler cone conditions are $d>3c/q$ and $3/q+d/c<0$.

\subsubsection{\label{CICY}CICY}

We will now systematically analyse BPS as well as non-BPS black hole
attractors in two-moduli Calabi-Yau models arising as complete intersection
of hypersurfaces in product of projective spaces. A few of these complete
intersection Calabi-Yau (CICY) manifolds were already treated in \cite{LSVY}%
. Intersection numbers and other relevant cohomology data for all 36 such
two-moduli CICY manifolds have been recently computed and reported in \cite%
{Ruehle} (cfr. App. A therein). We intend to examine extremal black holes in
all these CICY models in order to investigate the issue of \textit{%
non-uniqueness} of such solutions. In the following treatment, we will first
workout one model in full detail, and then summarise our results for the
remaining models in various Tables in App. \ref{cicymodel}.\medskip

We consider the CICY model with configuration matrix \cite{Ruehle}
\begin{equation}
\begin{pmatrix}
0 & 0 & 2 & 1\cr2 & 2 & 1 & 1%
\end{pmatrix}%
\ .  \label{cicym1}
\end{equation}%
The Calabi-Yau manifold constitutes of the intersection of four
hypersurfaces, which are given by the zero locus of polynomials with
bi-degrees $(0,2),(0,2),(2,1)$ and $(1,1)$ respectively in the product space
$\mathbb{P}^{2}\times \mathbb{P}^{5}$. Each of the columns of the
configuration matrix represents the bi-degree of the respective polynomial.
The K\"{a}hler cone consists of the positive quadrant in the $xy$-plane. The
intersection numbers of the Calabi-Yau manifold are given by\footnote{%
Our convention for the overall normalization of the triple intersection
numbers differs from \cite{LSVY,Ruehle} by a factor of 6.} $a=0,b=2/3,c=2$
and $d=4/3$. The volume of the Calabi-Yau manifold is given by
\begin{equation}
\mathcal{V}=2x^{2}y+6xy^{2}+\frac{4}{3}y^{3}\ .
\end{equation}

We will first consider the BPS equations:
\begin{eqnarray}
3q_{1}-2y(2x+3y)(q_{1}x+q_{2}y) &=&0\ ,  \label{bpsm1} \\
3q_{2}-2(x^{2}+6xy+2y^{2})(q_{1}x+q_{2}y) &=&0\ .
\end{eqnarray}%
In addition, we need to impose the constraing $\mathcal{V}=1$, which turns
out to be
\begin{equation}
2x^{2}y+6xy^{2}+\frac{4}{3}y^{3}=1\ .  \label{volr1}
\end{equation}%
To analyse the above equations we will do the rescaling $t=x/y$ and $%
q=q_{1}/q_{2}$. Solving \eqref{volr1} for $y$ as a function of $t$, we find
\begin{equation}
y=\left( \frac{3}{2(3t^{2}+9t+2)}\right) ^{1/3}  \label{volry1}
\end{equation}%
Substituting the above in \eqref{bpsm1} we find
\begin{equation}
qt^{2}+2(3q-1)t+(2q-3)=0\ .
\end{equation}%
Solving the above for $t$ we find the critical value
\begin{equation}
t_{\pm }=\frac{1}{q}\left( 1-3q\pm \sqrt{7q^{2}-3q+1}\right) \ .
\end{equation}%
%
%
%
%
%
%
%
%
%
%
%
%
%
%
%
%
%
%
%
%
%
%
We need to make sure that the solution lies in the K\"{a}hler cone, \textit{%
i.e.}, the critical value of $t$ must be positive. It can be observed that $%
t_{-}$ is negative for all values of $q$, whereas $t_{+}$ becomes positive
for $0<q<3/2$. Thus, the equations of motion admit a \textit{unique} BPS
attractor for $0<q<3/2$.

Though it is sufficient for our purpose to have the expression for $t$ as a
function of $q$, for the sake completeness, we reproduce in the following
the form of the moduli $x,y$ in terms of the ratio the electric charges $%
q=q_{1}/q_{2}$:
\begin{eqnarray}
x &=&\left( \frac{3}{2q}\right) ^{1/3}\frac{1-3q+\sqrt{7q^{2}-3q+1}}{\left(
(23q^{2}-18q+6)-(9q-6)\sqrt{7q^{2}-3q+1}\right) ^{1/3}}\ , \\
y &=&\left( \frac{3q^{2}}{2}\right) ^{1/3}\frac{1}{\left(
(23q^{2}-18q+6)-(9q-6)\sqrt{7q^{2}-3q+1}\right) ^{1/3}}\ .
\end{eqnarray}

We will now compute the black hole entropy for the above configuration. For
the BPS solution, the effective potential:
\begin{equation}
V=(q_{1}x+q_{2}y)^{2}=\left( \frac{3}{2}\right) ^{2/3}q_{2}^{2}\ \frac{%
(1+qt)^{2}}{(3t^{2}+9t+2)^{2/3}}\ .
\end{equation}%
Substituting $t=t_{+}$ in this result, we then find the entropy of the black
hole to be 
%
\begin{equation}
S=\frac{\pi q}{3}\sqrt{\frac{2q_{2}^{3}\ \big(2-3q+\sqrt{7q^{2}-3q+1}\big)%
^{3}}{(23q^{2}-18q+6)-(9q-6)\sqrt{7q^{2}-3q+1}}}\ .
\end{equation}

We will now turn our attention to the non-BPS attractors. The equations of
motion are given as
\begin{eqnarray}
q_{1}-\frac{2}{3}y(q_{1}x+q_{2}y)\left( (2x+3y)+\frac{8\big(x^{2}+3xy+7y^{2}%
\big)^{2}}{(2x+3y)\big(x^{2}+3xy-12y^{2}\big)}\right) &=&0,  \label{nbps1a}
\\
q_{2}-\frac{2}{3}(q_{1}x+q_{2}y)\left( \big(x^{2}+6xy+9y^{2}\big)-\frac{8x%
\big(x^{2}+3xy+7y^{2}\big)^{2}}{(2x+3y)\big(x^{2}+3xy-12y^{2}\big)}\right)
&=&0\ .
\end{eqnarray}%
To find the non-BPS black holes, we need to solve the above equations along
with the volume constraint \eqref{volr1}. In order to simplify \eqref{nbps1a}
we will once more introduce the variable $t=x/y$ and the charge ratio $%
q=q_{1}/q_{2}$. In terms of these quantities, the equations of motion become
\begin{eqnarray}
q-\frac{2y^{3}(qt+1)\big(12t^{4}+72t^{3}+181t^{2}+219t+284\big)}{%
3(2t+3)(t^{2}+3t-12)} &=&0\ , \\
1+\frac{2y^{3}(qt+1)\big(6t^{5}+27t^{4}+141t^{3}+444t^{2}+683t+72\big)}{%
3(2t+3)(t^{2}+3t-12)} &=&0\ .
\end{eqnarray}%
Substituting the expression for $y$ from \eqref{volry1} in the above we find
one linearly independent equation involving $t$ and $q$. For any given $q$
it can be numerically solved to obtain the corresponding value of $t$. We
need the verify whether this corresponds to a physical solution with $t$
lying within the K\"{a}hler cone. It is however much more instructive to
deal with the inverse problem, \textit{i.e.}, express $q$ as a function of $%
t $. Upon simplification, we find
\begin{equation}
q=-\frac{12t^{4}+72t^{3}+181t^{2}+219t+284}{%
6t^{5}+27t^{4}+141t^{3}+444t^{2}+638t+72}\ .
\end{equation}%
We find that the rhs of the above equation is a monotonically increasing
function of $t$. For $t=0$ the charge ratio $q$ takes the value $-71/18$,
and it vanishes as $t\rightarrow \infty $. Since the function is monotonic,
we have a \textit{unique} non-BPS black hole for every value of $q$ in the
range $-71/18<q<0$. This is in contrast to the BPS case, where we have a
\textit{unique} solution for $0<q<3/2$. Thus, both the solutions are
mutually exclusive.

The black hole effective potential $V$ takes the form
\begin{equation}
\frac{V}{ q_2^2} = \left(\frac{3}{2}\right)^{2/3} \big(3 t^2 + 9 t + 2\big)%
^{4/3} \frac{36 t^6 + 324 t^5 + 1557 t^4 + 4482 t^3 + 10329 t^2 + 15192 t +
12272} {\big(6 t^5 + 27 t^4 + 141 t^3 + 444 t^2 + 638 t + 72\big)^2}
\end{equation}
where $t$ is the critical value in the above equation.

As an example, consider the value $q=-48/83$. This gives the solution $t=1$.
Thus, we have
\begin{equation}
x=y=\left( \frac{3}{28}\right) ^{1/3}\ .  \label{t1xy}
\end{equation}%
The entropy for this configuration is
\begin{equation}
S=\frac{7\pi }{249\sqrt{83}}\left( \frac{1381q_{2}^{2}}{2}\right) ^{3/4}\ .
\end{equation}%
\medskip

A similar analysis can be carried out for all other 35 two-moduli CICY
models. In App. \ref{cicymodel} we summarise both BPS as well as non-BPS
solutions along with their respective ranges of validity for all such 35
two-moduli CICY models. In \textit{all} such models, both BPS and non-BPS
black hole attractors are \textit{unique}.

\subsubsection{\label{thcy}THCY}

In this subsection we will consider two-moduli Calabi-Yau manifolds arising
as a hypersurface in a toric variety. A toric variety is specified in terms
of a reflexive polytope\footnote{%
An integral polytope is called reflexive if its dual polytope is also
integral.} with a specific triangulation of its faces \cite{Hosono:1993qy}.
In the case of Calabi-Yau threefolds we need to consider reflexive polytopes
in four dimensions. There is a one-to-one map from the faces of the
reflexive polytope to the vertices of the dual polytope. Typically, the four
dimensional vectors corresponding the these vertices are not linearly
independent. A generic dual polytope with $n$ vertices $\vec{v}_{i}$ will
have $(n-4)$ linear relationships like ($r=1,...,n-4$)
\begin{equation}
\sum_{i=1}^{n}q_{i}^{r}\vec{v}_{i}=0\ .
\end{equation}%
The coefficients $q_{i}^{r}$ constitute a $(n-4)\times n$ matrix, which is
called as the weight matrix or charge matrix of the toric variety. Each of
these vertices is associated with a homogenenous coordinate $z_{i}\in
\mathbb{C}^{n}$ and the rows of the weight matrix give equivalence relations
among these homogeneous coordinates. Removing a fixed point set $F$ which is
determined by the given triangulation, and taking quotient with the above
mentioned equivalence relation gives rise to the four dimentional toric
variety. The number of rows of the weight matrix gives the Picard number of
the toric variety. Since we are interested in toric varieties with Picard
number two, we need to consider dual polytopes with six vertices.

A complete classification of all such reflexive polytopes in four dimensions
has been carried out \cite{Kreuzer:2000xy}. Hypersurfaces with vanishing
first Chern class in these toric varieties give rise to Calabi-Yau manifolds
(THCY). The intersection numbers as well as the K\"{a}hler cone and other
relevant cohomology data for the two-moduli THCY models have also been
recently computed \cite{Ruehle} (cfr. App. B therein), and the corresponding
cubic forms are available in the database \cite{CY-database} (see also \cite%
{Altman:2014bfa}). As done for the CICY's, in the following treatment we
will consider one specific model in detail, and then summarize the results
for the remaining THCY models in various Tables in App. \ref{thcys}.\medskip

The toric variety of our interest is described by the weight matrix%
\begin{equation}
\left(
\begin{array}{cccccc}
-1 & 1 & 1 & 0 & 2 & 3 \\
1 & 0 & 0 & 1 & 0 & -2%
\end{array}%
\right) .  \label{thcym1}
\end{equation}
The triangulation is specified in terms of the Stanley-Reisner ideal
\begin{equation}
\left\langle z_{0}z_{3},z_{1}z_{2}z_{4}z_{5}\right\rangle \ .
\end{equation}%
The triple intersection numbers of the corresponding THCY model in a basis
where the K\"{a}hler cone coincides with the first quadrant of the
Argand-Gauss plane are $a=1/3,b=1/2,c=1/2$ and $d=1/2.$ The volume of the
Calabi-Yau manifold is
\begin{equation}
{\mathcal{V}}=\frac{x^{3}}{3}+\frac{3x^{2}y}{2}+\frac{3xy^{2}}{2}+\frac{y^{3}%
}{2}\ .
\end{equation}

The BPS equations are
\begin{eqnarray}
q_{1}-\frac{1}{6}(q_{1}x+q_{2}y)(2x^{2}+3y^{2}+6xy) &=&0,  \label{bps2} \\
q_{2}-\frac{1}{2}(q_{1}x+q_{2}y)(x+y)^{2} &=&0\ .
\end{eqnarray}%
Once again we will use the rescaled coordinate $t=x/y$ and the charge ratio $%
q=q_{1}/q_{2}$. The constraint $\mathcal{V}=1$ gives
\begin{equation}
y=\left( \frac{6}{2t^{3}+9t^{2}+9t+3}\right) ^{1/3}\ .  \label{vconst}
\end{equation}%
Substituting the above expression for $y$ in \eqref{bps2}, we find
\begin{equation}
3q(t+1)^{2}-(2t^{2}+6t+3)=0\ .
\end{equation}%
It is straightforward to write down the solutions to the above equation. We
find
\begin{equation}
t_{\pm }=\frac{3(1-q)\pm \sqrt{3(1-q)}}{3q-2}\ .
\end{equation}%
Here $t_{+}$ corresponds to the physical solution lying inside the K\"{a}%
hler cone for $q$ taking values in the range $2/3<q<1$.

We will now compute the entropy for this configuration. The effective
potential for the BPS black hole is
\begin{equation}
V = (q_1 x + q_2 y)^2 = 6^{2/3} q_2^2 \frac{(1+qt)^2}{(2t^3+9t^2+9t+3)^{2/3}}%
,
\end{equation}

and thus the entropy reads
\begin{equation}
S=\left( \frac{2q_{2}}{3}\right) ^{3/2}\pi \sqrt{\frac{\big(3q^{2}-6q+2-q%
\sqrt{3(1-q)}\big)^{3}}{\big(-9q^{3}+36q^{2}-36q+8\big)+\sqrt{3(1-q)}\big(%
9q^{2}-16q+4\big)}}.
\end{equation}

We will now consider the non-BPS equations of motion. Substituting the
values of the intersection numbers in \eqref{nbps1b}, we find
\begin{eqnarray}
q_{1}-(q_{1}x+q_{2}y)\left( \frac{1}{6}(2x^{2}+6xy+3y^{2})+\frac{%
4x^{2}y(x+y)^{2}}{4x^{3}+9x^{2}y+9xy^{2}+3y^{3}}\right) &=&0, \\
q_{2}-(q_{1}x+q_{2}y)\left( \frac{1}{2}(x+y)^{2}-\frac{4x^{3}(x+y)^{2}}{%
4x^{3}+9x^{2}y+9xy^{2}+3y^{3}}\right) &=&0\ .~
\end{eqnarray}%
Substituting $x=ty,q_{1}=qq_{2}$ and using the constraint \eqref{vconst} in
the above equations, we obtain
\begin{equation}
3q(t+1)^{2}\left( 4t^{3}-9t^{2}-9t-3\right)
+8t^{5}+66t^{4}+132t^{3}+111t^{2}+45t+9=0\ .  \label{nbpseq1}
\end{equation}

For a given $q$, we can numerically solve the above equation to obtain the
value of $t$. To obtain a qualitative behaviour, we will instead solve the
above equation for $q$ as a function of $t$. We find
\begin{equation}
q=-\frac{8t^{5}+66t^{4}+132t^{3}+111t^{2}+45t+9}{3(t+1)^{2}\left(
4t^{3}-9t^{2}-9t-3\right) }
\end{equation}%
Note that, in the physical $t>0$ region the denominator of the rhs in the
above equation vanishes for $t=t_{\ast }\simeq 3.06$. At the $x=0$ boundary
of the moduli space $q$ takes the value $q=1$. For all values of $q$ in the
range $1<q<\infty $ the solution for $t$ lies in the region $0<t<t_{\ast }$.
Similarly, at the $y=0$ boundary of the moduli space $q=-2/3$. Thus, for $%
-\infty <q<-2/3$ the value of $t$ lies in the range $t_{\ast }<t<\infty $.
For $-2/3<q<1$, the equation \eqref{nbpseq1} does not admit any solution
with $t>0$. Beyond this region, there is a \textit{unique} non-BPS black
hole solution for any given value of $q$.

As an example, consider the value $q=371/204$. This gives rise to the
critical value $t=1$. The corresponding values for the moduli $x$ and $y$
are given by
\begin{equation}  \label{t2xy}
x = \left(\frac{6}{23}\right)^{1/3} = y \ .
\end{equation}

The black hole effective potential corresponding to the non-BPS black hole
as a function of $t$ at the critical point is given by
\begin{equation}
\frac{V}{q_{2}^{2}}=\frac{2^{2/3}\left( 2t^{3}+9t^{2}+9t+3\right)
^{4/3}\left( 112t^{6}+360t^{5}+441t^{4}+282t^{3}+135t^{2}+54t+9\right) }{3%
\sqrt[3]{3}(t+1)^{4}\left( 4t^{3}-9t^{2}-9t-3\right) ^{2}}  \label{vcrit1}
\end{equation}%
We can use the above expression to compute the entropy. For example, for the
case of $q=371/204$, the black hole entropy is given by
\begin{equation}
S=23\ \pi \ \frac{(1393\ q_{2}^{2})^{3/4}}{306\sqrt{102}}\ .
\end{equation}%
\medskip

A similar analysis can be carried out for all other 47 two-moduli THCY
models. In App. \ref{thcys} we summarise both BPS as well as non-BPS
solutions along with their respective ranges of validity for all such 47
two-moduli THCY models. In \textit{all} such models, both BPS and non-BPS
black hole attractors are \textit{unique}.\medskip

Thus, in \textit{all} $36+48=84$ two-moduli models (of CICY and THCY type)
we have considered, we have found that \textit{all} (\textit{both} BPS and
non-BPS) black hole (electric) attractors are \textit{unique}, confirming
and generalizing the results of \cite{LSVY}.

\subsubsection{\label{BH-rec}Non-BPS Black Holes: Recombination Factor and
Instability}

It is important to analyse the issue of stability for the non-BPS black
holes. In this context an important quantity, namely the \textit{%
recombination factor} $R$ has been introduced in \cite{LSVY}. It is given by
the ratio of the mass of the non-BPS black holes to that of the minimal
piecewise calibrated representative corresponding to the same homology
class. For $R>1$ the non-BPS black hole is unstable, and it decays into the
corresponding BPS-anti-BPS constituent which form the piecewise calibrated
representative. On the other hand, the value $R<1$ indicates that the
constituent BPS-anti-BPS pairs recombine, in order to give rise to a stable
non-BPS black hole in the spectrum \cite{LSVY}. In the following treatment,
we compute the recombination factor for non-BPS black holes in the
two-moduli THCY model treated above.

We consider an $M2$-brane of charge $q_{I}$ wrapped on the curve $C=\alpha
C_{1}+\beta C_{2}$. Thus, we have
\begin{equation}
q_{1}=J_{1}\cdot C=\alpha J_{1}\cdot C_{1}+\beta J_{1}\cdot C_{2}\ \mathrm{%
and}\ q_{2}=J_{2}\cdot C=\alpha J_{2}\cdot C_{1}+\beta J_{2}\cdot C_{2}.
\end{equation}%
For the basis where the K\"{a}hler cone coincides with the first quadrant,
we have
\begin{equation}
J_{1}=\frac{1}{2}D_{4}\ \mathrm{and}\ J_{2}=\frac{3}{4}D_{4}-\frac{1}{2}%
D_{5}\ .
\end{equation}%
By considering the two-moduli THCY model treated above, from \eqref{thcym1}
we find the intersection numbers%
\begin{eqnarray}
C_{1}\cdot D_{0} &=&-1,C_{1}\cdot D_{1}=C_{1}\cdot D_{2}=1,C_{1}\cdot
D_{3}=0,C_{1}\cdot D_{4}=2,C_{1}\cdot D_{5}=3,  \notag \\
C_{2}\cdot D_{0} &=&1,C_{2}\cdot D_{1}=C_{2}\cdot D_{2}=0,C_{2}\cdot
D_{3}=1,C_{2}\cdot D_{4}=0,C_{2}\cdot D_{5}=-2\ .
\end{eqnarray}
Thus, we have $q_{1}=\alpha $ and $q_{2}=\beta $.

The $M2$ brane wrapping the curve $C$ will give rise to a non-BPS black
hole. For simplicity's sake, we will here confine ourselves to deal with a
\textit{doubly-extremal} black hole of charge $q_{I}$, in which thus the
moduli are fixed to the respective attractor value\footnote{%
The generalization to non-doubly extremal but extremal non-BPS black holes
may be discussed by exploiting the so-called \textit{first order formalism},
as recently treated in \cite{Rudelius} (see also Refs. therein).}. The mass $%
M_{C}$ of the black hole is given by the square root of the critical value
of the black hole effective potential:
\begin{equation}
M_{C}=\sqrt{V}\ .
\end{equation}%
Let $C^{\cup }$ be the minimum volume piecewise calibrated representative of
the class $[C]$ and denote $M_{C^{\cup }}$ to be the mass of the $M2$ brane
wrapping $C^{\cup }$. We find
\begin{equation}
M_{C^{\cup }}=\int_{C^{\cup }}J=t^{1}|\alpha |+t^{2}|\beta |\
=t^{1}|q_{1}|+t^{2}|q_{2}|\ .
\end{equation}%
We can rewrite the above as
\begin{equation}
M_{C^{\cup }}=y|q_{2}|\big(1+|q|t\big)\ .
\end{equation}%
Thus, the recombination factor in the present case is given by
\begin{equation}
R=\frac{M_{C}}{M_{C^{\cup }}}=\left. \frac{\sqrt{V}}{y|q_{2}|\big(1+|q|t\big)%
}\right\vert _{t=t_{c}}\ ,
\end{equation}%
where $t_{c}$ denotes the critical value of $t$. Using the expression of the
effective potential in \eqref{vcrit1} we find
\begin{equation}
R=\left. \frac{2^{1/6}\left( 2t^{3}+9t^{2}+9t+3\right) ^{5/6}\sqrt{%
112t^{6}+360t^{5}+441t^{4}+282t^{3}+135t^{2}+54t+9}}{3^{5/6}(t+1)^{2}\left(
4t^{3}-9t^{2}-9t-3\right) (|q|t+1)}\right\vert _{t=t_{c}}\ .
\end{equation}%
Notice that the $q_{2}$ dependence on $V_{\mathrm{cr}}$ drops out in the
ratio, and hence the recombination factor $R$ only depends upon the value of
$q$.

From the discussion below \eqref{nbpseq1} we observe that the non-BPS
solution does not exist for $-2/3<q<1$. There are two branches of solutions
for $q>1$ and for $q<-2/3$. We can numerically evaluate $R$ in both the
branches. The value $q=1$ corresponds to $t=0$. This gives rise to the value
$R=1$. We find that $R$ increases monotonically in this branch. For the
second branch, $R=\sqrt{7}$ for $q=-2/3$. As we decrease the value of $q$
further, $R$ continues to decrease till $q\simeq -4.01$ where it takes the
minimum value $R\simeq 2.06$. It increases beyond this value and rises to $%
R\simeq 2.12$ as $q\rightarrow -\infty $. In both the branches the value of $%
R$ remains greater than $1$ throughout.

In Fig.\ref{m1r1f} we plot $R$ as a function of $q$ in the two branches $%
q<-2/3 $ and $q>1$. As we can see, $R$ increases monotonically in $q>1$
region. In the $q<-2/3$ region, it decreases rapidly till $q\simeq -4.01$
and then increases very slowly. To understand these results better, we also
plot $R$ as a function of the critical value $t_{c}$ in Fig.\ref{m1r2f}. The cusp at $%
t_{c}=t_{\star }\simeq 3.06$ separates the two branches of solutions. From
the graph we can clearly see that $R>1$ throughout, and thus we can conclude
that the non-BPS solution remains unstable for all values of $q$.

Thus, we can conclude that such black holes does \textit{not} enjoy
recombination, and they are unstable, decaying into BPS and anti-BPS
constituents; this confirms the results of \cite{LSVY}, in which \textit{all}
non-BPS black holes were found to be unstable.

\begin{center}
\begin{figure}
\begin{center}
\includegraphics[width=10cm]{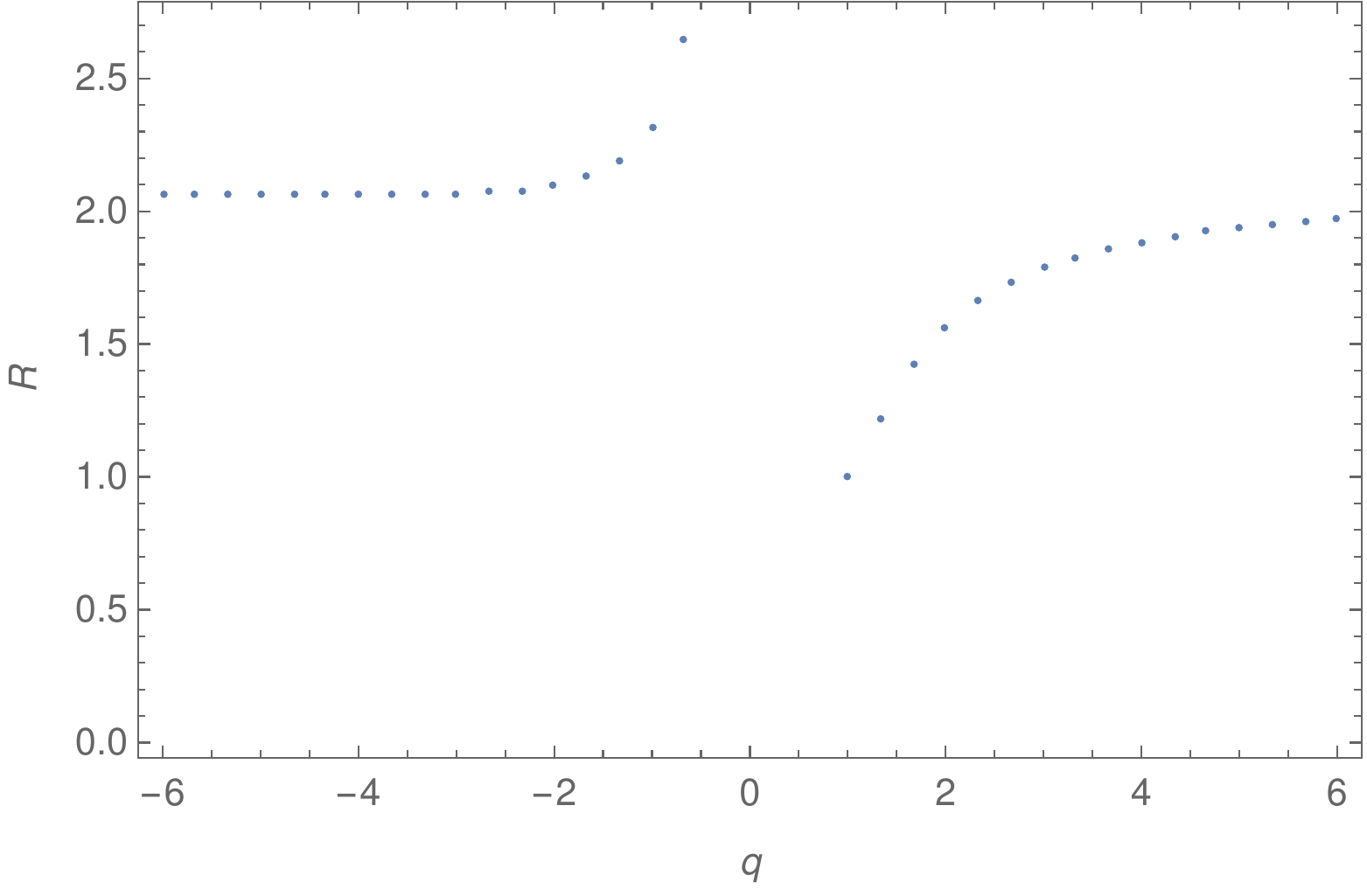}
\caption{Recombination factor for the foregoing non-BPS black hole for
different values of $q$.}
\label{m1r1f}
\end{center}
\end{figure}

\begin{figure}
\begin{center}
\includegraphics[width=10cm]{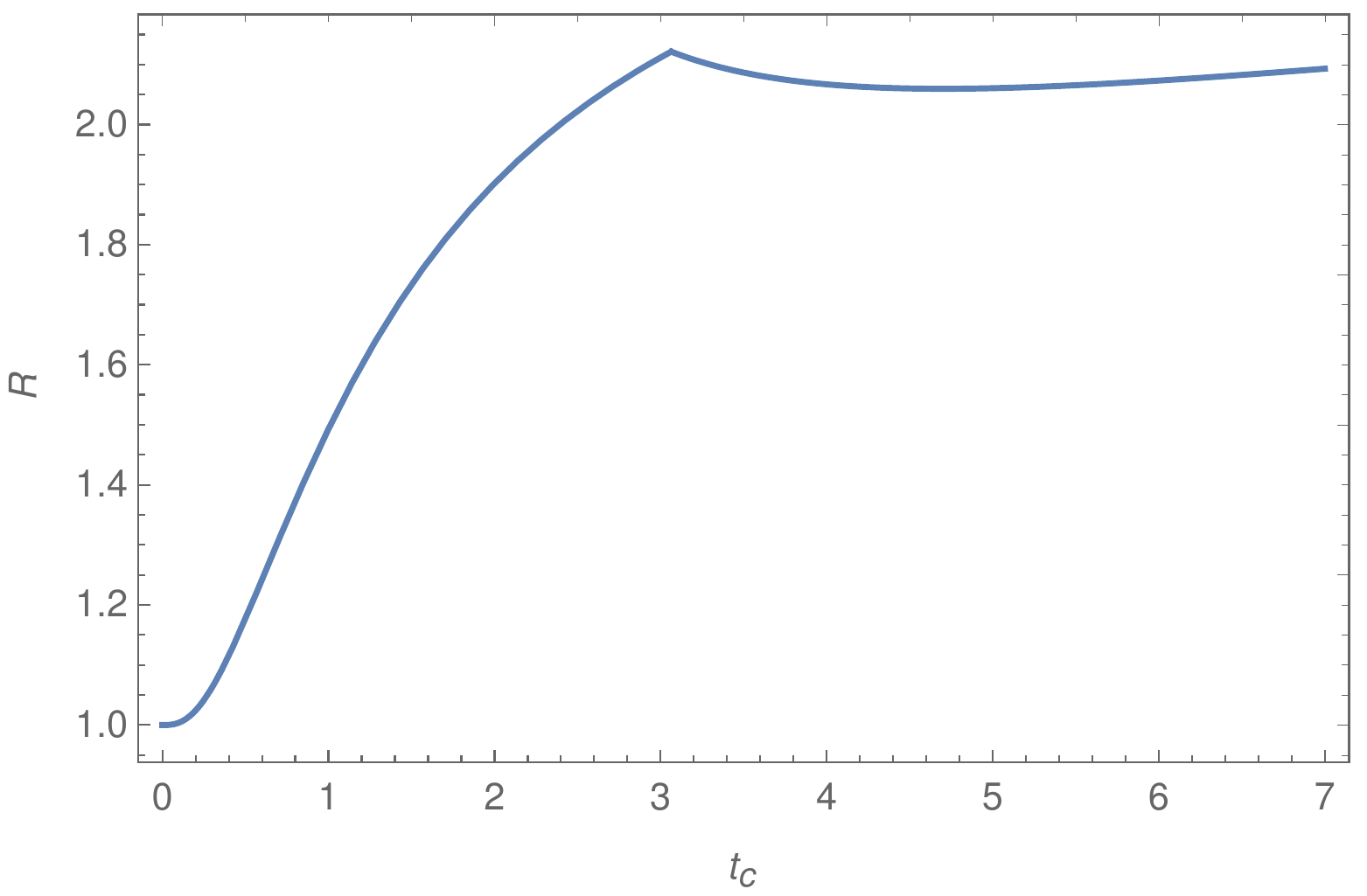}
\caption{Recombination factor for the same non-BPS black hole as a function
of the critical value $t_c$. }
\label{m1r2f}
\end{center}
\end{figure}
\end{center}

\section{\label{5D-BSs}5D Black String Attractors}

We will now turn our attention to BPS as well as non-BPS black string
configurations in five dimensions. We will first consider general analysis
for arbitrary number of moduli. The BPS condition in this case gives rise to
a \textit{unique} solution. We will then develope formalism to study the
non-BPS configurations. Here we will use the conventions of \cite%
{Chou:1997ba}. The black string effective potential in five dimensions is
given by (\ref{V}), with the central charge reading%
\begin{equation}
Z=C_{IJK}p^{I}t^{J}t^{K}.  \label{Zm}
\end{equation}%
This implies that%
\begin{equation}
\partial _{i}Z=2C_{IJK}p^{I}t^{J}\partial _{i}t^{K},
\end{equation}%
and thus, by recalling the definition (\ref{Pij}), the effective potential
can be written as
\begin{equation}
V=Z^{2}+6C_{IJK}p^{I}t^{J}C_{LMN}p^{L}t^{M}g^{ij}\partial _{i}t^{K}\partial
_{j}t^{N}=Z^{2}+6\Pi ^{KN}C_{IJK}C_{LMN}p^{I}p^{L}t^{J}t^{M}.
\end{equation}%
Then, by recalling (\ref{Pij-2}), one can write%
\begin{eqnarray}
V &=&G_{IJ}p^{I}p^{J}=Z^{2}-2\left( C^{KN}-t^{K}t^{N}\right)
C_{IJK}C_{LMN}p^{I}p^{L}t^{J}t^{M}  \notag \\
&=&3Z^{2}-2C_{LMN}p^{N}p^{L}t^{M}=3Z^{2}-2C_{IJ}p^{I}p^{J}.
\end{eqnarray}%
We will use the method of Lagrange multiplier to extremize this potential
subjected to the constraint $C_{IJK}t^{I}t^{J}t^{K}=1$. Extremizing
\begin{equation}
\breve{V}=3Z^{2}-2C_{IJ}p^{I}p^{J}+\lambda (C_{IJK}t^{I}t^{J}t^{K}-1)
\end{equation}%
with respect to $t^{I}$ and $\lambda $ we find
\begin{eqnarray}
0 &=&C_{IJK}t^{I}t^{J}t^{K}-1; \\
0 &=&12ZC_{IJK}p^{J}t^{K}-2C_{IJK}p^{J}p^{K}+3\lambda C_{IJK}t^{J}t^{K}
\notag \\
&=&C_{IJK}\left( 12Zp^{J}t^{K}-2p^{J}p^{K}+3\lambda t^{J}t^{K}\right) .
\end{eqnarray}%
Multiplying by $t^{I}$ in the second of the above, we find
\begin{equation}
3\lambda =-12Z^{2}+2C_{IJK}t^{I}p^{J}p^{K}=-12Z^{2}+C_{IJ}p^{I}p^{J}.
\end{equation}%
Using this value of $\lambda $ we find the equation of motion
\begin{eqnarray}
0
&=&12ZC_{IJK}p^{J}t^{K}-2C_{IJK}p^{J}p^{K}-12Z^{2}C_{IJK}t^{J}t^{K}+2C_{LM}p^{L}p^{M}C_{IJK}t^{J}t^{K}
\notag \\
&=&12ZC_{IJK}t^{K}\left( p^{J}-Zt^{J}\right) -2p^{J}p^{K}\left(
C_{IJK}-C_{MJK}t^{M}C_{ILN}t^{L}t^{N}\right) ,  \label{5deom-2-pre}
\end{eqnarray}%
along with the constraint (\ref{constraint}). Multiplying $C^{IJ}$ we find
\begin{equation}
6Z(p^{J}-Zt^{J})-C^{JK}C_{KLM}p^{L}p^{M}+C_{LM}p^{L}p^{M}t^{J}=0.
\end{equation}%
The equation of motion can also be rewritten in a compact form as
\begin{equation}
C_{ILM}p^{K}\left( 6Zt^{J}-p^{J}\right) \left( \frac{1}{2}\delta
_{J}^{L}\delta _{K}^{M}+\frac{1}{2}\delta _{K}^{L}\delta
_{J}^{M}-t^{L}t^{M}C_{JKN}t^{N}\right) =0.  \label{5deom-2}
\end{equation}%
The supersymmetric critical points correspond to
\begin{equation}
t^{I}=\frac{p^{I}}{Z}.  \label{j}
\end{equation}%
This equation can be rewritten as
\begin{equation}
t^{I}=\frac{p^{I}}{\big(C_{JKL}p^{J}p^{K}p^{L}\big)^{1/3}}.
\end{equation}%
Thus, for a given set of (supporting, magnetic) charges, there \textit{always%
} is a \textit{unique} BPS black string solution.

The non-BPS black holes can be found upon solving (\ref{5deom-2}) such that $%
Zt^{I}-p^{I}\neq 0$. It is worth exploring whether we can obtain an equation
analogous to Eq. (\ref{j}) for the non-BPS critical points. A na\"{\i}ve
analysis of (\ref{5deom-2}) might suggest the non-BPS solutions to the
equation of motion given in terms of $6Zt^{I}=p^{I}$. However, such for such
a solution $C_{IJK}t^{I}t^{J}t^{K}=1/6$ and hence it is not consistent with
the constraint (\ref{constraint}).

In order to obtain the algebraic equation corresponding non-BPS critical
points, we set
\begin{equation}
X^{I}\equiv p^{I}-Zt^{I}  \label{X^}
\end{equation}%
Note that the constraint (\ref{constraint}) implies
\begin{equation}
C_{IJK}t^{I}t^{J}X^{K}=0  \label{jjj}
\end{equation}%
Substituting $p^{I}=X^{I}+Zt^{I}$ in (\ref{5deom-2-pre}) we find
\begin{eqnarray}
0 &=&12ZC_{IJK}X^{J}t^{K}-2\left( X^{J}+Zt^{J}\right) \left(
X^{K}+Zt^{K}\right) \left( C_{IJK}-C_{MJK}t^{M}C_{INP}t^{N}t^{P}\right)
\notag \\
&=&12ZC_{IJK}X^{J}t^{K}  \notag \\
&&-2\left( X^{J}X^{K}+ZX^{J}t^{K}+Zt^{J}X^{K}+Z^{2}t^{J}t^{K}\right) \left(
C_{IJK}-C_{MJK}t^{M}C_{INP}t^{N}t^{P}\right)  \notag \\
&=&12ZC_{IJK}X^{J}t^{K}-2X^{J}X^{K}C_{IJK}-4ZX^{J}t^{K}C_{IJK}-2Z^{2}t^{J}t^{K}C_{IJK}
\notag \\
&&+2X^{J}X^{K}C_{MJK}t^{M}C_{INP}t^{N}t^{P}  \notag \\
&&+4ZX^{J}t^{K}C_{MJK}t^{M}C_{INP}t^{N}t^{P}+2Z^{2}t^{J}t^{K}C_{MJK}t^{M}C_{INP}t^{N}t^{P}
\notag \\
&=&8ZC_{IJ}X^{J}-2C_{IJK}X^{J}X^{K}+2C_{MN}X^{M}X^{N}C_{IJ}t^{J}.  \label{jj}
\end{eqnarray}%
This equation can be simplified a bit upon multiplying both sides of (\ref%
{jj}) with $C^{KI}$ to obtain
\begin{eqnarray}
0
&=&8ZC^{KI}C_{IJ}X^{J}-2C^{KI}C_{IJM}X^{J}X^{M}+2C_{MN}X^{M}X^{N}C^{KI}C_{IJ}t^{J}
\notag \\
&=&8ZX^{K}-2C^{KI}C_{IJM}X^{J}X^{M}+2C_{LM}X^{L}X^{M}t^{K},
\end{eqnarray}%
which can be rewritten as
\begin{equation}
4ZX^I + X^JX^KC_{JKL}\big(t^Lt^I - C^{LI}\big) = 0 .  \label{nbpsbs}
\end{equation}
Solving the above along with the constraint $C_{IJK}t^{I}t^{J}X^{K}=0$ we
can obtain the expression for $X^{I}$ as a function of $t^{I}$ and $p^{I}$.
It might be easier to solve $C_{IJK}t^{I}t^{J}X^{K}=0$ first. This will give
a possible solution for $X^{I}$ up to an overall multiplicative factor. We
can determine it as follows. Multiply both sides of \eqref{jj} with $X^I$
and use the constraint \eqref{jjj} to obtain
\begin{equation}
4ZC_{IJ}X^IX^J - C_{IJK}X^IX^JX^K = 0 .  \label{scal}
\end{equation}
This will determine the overall multiplicative factor in $X^{I}$. We can
substitute the resulting expression back in (\ref{nbpsbs}) to verify that it
holds. The trivial solution $X^{I}=0$ corresponding to BPS critical points
whereas any non-zero solution for $X^{I}$ in the above equation will
correspond to a non-BPS black string.

Note that, from the \textquotedblleft magnetic" version of the
\textquotedblleft new attractor" approach to 5D attractors treated in \cite%
{CFMZ1-D=5, FG-D=5} (see App. \ref{App-1}), (\ref{X^}) can equivalently be
rewritten as%
\begin{equation}
X^{I}=\frac{3^{3/2}}{2^{5/2}}\frac{1}{Z}T^{ijk}\partial _{j}Z\partial
_{k}Z\partial _{i}t^{I}.
\end{equation}

\section{\label{2mbs}Two-moduli Models}

We will now focus our attention to two-moduli models. As before, we will use
the notation $t^{1}=x,t^{2}=y$. From Sec. \ref{2-moduli BHs}, we here report
\begin{equation}
C_{IJ}=\left(
\begin{array}{cc}
ax+by & bx+cy \\
bx+cy & cx+dy%
\end{array}%
\right) ,
\end{equation}%
and its inverse
\begin{equation}
C^{IJ}=\frac{1}{Lxy-Nx^{2}-My^{2}}\left(
\begin{array}{cc}
cx+dy & -(bx+cy) \\
-(bx+cy) & ax+by%
\end{array}%
\right) .
\end{equation}%
Furthermore\footnote{%
The $P$'s are the (un-normalized) \textquotedblleft Jordan
duals\textquotedblright\ \cite{JD} of the magnetic charges $p$'s.}, $%
C_{IJK}p^{J}p^{K}=:(P_{1},P_{2})^{T}$, where we have introduced the notation
\begin{equation}
P_{1}=a(p^{1})^{2}+2bp^{1}p^{2}+c(p^{2})^{2},\
P_{2}=b(p^{1})^{2}+2cp^{1}p^{2}+d(p^{2})^{2}.
\end{equation}%
Thus, we find
\begin{equation}
C^{JK}C_{KLM}p^{L}p^{M}=\frac{1}{(Lxy-Nx^{2}-My^{2})}%
\begin{pmatrix}
P_{1}(cx+dy)-P_{2}(bx+cy) \\
-P_{1}(bx+cy)+P_{2}(ax+by)%
\end{pmatrix}%
.
\end{equation}

The equations of motion can now be expressed as
\begin{eqnarray}
6Z(p^{1}-Zx)+x(xP_{1}+yP_{2})-\frac{P_{1}(cx+dy)-P_{2}(bx+cy)}{%
(Lxy-Nx^{2}-My^{2})} &=&0; \\
6Z(p^{2}-Zy)+y(xP_{1}+yP_{2})+\frac{P_{1}(bx+cy)-P_{2}(ax+by)}{%
(Lxy-Nx^{2}-My^{2})} &=&0.
\end{eqnarray}%
These give rise to two coupled degree seven equations in variables $x$ and $%
y $. They give rise to both BPS as well as non-BPS critical points. The BPS
critical points are obtained upon solving $(p^{1}-Zx)=0=(p^{2}-Zy)$, and
have the exact expression
\begin{eqnarray}
x &=&\frac{p^{1}}{(p^{1}P_{1}+p^{2}P_{2})^{1/3}}; \\
y &=&\frac{p^{2}}{(p^{1}P_{1}+p^{2}P_{2})^{1/3}}.
\end{eqnarray}%
This result can be regarded as the generalization of the treatment of \cite%
{Kallosh:1999mz} (done for the \textit{electric} black holes) to \textit{%
magnetic} black strings.

It is however in general not possible to obtain the non-BPS critical points.
The formulation discussed in the previous section gives rise to a somewhat
simpler set of equations for the non-BPS critical points. We need to solve %
\eqref{nbpsbs} along with the constraint \eqref{jjj}. The constraint %
\eqref{jjj} can be solved for $X^{I}$ in terms of the moduli $t^{I}$ up to
an overall multiplicative factor. To find its value, let $\tilde{X}^{I}$ be
a solution of \eqref{jjj}. Then, substitute $X^{I}=\check{X}\tilde{X}^{I}$
in \eqref{scal} to obtain the value of the multiplicative factor $\check{X}$
as
\begin{equation}
\check{X}=\frac{4ZC_{MN}\tilde{X}^{M}\tilde{X}^{N}}{C_{IJK}\tilde{X}^{I}%
\tilde{X}^{J}\tilde{X}^{K}}.
\end{equation}

For two-moduli models, it is easy to solve the constraint \eqref{jjj}.
Recall from \eqref{cijtj} we have $C_{IJ}t^{J}=(A_{1},A_{2})^{T}$, with $%
A_{1},A_{2}$ given in \eqref{a1a2}. Thus, \eqref{jjj} can be solved to
obtain $X^{I}=\check{X}\tilde{X}^{I}$ with
\begin{equation}
\tilde{X}^{1}=-A_{2}=A^{1},\ \mathrm{and}\ \tilde{X}^{2}=A_{1}=A^{2}\ .
\end{equation}%
Upon using $C_{IJK}t^{I}t^{J}t^{K}=1$ we find that $C_{IJ}A^{I}A^{J}=\mathrm{%
det}C$ and hence
\begin{equation}
\check{X}=\frac{4Z\ {\mathrm{det}C}}{C_{IJK}A^{I}A^{J}A^{K}}.
\end{equation}

Thus, the non-BPS critical point corresponding to black strings for an
arbitrary two-moduli model is given by
\begin{equation}
p^{I}-Zt^{I}-\frac{4Z\ {\mathrm{det}C}}{C_{IJK}A^{I}A^{J}A^{K}}A^{I}=0\ .
\label{bseom}
\end{equation}%
For a given value of $C_{IJK}$ and for a given set of charges, this equation
can be solved numerically to obtain the values of the moduli $t^{I}$
corresponding to a non-BPS critical point.

The effective black hole potential has the expression
\begin{eqnarray}
V &=&3\left[ p^{1}(ax^{2}+2bxy+cy^{2})+p^{2}(bx^{2}+2cxy+dy^{2})\right] ^{2}
\notag \\
&-&2\left[ (p^{1})^{2}(ax+by)+2p^{1}p^{2}(bx+cy)+(p^{2})^{2}(cx+dy)\right] \
.
\end{eqnarray}

By adopting the normalization of \cite{LSVY}, the black string tension $T$
can be determined from the critical value of the effective black hole
potential as\footnote{%
Actually, the effective black string potential discussed in \cite{LSVY}
differs from ours by a factor of $3/2$ and hence we have $\sqrt{V}$ instead
of $\sqrt{\frac{3}{2}V}$.}%
\begin{equation}
T=\sqrt{V}.
\end{equation}%
For BPS solution :
\begin{gather}
V=Z^{2}=\left[ p^{1}(ax^{2}+2bxy+cy^{2})+p^{2}(bx^{2}+2cxy+dy^{2})\right]
^{2}\ ; \\
\Downarrow  \notag \\
T=\left\vert Z\right\vert =\left\vert
p^{1}(ax^{2}+2bxy+cy^{2})+p^{2}(bx^{2}+2cxy+dy^{2})\right\vert .
\end{gather}%
\medskip

We anticipate here that in \textit{all} 36 two-moduli models of CICY type we
will find that \textit{all} non-BPS black string (magnetic) attractors are
\textit{unique}, confirming and generalizing the results of \cite{LSVY}. On
the other hand, in most of the 48 two-moduli models of THCY type we will
find that there exist \textit{multiple} non-BPS black string (magnetic)
attractors, a phenomenon which was not observed in \cite{LSVY} Moreover, by
analising the so-called \textit{recombination factor}, we will also find
evidence for the existence of non-BPS black strings which enjoy \textit{%
recombination}, and thus that are actually stable against the decay into
their BPS/anti-BPS constituent pairs, again confirming and generalizing the
results of \cite{LSVY}.

\subsection{\label{Uniqueness-BSs}Uniqueness of Attractors}

\subsubsection{\label{cdzero-2}$c=d=0$}

Before considering specific models in detail, we would like to note that
also for black string we have \textit{unique} solution for the special case
of $c=d=0$. Upon setting $c=0=d$ and introducing $t=x/y$ and $p=p^{1}/p^{2}$%
, rescaling and solving the non-BPS equation we find
\begin{equation}
t=-\frac{3bp}{3b+2ap}\ .
\end{equation}%
Thus, the black string solution in this case is given by
\begin{eqnarray}
x &=&-\frac{3b+2ap}{3bp^{1/3}(3b+ap)^{2/3}}\ , \\
y &=&\frac{3b+2ap}{3bp^{2/3}(3b+ap)^{1/3}}\ .
\end{eqnarray}%
The K\"{a}hler cone condition is given by $3/p+2a/b<0$ and $a+3b/p<0$. The
tension of the black string is
\begin{equation}
T=\left\vert p^{2}p^{2/3}(3b+ap)^{1/3}\right\vert \ .
\end{equation}

\subsubsection{\label{cdzero-3}$a=b=0$}

A similar analysis can be done for the case $a=b=0$. We find
\begin{equation}
t=-\frac{2d+3cp}{3c}
\end{equation}%
and hence
\begin{equation}
x=\frac{2d+3cp}{3c(d+3cp)^{1/3}}\ ,\ y=-\frac{1}{(d+3cp)^{1/3}}\ .
\end{equation}%
The string tension is
\begin{equation}
T=\left\vert p_{2}(d+3cp)^{1/3}\right\vert \ .
\end{equation}%
The K\"{a}hler cone condition is $d+3cp<0$ and $(2d+3cp)/c<0$.

\subsubsection{\label{mcicy}CICY}

We will now consider black string attractor solutions in the 36 two-moduli
CICY models, already treated in Sec. \ref{CICY}. Since BPS black string
attractors are always \textit{unique}, we will henceforth only analyse the
corresponding non-BPS equations. Once again, we will workout in some detail
the two-moduli CICY model considered in\ Sec. \ref{CICY}.\medskip\
Substituting the values of the intersection numbers for the hypersurface %
\eqref{cicym1} in the non-BPS equations\eqref{bseom} we find%
\begin{eqnarray}
0 &=&p^{1}-\frac{2\left(
6x^{4}+45x^{3}y+93x^{2}y^{2}+156xy^{3}+56y^{4}\right) \left(
p^{1}y(2x+3y)+p^{2}\left( x^{2}+6xy+2y^{2}\right) \right) }{3(2x+3y)\left(
x^{2}+3xy-12y^{2}\right) }, \\
0 &=&2p^{1}y^{2}(2x+3y)\left( 3x^{2}+9xy+40y^{2}\right)  \notag \\
&&+p^{2}\left[ 6x^{4}y+54x^{3}y^{2}+4y^{2}\left( -9+40y^{3}\right)
+x^{2}\left( 3+200y^{3}\right) +x\left( 9y+516y^{4}\right) \right] .
\end{eqnarray}%
Substituting $x=yt$, $p^{1}=p^{2}p$ and then using the constraint $\mathcal{V%
}=1$ from \eqref{vconst}, we find that the above equations take the simple
form
\begin{equation}
\left( 6t^{4}+45t^{3}+93t^{2}+156t+56\right) +p\left(
6t^{3}+27t^{2}+107t+120\right) =0.
\end{equation}%
This is a quartic equation and hence we can write down the exact solution
for $t$ in terms of $p$. However, before doing so we will analyse the above
equation qualitatively. Solving the above for $p$ as a function of $t$ we
find
\begin{equation}
p=-\frac{6t^{4}+45t^{3}+93t^{2}+156t+56}{6t^{3}+27t^{2}+107t+120}
\end{equation}%
As $t\rightarrow 0$ the r.h.s. goes to $-7/15$, and it goes to $-\infty $ in
the limit $t\rightarrow \infty $. Thus, for positive $t$ the value of $p$
must be less than $-7/15$. The r.h.s. is a monotonic function, and hence we
have a \textit{unique} solution with $t>0$ for $p<-7/15$. The string tension
$T=\sqrt{V}$ is given by
\begin{equation}
T=\left( \frac{3}{2}\right) ^{-1/3}|p^{2}|\frac{\sqrt{p^{2}\left(
6t^{2}+18t+23\right) +2p\left( 9t^{2}+8t+6\right)
+3t^{4}+18t^{3}+54t^{2}+24t+4}}{\left( 3t^{2}+9t+2\right) ^{2/3}}.
\end{equation}

As an example, let us consider the value $t=1$. It corresponds to $p=-89/65$%
. The moduli are same as \eqref{t1xy}:
\begin{equation}
x=y=\left( \frac{3}{28}\right) ^{1/3}\ .
\end{equation}%
The corresponding tension of the black string $T=\sqrt{V}$ is given by
\begin{equation}
T=|p^{2}|\ \frac{2^{7/6}\sqrt{1381}}{65}\ \left( \frac{7}{3}\right)
^{1/3}\simeq 1.7\ |p^{2}|\ .
\end{equation}

We will now consider the exact solution. We find
\begin{eqnarray}
t &=&\sqrt{\frac{133\left( p^{2}+3p+7\right) }{4{\mathcal{D}}_{2}(p)}-\frac{%
\left( 24p^{3}+108p^{2}+2746p+3957\right) }{32\sqrt{3{\mathcal{D}}_{3}(p)}}-%
\frac{{\mathcal{D}}_{2}(p)}{16}+\frac{{\mathcal{D}}_{3}(p)}{96}}  \notag \\
&&+\frac{1}{8}\sqrt{\frac{{\mathcal{D}}_{3}(p)}{3}}-\frac{p}{4}-\frac{15}{8}%
\ ,
\end{eqnarray}%
where we have used the notation
\begin{eqnarray}
{\mathcal{D}}_{1}(p) &:&=19\left(
1372p^{6}+12348p^{5}+172731p^{4}+851166p^{3}+1565367p^{2}+1032552p+518096%
\right) \ ,  \notag \\
\mathcal{D}_{2}(p) &:&=76^{1/3}\left( 1425p^{2}+4275p+950+\sqrt{\mathcal{D}%
_{1}(p)}\right) ^{1/3}\ ,  \notag \\
\mathcal{D}_{3}(p) &:&=12p^{2}+36p+179+4{\mathcal{D}}_{2}(p)-\frac{%
2128\left( p^{2}+3p+7\right) }{{\mathcal{D}}_{2}(p)}\ .
\end{eqnarray}%
\medskip

A similar analysis can be carried out for all other 35 two-moduli CICY
models. In App. \ref{cicys2} we report the non-BPS solutions along with
their respective ranges of validity for all such 35 two-moduli CICY models.
In \textit{all} such models, non-BPS black string attractors are \textit{%
unique}.

\subsection{\label{thcym}THCY \textit{Multiple} Black Strings}

In this section we will consider non-BPS black string attractors in the
various two-moduli THCY models, already introduced in\ Sec. \ref{thcy}. Once
again, we will workout in some detail the two-moduli THCY model considered
in\ Sec. \ref{thcy}. Setting $x=ty$ and $p^{1}=pp^{2}$, we obtain the
quartic equation
\begin{equation}
p(8t^{4}+28t^{3}+27t^{2}+3t-3)+t\left( 16t^{3}+45t^{2}+45t+15\right) =0\ .
\label{multstr}
\end{equation}%
To understand the qualitative features of the equation we solve the above
for $p$:
\begin{equation}
p=-\frac{t\left( 16t^{3}+45t^{2}+45t+15\right) }{8t^{4}+28t^{3}+27t^{2}+3t-3}%
.
\end{equation}%
Note that the coefficients in the denominator change sign once. Thus,
according to Descartes' rule, it must admit \textit{at least} one positive
root where the rational polynomial function at the r.h.s. diverges.
Numerically solving the equation
\begin{equation}
8t^{4}+28t^{3}+27t^{2}+3t-3=0,
\end{equation}%
we find that it admits only one positive root $t=t_{\star }\simeq 0.25$. The
rational function is monotonic in the region $0<t<t_{\star }$. It vanishes
at $t=0$, diverges at $t=t_{\star }$ and takes positive values in the
interval $(0,t_{\star })$. Thus, for all $p>0$, we have a \textit{unique}
solution for $t\in (0,t_{\star })$. On the other hand, the rational function
is no longer monotonic for $t>t_{\star }$. Differentiating it with respect
to $t$, we find that the extrema occurs for
\begin{equation}
88t^{6}+144t^{5}-261t^{4}-762t^{3}-675t^{2}-270t-45=0\ .
\end{equation}%
Using Descartes' rule, we find that this also has \textit{at least} one
positive root. Solving numerically, we find the only positive root of the
above equation is at $t=t_{0}\simeq 2.24$. This corresponds to a local
maximum of the rational function. Its value at $t=t_{0}$ is given by $%
p_{0}\simeq -1.78$. On the other hand, $p$ takes the value $-2$ as $%
t\rightarrow \infty $. Numerically solving \eqref{multstr} with $p=-2$, we
find $t=t_{m}\simeq 0.86$. Thus, for all $p<-2$, we have \textit{unique}
black string solutions in the narrow window $t_{\star }<t<t_{m}$. However,
for all $-2<p<p_{0}$ we have double roots for the equation \eqref{multstr}
in the region $t>t_{m}$, thereby leading to \textit{multiple critical points
for the black string attractor}. The values of the moduli $x,y$ in terms of $%
t$ are given by \eqref{vconst}:
\begin{equation}
x=\left( \frac{6t^{3}}{2t^{3}+9t^{2}+9t+3}\right) ^{1/3}\ ,\ y=\left( \frac{6%
}{2t^{3}+9t^{2}+9t+3}\right) ^{1/3}\ .
\end{equation}%
The tension of the black string given by
\begin{equation}
T=\frac{|p^{2}|}{6^{1/3}\left( 2t^{3}+9t^{2}+9t+3\right) ^{2/3}}\left(
\begin{array}{l}
p^{2}\left( 4t^{4}+24t^{3}+54t^{2}+42t+9\right) \\
+6p\left( 2t^{4}+8t^{3}+15t^{2}+12t+3\right) \\
+15t^{4}+42t^{3}+54t^{2}+36t+9%
\end{array}%
\right) ^{1/2}\ .  \label{TT}
\end{equation}%
Here $t$ takes the critical value for a given $p$.

As an example, consider the value $p=-121/63$. It admits two solutions for $%
t $, namely $t_{1}=1$ and $t_{2}\simeq 14.5$. The corresponding values of $%
(x,y)$ are
\begin{equation}
(x_{1},y_{1})=\left( \left( \frac{6}{23}\right) ^{1/3},\left( \frac{6}{23}%
\right) ^{1/3}\right) ,\ (x_{2},y_{2})\simeq (1.31,0.09)\ .
\end{equation}%
The tension for these two solutions are
\begin{equation}
T_{1}=\sqrt{\frac{199}{567}}\left( \frac{23}{6}\right) ^{1/3}\ |p^{2}|\simeq
0.93\text{ }|p^{2}|,
\end{equation}%
and
\begin{equation}
T_{2}\simeq 0.89\text{ }|p^{2}|\ .
\end{equation}

We will now write down the exact expression for both these solutions. They
are given by%
\begin{eqnarray}
2t_{\pm } &=&\sqrt{\mathcal{D}_{3}(p)}-\frac{28p+45}{16(p+2)}  \notag \\
&&\pm \sqrt{\frac{3\left( 208p^{2}+408p+105\right) }{256(p+2)^{2}}+\frac{%
11\left( 64p^{3}+720p^{2}+1476p+705\right) }{2048(p+2)^{3}\sqrt{\mathcal{D}%
_{3}(p)}}-\mathcal{D}_{3}(p)},
\end{eqnarray}
where we have introduced the notation%
\begin{eqnarray}
\mathcal{D}_{1}(p)
&:&=-1616p^{6}+10152p^{5}+60777p^{4}+104634p^{3}+84375p^{2}+33750p+5625,
\notag \\
\mathcal{D}_{2}(p) &:&=\left( \frac{3}{2}\right) ^{1/3}\left(
50p^{3}+225p^{2}+225p+75+\sqrt{\mathcal{D}_{1}(p)}\right) ^{1/3},  \notag \\
\mathcal{D}_{3}(p) &:&=\frac{21p(p+1)}{8(p+2)\mathcal{D}_{2}(p)}+\frac{%
\mathcal{D}_{2}(p)}{8(p+2)}+\frac{208p^{2}+408p+105}{256(p+2)^{2}}.
\end{eqnarray}%
\medskip

A similar analysis can be carried out for all other 37 two-moduli THCY
models. The results are summarized in App. \ref{thcy2}, and they provide
evidence for a remarkable difference with respect to the results of \cite%
{LSVY}; indeed, interestingly, \textit{we find multiple black string non-BPS
attractor solutions for most of the models}. Intriguingly, we also find that
the (local) minimum value of the effective black string potential (and thus,
the tension of the non-BPS black string) is \textit{different }for \textit{%
different}, \textit{multiple} attractors; again, this fact highlights a new
phenomenon with respect to the findings of \cite{LSVY} : at a geometrical
level, this should correspond to connected locally volume-minimizing
representatives of the (non-BPS) homology class having different values of
their (local) minimal volumes as a function of the moduli.

\subsection{\label{BS-rec}Non-BPS Black Strings: Recombination Factor and
Stability}

As done in Sec. \ref{BH-rec} for black hole attractors, here also we can
introduce the recombination factor in order to study the stability of
non-BPS black string doubly-extremal. Analogously, the recombinaton factor $R
$ is defined by the ratio of the black string tension to that of the minimum
piecewise calibrated representative in the same homology class as the black
string. For $R>1$ the non-BPS black string decays into constituent
BPS-anti-BPS pairs, whereas for $R<1$ we have a stable black string as a
result of recombination.

A black string of charge $p^{I}=(p^{1},p^{2})$ is obtained upon wrapping an $%
M5$ brane on the divisor $D=p^{1}J_{1}+p^{2}J_{2}$. For a double extremal
solution, the string tension $T$ is given by the square root of the
corresponding effective potential
\begin{equation}
T=\sqrt{V}|_{t=t_{c}}\ .
\end{equation}%
Let $D^{\cup }$ be the minimum volume piecewise calibrated representative of
the class $[D]$ with volume $V_{D^{\cup }}$. Then,
\begin{equation}
V_{D^{\cup }}=|p^{1}|\int_{J_{1}}J\wedge J+|p^{2}|\int_{J_{2}}J\wedge J\ ,
\end{equation}%
which gives rise to
\begin{equation}
V_{D^{\cup }}=|p^{2}|(A_{2}+|p|A_{1})|_{t=t_{c}},  \label{vdu}
\end{equation}%
where $A_{1}$ and $A_{2}$ are defined in \eqref{a1a2}. The recombination
factor is given by the ratio
\begin{equation}
R=\left. \frac{\sqrt{V}}{V_{D^{\cup }}}\right\vert _{t=t_{c}}\ .
\end{equation}

In the following treatment, we compute the recombination factor for non-BPS
black strings in the two-moduli THCY model treated above. We start and
recall here Eq. (\ref{TT}) :
\begin{equation}
T=\frac{|p^{2}|}{6^{1/3}\left( 2t^{3}+9t^{2}+9t+3\right) ^{2/3}}\cdot \left(
\begin{array}{l}
p^{2}\left( 4t^{4}+24t^{3}+54t^{2}+42t+9\right) \\
+6p\left( 2t^{4}+8t^{3}+15t^{2}+12t+3\right) \\
+15t^{4}+42t^{3}+54t^{2}+36t+9%
\end{array}%
\right) ^{1/2}\ .  \label{eq1}
\end{equation}%
On the other hand, substituting the value of the intersection numbers in %
\eqref{vdu} we find
\begin{equation}
V_{D^{\cup }}=\frac{|p^{2}|\left( \left( 2t^{2}+6t+3\right) \left\vert
p\right\vert +3(t+1)^{2}\right) }{6^{1/3}\left( 2t^{3}+9t^{2}+9t+3\right)
^{2/3}}\ .  \label{eq2}
\end{equation}%
Taking the ratio, we find the expression for the recombination factor
\begin{equation}
R=\frac{1}{\left( 2t^{2}+6t+3\right) \left\vert p\right\vert +3(t+1)^{2}}%
\left(
\begin{array}{l}
p^{2}\left( 4t^{4}+24t^{3}+54t^{2}+42t+9\right) \\
+6p\left( 2t^{4}+8t^{3}+15t^{2}+12t+3\right) \\
+15t^{4}+42t^{3}+54t^{2}+36t+9%
\end{array}%
\right) ^{1/2}.  \label{eq3}
\end{equation}%
Note that in Eqs. (\ref{eq1})-(\ref{eq3}) $t$ takes its critical value. The
plot for $R$ in the range $p<-2$ and $p>0$ are shown below in Figs.\ref{bsr1f}-\ref{bsr3f}.

\begin{figure}\centering
\includegraphics[width=4in]{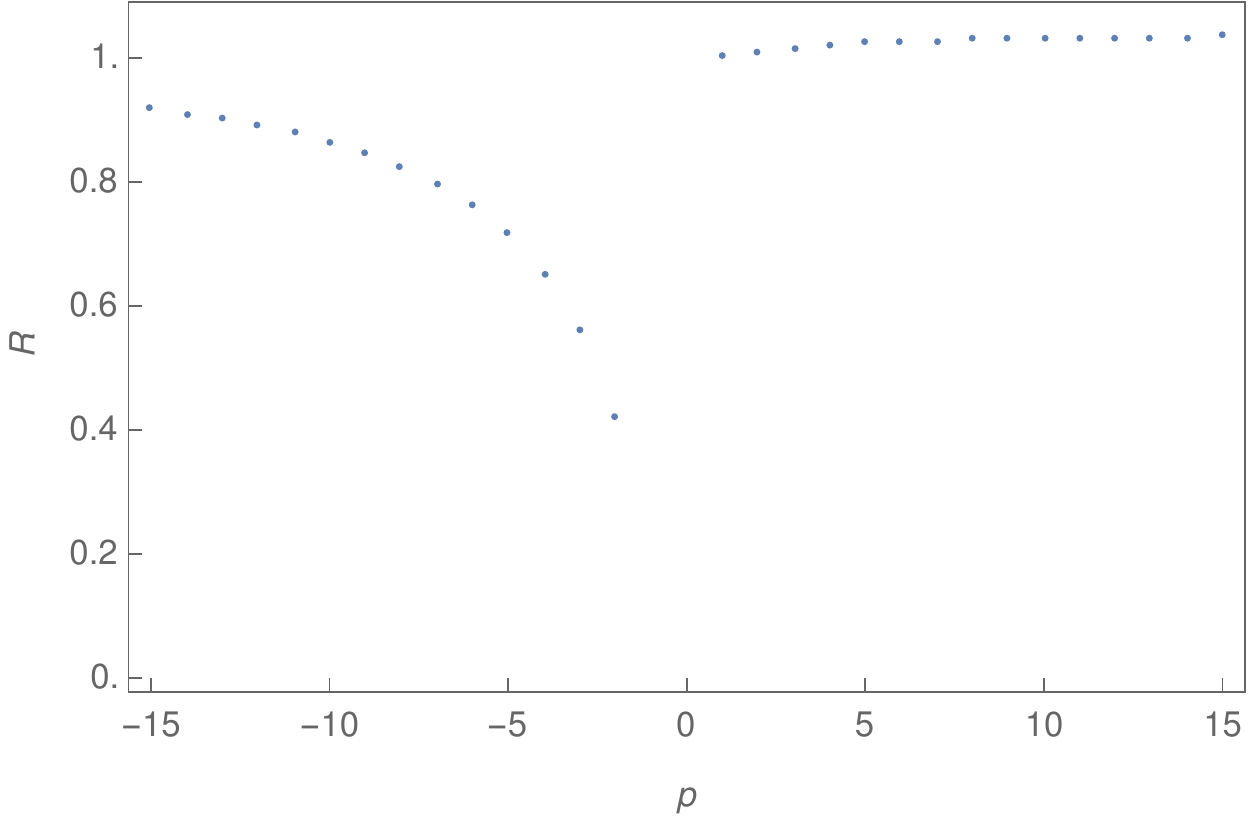}
\caption{The recombination factor of the foregoing non-BPS black string for various values
of $p$.}
\label{bsr1f}
\end{figure}

\begin{figure}\centering
\includegraphics[width=4in]{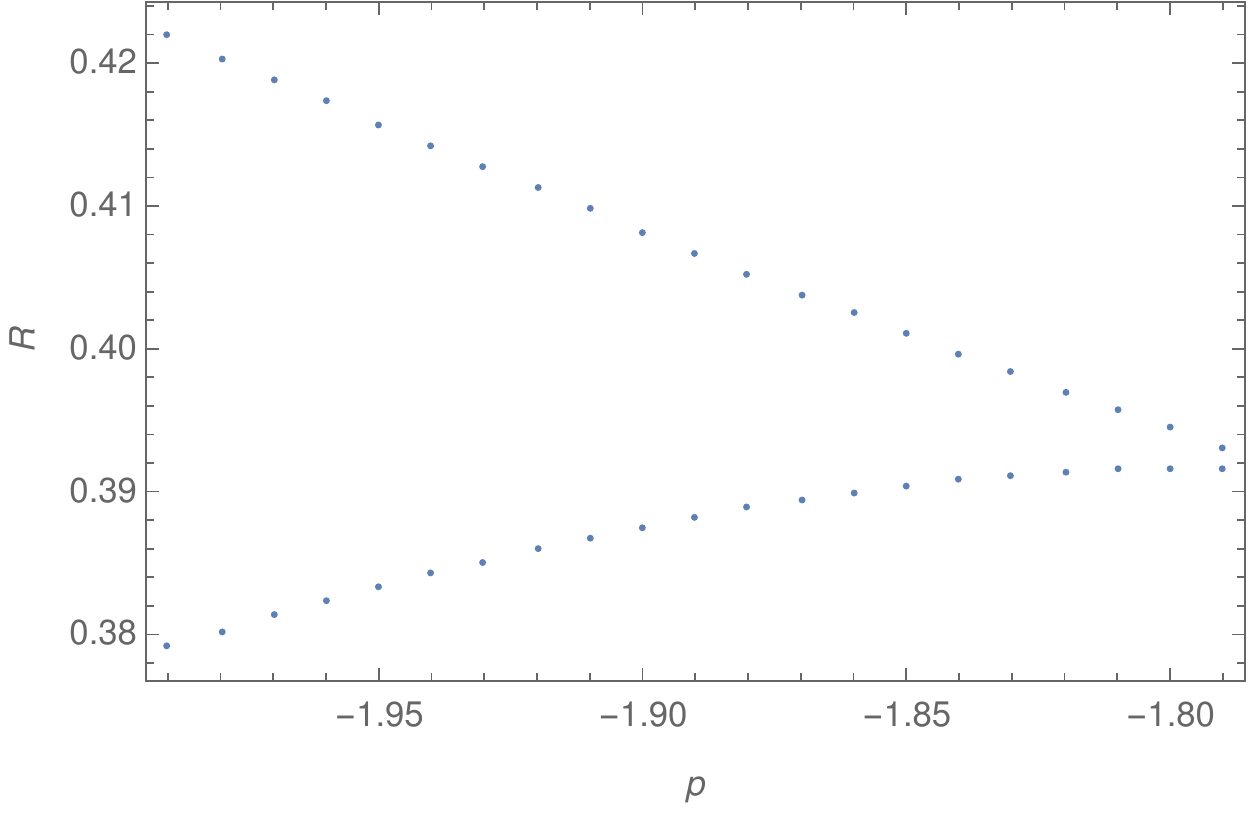}
\caption{The recombination factor for various values of $p$ supporting multiple non-BPS black strings. }
\label{bsr2f}
\end{figure}

\begin{figure}\centering
\includegraphics[width=4in]{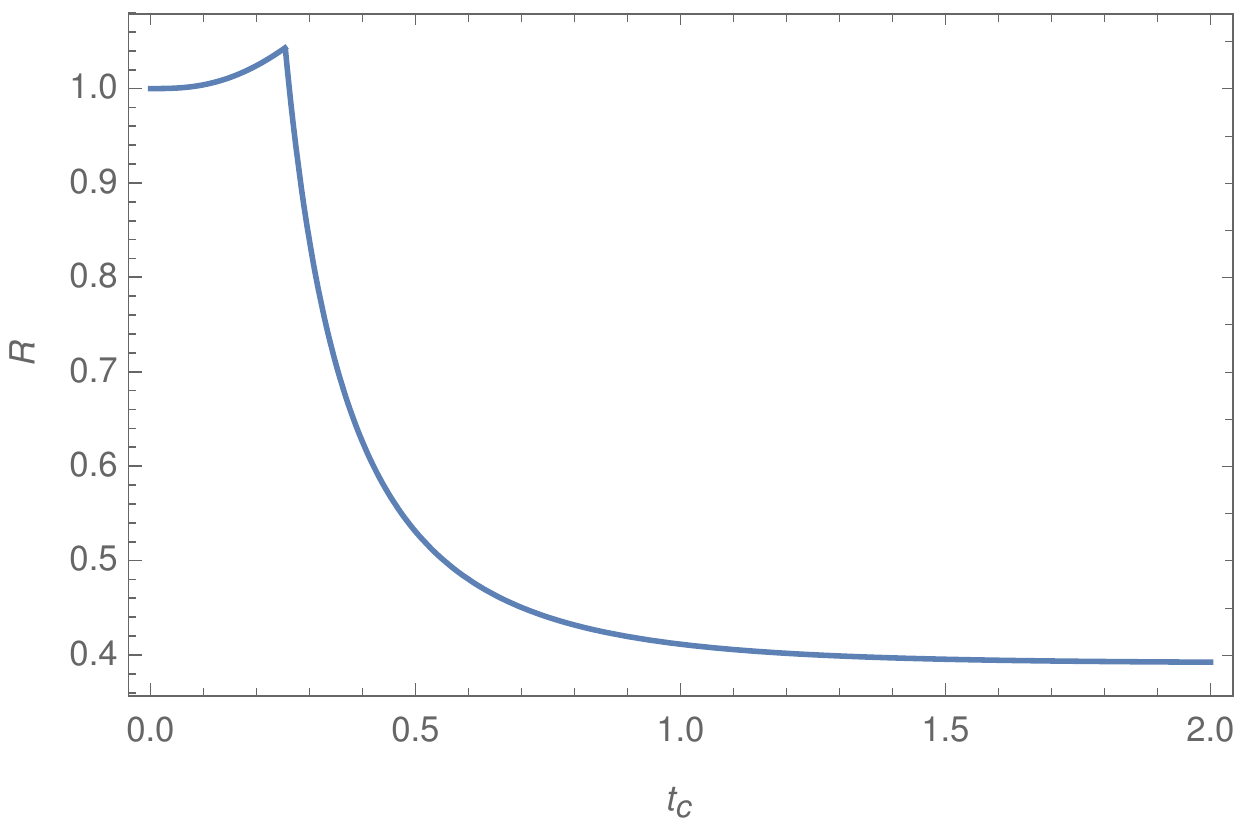}
\caption{The recombination factor of the same non-BPS black string(s) as a function of the critical value $t_c$.}
\label{bsr3f}
\end{figure}

For $p>0$ and for large negative $p$ (i.e. $p=p_{m}\lesssim -44.856$) the
value of $R$ is greater than $1$ and the solution remains unstable. Whereas
for $p$ in the range $p_{m}<p<p_{0}$ we have stable non-BPS attractor. Thus,
in some, suitable ranges of the supporting magnetic charges, black strings
do enjoy \textit{recombination}, and they are thus stable against the
decaying into their BPS/anti-BPS constituents.

It is here worth remarking that there exist \textit{at least} some subsector
of the supporting magnetic charge ratio $p$ for which the \textit{multiple}
non-BPS black string attractors (if any) are \textit{stable}. This adds
interest and physical relevance to the discovery of \textit{multiple}
non-BPS black string attractors, which is a result of the present paper (and
which, for instance, was not observed in \cite{LSVY}). Indeed, in \cite{LSVY}
stable non-BPS (doubly-extremal) black strings were observed, but they were
unique. Here, we have discovered multiple (and, in the doubly-extremal case,
\textit{stable}) non-BPS black strings in most of two-moduli THCY models.

\section{\label{Conclusions}Conclusions}

Motivated by the relevance of extremal non-BPS black holes and black strings
for the Weak Gravity Conjecture (WGC) \cite{Arkani-Hamed:2006emk} and by the
recently established \cite{LSVY} evidence for stable non-BPS remnants of
black strings in minimal, $N=2$ five-dimensional supergravity, we have
investigated non-BPS attractors in the low-energy limit of compactifications
of M-theory on Calabi-Yau threefolds with $h_{1,1}=2$ moduli.

On one hand, by computing the so-called \textit{recombination factor} for
doubly-extremal non-BPS black holes, we have confirmed the results of \cite%
{LSVY} in the whole set of CICY and THCY two-moduli models (respectively
made of 36 and 48 models) : for a given, supporting electric charge
configuration, non-BPS extremal black holes are always \textit{unstable} and
\textit{unique}. This means that such solutions correspond to local, but not
global, volume minimizers of the corresponding curve classes, as there is
always a disconnected, piecewise-calibrated representative (union of
holomorphic and anti-holomorphic curves) which corresponds to the
BPS/anti-BPS black hole constituents, and whose smaller volume implies that
the WGC is satisfied, yielding to the decay of non-BPS black holes into
widely-separated BPS and anti-BPS particles. In other words, non-BPS black
holes are bound to decay to BPS and anti-BPS constituents, and, by sweeping
all two-moduli CICY and THCY models, we have found, as in \cite{LSVY}, no
examples of macroscopic black holes whose mass predicts a stable remnant
microscopic black hole coming from Calabi-Yau threefolds.

On the other hand, we have extended the results of \cite{LSVY} concerning
the stability of some non-BPS (doubly-extremal) black string solutions,
confirming the existence of a phenomenon called \textit{\textquotedblleft
recombination\textquotedblright } for a large fraction of the whole set of
CICY and THCY two-moduli models : in such a phenomenon,\ holomorphic and
anti-holomorphic constituents of the same homology class fuse together to
make a smaller cycle, and by the WGC this yields to the prediction that
there should be microscopic and \textit{stable}, non-BPS black strings (with
small charge) in the spectrum of the resulting supergravity theory. Thus,
extremal non-BPS configurations, at least for large charges, may have robust
features similar to what one sees for supersymmetric, BPS states. However,
it should be here remarked that, for a given supporting magnetic charge
configuration, in all CICY two-moduli models all black string solutions have
been found to be \textit{unique}.

A new evidence, which constitute the novel contribution of the present
investigation to the study of non-BPS attractors within the WGC, is the
\textit{non-uniqueness} of non-BPS, extremal (\textit{stable}) black strings
in most of two-moduli THCY models : for a given, supporting magnetic charge
configurations, in many models there exist multiple non-BPS black string
attractors with different tensions, which are stable : in these models,
recombination occurs in presence of connected, locally volume-minimizing,
representatives of the same (non-BPS) homology class having different values
of their (local) minimal volumes as a function of the moduli. All this begs
for a clearer mathematical explanation. From a physical perspective, by the
WGC, a given (small) magnetic charge configuration may support multiple
non-BPS, extremal black string solutions which are stable against the decay
into constituent BPS/anti-BPS black string constituents.

As speculated in \cite{LSVY}, a possible explanation for such a difference
between non-BPS black holes and non-BPS black strings may lie in the fact
that black holes correspond to (M2 branes wrapping) 2-cycles, which are thus
less than half of the dimension of the Calabi-Yau threefold, whereas black
strings correspond to (M5 branes wrapping) 4-cycles, with dimension bigger
than half of the dimension of the Calabi-Yau threefold. This might hint,
\textit{at least} for the class of two-moduli THCY models which has been
investigated here, for an intersection for higher dimensional cycles due to
local instability modes localized where holomorphic and anti-holomorphic
cycles intersect.

Many different developments may be considered, starting from the present
paper. For instance, it would be interesting to compute the recombination
factor for non-BPS black holes and black strings which are extremal but not
douby-extremal; as pointed out along the treatment, this can be done by
exploiting the so-called first order formalism for extremal solutions, as
recently done in \cite{Rudelius}. Also, one could consider to carry out an
extensive analysis over other classes of $h_{1,1}\geqslant 3$-moduli
Calabi-Yau threefolds, starting from the topological data provided by
existing classifications (see e.g. \cite{CY3-1} and \cite{CY3-2} for recent
studies). Concerning non-uniqueness of attractors (in the same basin of
attraction of the moduli space), by exploiting a Kaluza-Klein
compactification from 5 to \ 4 space-time dimensions, it would be
interesting to relate the multiple non-BPS, extremal and stable black
strings found in the present work with the multiple four-dimensional
extremal black hole attractors related to non-trivial involutory matrices,
as found in \cite{multiple-1, multiple-2}.

\section*{Acknowledgments}

We would like to thank Cody Long for useful comments, and Cumrun Vafa for correspondence. The work of A. Marrani is supported by a \textquotedblleft Maria Zambrano" distinguished researcher fellowship, financed by the European
Union within the NextGenerationEU program.


\appendix

\section{5D Electric \textquotedblleft New Attractor\textquotedblright
Approach \label{App-1}}

In this Appendix we will recall the so-called \textquotedblleft new
attractor\textquotedblright\ approach to the attractor equations of extremal
(electric) black holes in $N=2$, $D=s+t=4+1$ Maxwell-Einstein supergravity
coupled to $n$ Abelian vector multiplets \cite{CFMZ1-D=5,FG-D=5}.

We start and observe that the metric $G_{IJ}$ of the $\left( n+1\right) $%
-dimensional \textquotedblleft ambient space\textquotedblright , which is
the pull-back of the metric $g_{ij}$ of the scalar manifold $\mathcal{M}$ as
well as the canonical metric associated to the cubic form $%
C_{IJK}t^{I}t^{J}t^{K}$, reads%
\begin{equation}
G_{IJ}=t_{I}t_{J}+\frac{3}{2}g^{ij}\partial _{i}t_{I}\partial _{j}t_{J},
\end{equation}%
and its inverse reads%
\begin{equation}
G^{IJ}=t^{I}t^{J}+\frac{3}{2}g^{ij}\partial _{i}t^{I}\partial _{j}t^{J},
\end{equation}%
where%
\begin{equation}
t_{I}=C_{IJK}t^{J}t^{K}.
\end{equation}%
Thus, the following identity holds in projective special real geometry :%
\begin{equation}
\delta _{J}^{I}=t_{J}t^{I}+\frac{3}{2}g^{ij}\partial _{i}t^{I}\partial
_{j}t_{J}.  \label{id-id}
\end{equation}%
By contracting such an identity with $q_{I}$ and defining the (electric)
central charge function as
\begin{equation}
Z:=t^{I}q_{I}\Rightarrow \partial _{i}Z=\partial _{i}t^{I}q_{I},
\end{equation}%
one obtains the following identity\footnote{%
The \textquotedblleft $+$\textquotedblright\ in front of the second term of
the r.h.s. of (\ref{deliz}) corrects a typo e.g. in (3.4) of \cite{CFMZ1-D=5}%
.} :
\begin{equation}
q_{I}=t_{I}Z+\frac{3}{2}g^{ij}\partial _{i}t_{I}\partial _{j}Z\ ,
\label{deliz}
\end{equation}%
holding in the background of an extremal black hole.

For BPS black hole attractors,%
\begin{equation}
\partial _{j}Z=0\Rightarrow \partial _{j}V=0,  \label{BPS-AEs}
\end{equation}%
and hence, from the identity (\ref{deliz}) one finds
\begin{equation}
q_{I}=t_{I}Z,
\end{equation}%
which is an algebraic, equivalent re-writing of the electric BPS attractor
equations (\ref{BPS-AEs}) (cfr. e.g. (3.14) of \cite{CFMZ1-D=5}).

On the other hand, for non-BPS black hole attractors,%
\begin{equation}
\left.
\begin{array}{r}
\partial _{j}V=0 \\
\partial _{j}Z\neq 0~\text{for~some~}j%
\end{array}%
\right\} \Rightarrow \partial _{j}Z=\frac{1}{2Z}\sqrt{\frac{3}{2}}%
T_{jkl}g^{km}g^{lp}\partial _{m}Z\partial _{p}Z,  \label{non-BPS-AEs}
\end{equation}%
where%
\begin{equation}
T_{ijk}=-\left( \frac{3}{2}\right) ^{3/2}\partial _{i}t^{I}\partial
_{j}t^{J}\partial _{k}t^{K}C_{IJK}\ .
\end{equation}%
Thus, from the identity (\ref{deliz}) one finds\footnote{%
The \textquotedblleft $+$\textquotedblright\ in front of the second term of
the r.h.s. of (\ref{deliz2}) corrects a typo e.g. in (3.15) of \cite%
{CFMZ1-D=5}.}%
\begin{equation}
q_{I}=t_{I}Z+\frac{1}{2}\left( \frac{3}{2}\right) ^{3/2}\frac{1}{Z}%
T^{imp}\partial _{i}t_{I}\partial _{m}Z\partial _{p}Z\ ,  \label{deliz2}
\end{equation}%
where $T^{imp}:=g^{ij}g^{km}g^{lp}T_{jkl}$. Clearly, (\ref{deliz2}) is an
equivalent re-writing of the non-BPS electric attractor equations (\ref%
{non-BPS-AEs}).

\section{5D Magnetic \textquotedblleft New Attractor\textquotedblright
Approach \label{App-1-magn}}

In this Appendix we will present recall the so-called \textquotedblleft new
attractor\textquotedblright\ approach to the attractor equations of extremal
(magnetic) black strings in $N=2$, $D=s+t=4+1$ Maxwell-Einstein supergravity
coupled to $n$ Abelian vector multiplets, which has not been considered e.g.
in \cite{CFMZ1-D=5,FG-D=5}, and which for non-BPS attractors, as far as we
know, has never been presented in the literature.

By contracting the identity (\ref{id-id}) with $p^{J}$ and defining the
(magnetic) central charge function as
\begin{equation}
Z:=t_{I}p^{I}\Rightarrow \partial _{i}Z=\partial _{i}t_{I}p^{I},
\end{equation}%
one obtains the following identity :
\begin{equation}
p^{I}=t^{I}Z+\frac{3}{2}g^{ij}\partial _{i}t^{I}\partial _{j}Z\ ,
\label{deliz3}
\end{equation}%
holding in the background of an extremal black string solution.

For BPS black string attractors,%
\begin{equation}
\partial _{j}Z=0\Rightarrow \partial _{j}V=0,  \label{BPS-AEs-2}
\end{equation}%
and hence, from the identity (\ref{deliz3}) one finds
\begin{equation}
p^{I}=t^{I}Z,
\end{equation}%
which is an algebraic, equivalent re-writing of the magnetic BPS attractor
equations (\ref{BPS-AEs-2}).

On the other hand, for non-BPS black string attractors,%
\begin{equation}
\left.
\begin{array}{r}
\partial _{j}V=0 \\
\partial _{j}Z\neq 0~\text{for~some~}j%
\end{array}%
\right\} \Rightarrow \partial _{j}Z=\frac{1}{2Z}\sqrt{\frac{3}{2}}%
T_{jkl}g^{km}g^{lp}\partial _{m}Z\partial _{p}Z.  \label{non-BPS-AEs-2}
\end{equation}%
Thus, from the identity (\ref{deliz3}) one finds%
\begin{equation}
p^{I}=t^{I}Z+\frac{1}{2}\left( \frac{3}{2}\right) ^{3/2}\frac{1}{Z}%
T^{imp}\partial _{i}t^{I}\partial _{m}Z\partial _{p}Z\ ,  \label{deliz4}
\end{equation}%
which is an equivalent re-writing of the non-BPS magnetic attractor
equations (\ref{non-BPS-AEs-2}).

\section{\label{cicymodel}CICY Black Holes}

In this Appendix we will report the results of the study of extremal black
hole attractors in two-moduli complete intersection Calabi-Yau (CICY)
models, recently discussed in \cite{Ruehle} (cfr. App. A therein). For two
moduli, CICY's are given by intersections of hypersurfaces in an ambient
space of the form $\mathcal{A}=\mathbb{P}^{n_{1}}\times \mathbb{P}^{n_{2}}$.
As we have mentioned at the start of Sec. \ref{CICY}, the Calabi-Yau
manifold is specified by a configuration matrix. Each column of the
configuration matrix represents the bi-degree of a polynomial whose zero
\textit{locus} defines a hypersurface in $\mathcal{A}$. The common zero
\textit{locus} of all these polynomials becomes a Calabi-Yau manifold
provided the sum of the $i^{\mathrm{th}}$ row elements of the configuration
matrix takes the value $n_{i}+1$.

In Tables \ref{cicyt1}-\ref{cicyt3}, we report the extremal black hole
attractor solutions for 20 two-moduli CICY models. In particular, the first
model of Table \ref{cicyt1} is the one explicitly treated in Sec. \ref{CICY}%
. The first column of Tables \ref{cicyt1}-\ref{cicyt3} indicates the CICY
label as well as its configuration matrix, as from \cite{Ruehle}. The
superscripts in the configuration matrix are the Hodge numbers $h_{1,1}=2$
and $h_{2,1}$, respectively, whereas the subscript is the Euler number $\chi
:=2\left( h_{1,1}-h_{2,1}\right) $ of the CICY model. On the other hand, as
indicated in the first row of Tables \ref{cicyt1}-\ref{cicyt3}, in the
various partitions of their second column we respectively specify : the
intersection numbers of the CICY model, the critical values of $t=x/y$ as a
function of the charge ratio $q=q_{1}/q_{2}$ for the BPS black hole
attractors, along with the range of $q$ for which the BPS moduli lie inside
the K\"{a}hler cone, the expression for $q$ as a rational polynomial
function of $t$ for the non-BPS black hole attractors, along with the range
of $q$ for which the non-BPS moduli lie within the K\"{a}hler cone, and
finally the expression for the recombination factor. Note that in all such
Tables $t$ takes its critical value.

Our findings show that \textit{there are no multiple extremal BPS black holes%
} for a given value of (supporting electric charge ratio) $q$. Analogously,
for extremal non-BPS attractors, for any given $q$ within the specified
range there is a \textit{unique} $t$ inside the K\"{a}hler cone; thus, all
20 two-moduli CICY models under consideration give \textit{unique} non-BPS
black hole attractors. Moreover, we should remark that all BPS and non-BPS
solutions are \textit{mutually exclusive}. Except for the bi-cubic model
(which is the last model treated in Table \ref{cicyt3}), in which any given $%
q$ supports either a BPS or a non-BPS black hole attractor, the range of
allowed values of $q$ is finite.

We have numerically evaluated the recombination factor for the non-BPS black
hole attractors in the entire moduli space for all allowed values of $q$; we
have found that the recombination factor is always greater than $1$ for all
the 20 two-moduli CICY models under consideration : therefore, all non-BPS
black holes in the 20 two-moduli CICY models listed in Tables \ref{cicyt1}-%
\ref{cicyt3} are \textit{unstable}.\medskip

Since the two-moduli CICY models have been classified in a set of 36 models
(cfr. e.g. \cite{Ruehle}), a natural question arises : What about the
remaining 16 two-moduli CICY models, not reported in Tables \ref{cicyt1}-\ref%
{cicyt3}? All such unlisted models have either $c=d=0$ or $a=b=0$, and thus
the uniqueness of their BPS and non-BPS black hole attractors has been
discussed in Secs. \ref{cdzero} and \ref{cdzero2}, respectively. Moreover,
in such models the recombination factor of non-BPS black holes can be
computed exactly. For $a=b=0$, we find
\begin{equation}
R=3\left( \theta (-q)\left\vert \frac{3c-dq}{9c+dq}\right\vert +\theta
(q)\right) \ ,
\end{equation}%
where $\theta (q)$ is the Heavyside step function. As we have noticed in
Sec. \ref{cdzero2}, for the attractor solution to lie within the K\"{a}hler
cone, one must have $d>3c/q$ and $3/q+d/c<0$. Thus, for $c,d>0$, which is
the case in all models, the first condition $d>3c/q$ is automatically
satisfied, whereas the second condition implies that $q$ must be negative.
In addition, we must have $-3<dq/c<0$. Using this, we find that $1<R<3$, and
hence \textit{all} non-BPS black hole attractor solutions are \textit{%
unstable}.

Similarly, for $c=d=0$, we find
\begin{equation}
R=3\left( \theta (-q)\left\vert \frac{a-3bq}{a+9bq}\right\vert +\theta
(q)\right) \ .
\end{equation}%
Once again, by imposing the K\"{a}hler cone conditions $a/b+3q<0$ and $a>3bq$%
, we obtain that $R>1$, and hence \textit{all} non-BPS black hole attractor
solutions are \textit{unstable}.

Let us observe that :

\begin{itemize}
\item The two-moduli CICY models labelled by 7643 and 7668, i.e. the first
and the third model of Table \ref{cicyt1}, have the matrices of intersection
numbers reciprocally proportional, and they also share the same entries in
all partitions of the second column, such as the same expressions for $t$ as
a function of $q$ for BPS attractors, as well as the same expressions of $q$
as a rational polynomial function of $t$ for non-BPS attractors. However,
the fact that their intersection numbers are different (notwithstanding
being proportional) implies that the cubic constraint defining the 5D scalar
manifold $\mathcal{M}$ (i.e. $C_{IJK}t^{I}t^{J}t^{K}=1$) yields different
expressions for the moduli, and the similarity of the various expressions
observed here is a mere coincidence. Furthermore, such models also have
different $h_{2,1}$ (and thus different $\chi $), as well as different
configuration matrices.
\end{itemize}

\afterpage{
\clearpage
\thispagestyle{empty}
\begin{table}
  \centering
    \begin{tabular}{ || c | c |c| c|| }
\hline\multirow{3}{*} {$\begin{matrix}  \cr {\rm CICY\ label,} \cr
{\rm Configuration\ Matrix}
\end{matrix}$}
& $\begin{pmatrix}
c&d\cr b&a
\end{pmatrix}$ & BPS solution  & Non-BPS solution  \\
\cline{3-4} & & Range of validity & Range of validity \\
\cline{2-4} &\multicolumn{3}{|c||}{ Recombination factor $
\begin{matrix} {} \cr {} \cr {} \end{matrix}$} \\
\hline\hline
\multirow{3}{*} {$\begin{matrix}  \cr 7643 \cr
\begin{pmatrix}
 0 &0&2 & 1 \cr  2 & 2 & 1 & 1
\end{pmatrix}^{2,46}_{-88}\end{matrix}$}
& $\frac{2}{3}
\begin{pmatrix}
3&2\cr 1&0
\end{pmatrix}$ & $\frac{1-3q+\sqrt{7 q^2-3 q+1}}{q}$
 & $-\frac{12 t^4+72 t^3+181 t^2+219 t+284}{6 t^5+27 t^4+141 t^3+444 t^2+638 t+72}$
\\ \cline{3-4} & & $0<q<\frac{3}{2}$  & $-\frac{71}{18}<q<0$ \\
\cline{2-4} &\multicolumn{3}{|c||}{ $ \frac{\left(3 t^2+9 t+2\right) \sqrt{36 t^6+324 t^5+1557 t^4+4482 t^3+10329 t^2+15192 t+12272}}{(6 t^5+27 t^4+141 t^3+444 t^2+638 t+72)
(1- q t)}
\begin{matrix} {} \cr {} \cr {} \end{matrix}$} \\  \hline
\multirow{3}{*} {$\begin{matrix}  \cr 7644 \cr
\begin{pmatrix}
2 & 0 & 1 & 1 & 1 \cr 0 & 2 & 1 & 1 & 1
\end{pmatrix}^{2,46}_{-88} \end{matrix}$}
& $\frac{2}{3} \begin{pmatrix}
3&1\cr 3&1
\end{pmatrix}$ &
$\frac{3(1-q)+\sqrt{6q^2-8q+6}}{3q-1}$
 &
$-\frac{7t^5+123t^4+258t^3+238t^2+123t+51}{51t^5+123t^4+258t^3+238t^2+123t+7}$
\\ \cline{3-4} & & $\frac{1}{3}<q<3$ &
$-\frac{51}{7}<q<-\frac{7}{51}$  \\
\cline{2-4} &\multicolumn{3}{|c||}{ $\frac{\left(t^3+9 t^2+9 t+1\right) \sqrt{481 t^6+1854 t^5+3555 t^4+4220 t^3+3555 t^2+1854 t+481}}{(51 t^5+123 t^4+238 t^3+258 t^2+123 t+7(1-q t)}
\begin{matrix} {} \cr {} \cr {} \end{matrix}$} \\  \hline
\multirow{3}{*} {$\begin{matrix} 7668 \cr {} \cr
\begin{pmatrix}
  0 & 2 & 1 \cr 3 & 1 & 1
\end{pmatrix}^{2,47}_{-90}
\end{matrix}$}   & $\frac{1}{2}
\begin{pmatrix}
3&2\cr 1&0
\end{pmatrix} $ &
$\frac{ 1 - 3 q +\sqrt{7 q^2 - 3 q + 1}}{q}$    &
$- \frac{12 t^4 + 72 t^3 + 181 t^2 + 219 t + 284}
{6 t^5 + 27 t^4 + 141 t^3 + 444 t^2 + 638 t + 72} $ \\
\cline{3-4} &  & $0<q<\frac{3}{2}$ &
$-\frac{71}{18}<q<0$   \\
\cline{2-4} & \multicolumn{3}{|c||}{
$\frac{\left(3 t^2+9 t+2\right) \sqrt{36 t^6+324 t^5+1557 t^4+4482 t^3+10329 t^2+15192 t+12272}}{(6 t^5+27 t^4+141 t^3+444 t^2+638 t+72)(1-q t)}\begin{matrix} {} \cr {} \cr{} \end{matrix}$} \\
\hline
\multirow{3}{*} {$\begin{matrix}  \cr 7725 \cr
\begin{pmatrix}
0 & 0 & 1 & 1 & 1 \cr 2 & 2 & 1 & 1 & 1
\end{pmatrix}^{2,50}_{-96} \end{matrix}$}
& $\frac{2}{3}
\begin{pmatrix}
3&3\cr 1&0
\end{pmatrix} $
 & $\frac{ 1 - 3 q +\sqrt{6 q^2 - 3 q + 1}}{q}$ &
$-\frac{4t^4+24t^3+59t^2+69t+69}{2t^5+9t^4+39t^3+114t^2+159t+27}$
\\ \cline{3-4} & & $0<q<1$ & $-\frac{23}{9}<q<0$  \\
\cline{2-4} &\multicolumn{3}{|c||}{ $\frac{3 \left(t^2+3 t+1\right) \sqrt{4 t^6+36 t^5+165 t^4+450 t^3+915 t^2+1206 t+849}}{(2 t^5+9 t^4+39 t^3+114 t^2+159 t+27)(1-q t)}
\begin{matrix} {} \cr {} \cr {} \end{matrix}$} \\  \hline
\multirow{3}{*} {$\begin{matrix}  \cr 7726 \cr
\begin{pmatrix}
0&1&1&1&1\cr 2&1&1&1&1
\end{pmatrix}^{2,50}_{-96}
\end{matrix}$}
& $\frac{1}{3}\begin{pmatrix}
6&4\cr 4&1
\end{pmatrix}$
& $\frac{4-6q+\sqrt{10(2q^2-2q+1)}}{4q-1}$ & $-\frac{3t^5+70t^4+220t^3+300t^2+220t+112}
{4(7t^5+25t^4+65t^3+100t^2+70t+8)}$ \\
\cline{3-4}&  & $\frac{1}{4}<q<\frac{3}{2}$ & $-\frac{7}{2}<q<-\frac{3}{28}$ \\
\cline{2-4} & \multicolumn{3}{|c||}{$\frac{\left(t^3+12 t^2+18 t+4\right) \sqrt{89 t^6+516 t^5+1440 t^4+2440 t^3+2820 t^2+2016 t+704}}{4 \left(7 t^5+25 t^4+65 t^3+100 t^2+70
t+8\right)(1- q t)}
\begin{matrix} {} \cr {} \cr{} \end{matrix}$} \\ \hline
\multirow{3}{*} {$\begin{matrix}  \cr 7758 \cr
\begin{pmatrix}
0 & 2 & 1 \cr 2 & 1 & 2
\end{pmatrix}^{2,52}_{-100} \end{matrix}$} & $ \frac{1}{3}\begin{pmatrix}
5&2\cr 2&0
\end{pmatrix}$
& $\frac{2-5q+\sqrt{21q^2-10q+4}}{2q}$  & $-\frac{96t^4+480t^3+1018t^2+1045t+1289}
{48t^5+180t^4+858t^3+2320t^2+2816t+190}$
\\ \cline{3-4} &
&  $0<q<\frac{5}{2}$ & $-\frac{1289}{190}<q<0$ \\
\cline{2-4} &\multicolumn{3}{|c||}{ $\frac{\left(6 t^2+15 t+2\right) \sqrt{576 t^6+4320 t^5+17748 t^4+43740 t^3+89637 t^2+115530 t+83113}}{2 \left(24 t^5+90 t^4+429 t^3+1160
t^2+1408 t+95\right)(1-q t)}
\begin{matrix} {} \cr {} \cr {} \end{matrix}$}\\ \hline
\multirow{3}{*} {$\begin{matrix}  \cr 7759 \cr
\begin{pmatrix}
0&2&1&1\cr 2&1&1&1
\end{pmatrix}^{2,52}_{-100} \end{matrix}$} & $ \frac{1}{3}\begin{pmatrix}
5&2\cr 4&1
\end{pmatrix}$
& $\frac{4-5q+\sqrt{17q^2-18q+11}}{4q-1}$ &
$-\frac{35 t^5+821 t^4+2146 t^3+2458 t^2+1567 t+821}
{2 \left(172 t^5+503 t^4+1214 t^3+1646 t^2+982 t+67\right)}$
\\ \cline{3-4} &  &$\frac{1}{4}<q<\frac{5}{2}$ &  $-\frac{821}{134}<q<-\frac{35}{344}$ \\
\cline{2-4} &\multicolumn{3}{|c||}{ $\frac{\left(t^3+12 t^2+15 t+2\right) \sqrt{11873 t^6+56262 t^5+133323 t^4+195668 t^3+205611 t^2+134094 t+43793}}{2 \left(172 t^5+503 t^4+1214
t^3+1646 t^2+982 t+67\right)(1- qt)}
\begin{matrix} {} \cr {} \cr {} \end{matrix}$} \\ \hline
\end{tabular}
\caption{BPS and non-BPS extremal black holes in two-moduli CICY models, 1/3.}
\label{cicyt1}
\end{table}
}

\afterpage{
\clearpage
\thispagestyle{empty}
\begin{table}
  \centering
    \begin{tabular}{ || c | c |c| c|| }
\hline\multirow{3}{*} {$\begin{matrix}  \cr {\rm CICY\ label} \cr
{\rm Configuration\ Matrix}
\end{matrix}$}
& $\begin{pmatrix}
c&d\cr b&a
\end{pmatrix}$ & BPS solution  & Non-BPS solution  \\
\cline{3-4} & & Range of validity & Range of validity \\
\cline{2-4} &\multicolumn{3}{|c||}{ ${\rm Recombination\ Factor}
\begin{matrix} {} \cr {} \cr {} \end{matrix}$} \\
\hline\hline
\multirow{3}{*} {$\begin{matrix}  \cr 7761 \cr
\begin{pmatrix}
1&1&1&1&1\cr 1&1&1&1&1
\end{pmatrix}^{2,52}_{-100} \end{matrix}$} & $\frac{5}{6}\begin{pmatrix}
2&1\cr 2&1
\end{pmatrix}$
 &
$\frac{2(1-q)+\sqrt{2 q^2-3 q+2}}{2 q-1}$ &
 $-\frac{5 t^5+58 t^4+124 t^3+119 t^2+64 t+22}{22 t^5+64 t^4+119 t^3+124
t^2+58 t+5}$
\\ \cline{3-4} & & $\frac{1}{2}<q<2$ & $-\frac{22}{5}<q<-\frac{5}{22}$ \\
\cline{2-4} &\multicolumn{3}{|c||}{ $\frac{\left(t^3+6 t^2+6 t+1\right) \sqrt{281 t^6+1212 t^5+2472 t^4+3046 t^3+2472 t^2+1212 t+281}}{(22 t^5+64 t^4+119 t^3+124 t^2+58 t+5)(1-q t)}
\begin{matrix} {} \cr {} \cr {} \end{matrix}$} \\ \hline
\multirow{3}{*} {$\begin{matrix}  \cr 7799 \cr
\begin{pmatrix}
2&1&1\cr 1&2&1
\end{pmatrix}^{2,55}_{-106} \end{matrix}$} & $\frac{1}{6} \begin{pmatrix}
7&2\cr 7&2
\end{pmatrix}
$  &
$\frac{7(1-q)+\sqrt{5(7 q^2-9 q+7)}}{7 q-2}$ &
$-\frac{32 t^5+658 t^4+1372 t^3+1253 t^2+637 t+280}{280
t^5+637 t^4+1253 t^3+1372 t^2+658 t+32}$
\\ \cline{3-4} &
& $\frac{2}{7}<q<\frac{7}{2}$ & $-\frac{35}{4}<q<-\frac{4}{35}$ \\
\cline{2-4} &\multicolumn{3}{|c||}{ $\frac{\left(2 t^3+21 t^2+21 t+2\right) \sqrt{12256 t^6+45696 t^5+86205 t^4+101030 t^3+86205 t^2+45696 t+12256}}{\sqrt{5} \left(280 t^5+637
t^4+1253 t^3+1372 t^2+658 t+32\right)(1-q t)}
\begin{matrix} {} \cr {} \cr {} \end{matrix}$} \\ \hline
\multirow{3}{*} {$\begin{matrix}  \cr 7807 \cr
\begin{pmatrix}
0&1&1&1\cr 2&2&1&1
\end{pmatrix}^{2,56}_{-108} \end{matrix}$} &$\frac{1}{3}\begin{pmatrix}
5&4\cr 2&0
\end{pmatrix}$  &
$\frac{2-5q+\sqrt{17 q^2-10 q+4}}{2 q}$ &
$-\frac{96 t^4+480 t^3+986 t^2+965 t+831}{2 \left(24 t^5+90 t^4+333 t^3+820
t^2+957 t+130\right)}$
\\ \cline{3-4} &
& $0<q<\frac{5}{4}$ & $-\frac{831}{260}<q<0$ \\
\cline{2-4} &\multicolumn{3}{|c||}{ $\frac{\left(6 t^2+15 t+4\right) \sqrt{576 t^6+4320 t^5+16596 t^4+37980 t^3+65373 t^2+72870 t+43529}}{2 \left(24 t^5+90 t^4+333 t^3+820
t^2+957 t+130\right)(1- q t)}
\begin{matrix} {} \cr {} \cr {} \end{matrix}$} \\ \hline
\multirow{3}{*} {$\begin{matrix}  \cr 7808 \cr
\begin{pmatrix}
0&1&1&1\cr 3&1&1&1
\end{pmatrix}^{2,56}_{-108}\end{matrix}$}
&$\frac{1}{2}\begin{pmatrix}
3&3\cr 1&0
\end{pmatrix}$ &
$\frac{1-3q+\sqrt{6 q^2-3 q+1}}{q}$  &
$-\frac{4 t^4+24 t^3+59 t^2+69 t+69}{2 t^5+9 t^4+39 t^3+114 t^2+159 t+27}$
\\ \cline{3-4} & & $0<q<1$ & $-\frac{23}{9}<q<0$ \\
\cline{2-4} &\multicolumn{3}{|c||}{ $\frac{3 \left(t^2+3 t+1\right) \sqrt{4 t^6+36 t^5+165 t^4+450 t^3+915 t^2+1206 t+849}}{(2 t^5+9 t^4+39 t^3+114 t^2+159 t+27)(1 - q t)}
\begin{matrix} {} \cr {} \cr {} \end{matrix}$} \\ \hline
\multirow{3}{*} {$\begin{matrix}  \cr 7809 \cr
\begin{pmatrix}
1&1&1&1\cr 2&1&1&1
\end{pmatrix}^{2,56}_{-108} \end{matrix}$} &$\frac{1}{6}\begin{pmatrix}
9&5\cr 7&2
\end{pmatrix}$  &
$\frac{7-9q+\sqrt{46 q^2-53 q+31}}{7 q-2}$ &
$-\frac{656 t^5+13402 t^4+36092 t^3+42197 t^2+26617 t+11861}{5392
t^5+16457 t^4+37057 t^3+49162 t^2+29567 t+2815}$
\\ \cline{3-4} &
& $\frac{2}{7}<q<\frac{9}{5}$ & $-\frac{11861}{2815}<q<-\frac{41}{337}$ \\
\cline{2-4} &\multicolumn{3}{|c||}{ $\frac{\left(2 t^3+21 t^2+27 t+5\right)  \sqrt{3(353632 t^6+1752528 t^5+4196475 t^4+6114530 t^3+6107025 t^2+3776778 t+1143907)}}{(5392
t^5+16457 t^4+37057 t^3+49162 t^2+29567 t+2815)(1 - q t)}
\begin{matrix} {} \cr {} \cr {} \end{matrix}$} \\ \hline
\multirow{3}{*} {$\begin{matrix}  \cr 7821 \cr
\begin{pmatrix}
1&1&1\cr 2&2&1
\end{pmatrix}^{2,58}_{-112}\end{matrix}$} & $\frac{1}{6}\begin{pmatrix}
8&5\cr 4&0
\end{pmatrix}$
  &
$\frac{2(1-2q)+\sqrt{11 q^2-8 q+4}}{2 q}$  &
$-\frac{4 \left(24 t^4+96 t^3+158 t^2+124 t+87\right)}{48 t^5+144 t^4+432
t^3+856 t^2+801 t+85}$
\\ \cline{3-4} &
& $0<q<\frac{8}{5}$ & $-\frac{348}{85}<q<0$ \\
\cline{2-4} &\multicolumn{3}{|c||}{ $\frac{\left(12 t^2+24 t+5\right) \sqrt{144 t^6+864 t^5+2664 t^4+4896 t^3+6801 t^2+6114 t+2951}}{(48 t^5+144 t^4+432 t^3+856 t^2+801 t+85)(1 - q t)}
\begin{matrix} {} \cr {} \cr {} \end{matrix}$} \\ \hline
\multirow{3}{*} {$\begin{matrix}  \cr 7833 \cr
\begin{pmatrix}
2&1\cr 1&3
\end{pmatrix}^{2,59}_{-114}\end{matrix}$} &$\frac{1}{6}\begin{pmatrix}
7&2\cr 3&0
\end{pmatrix} $ &
$\frac{3-7q+\sqrt{43 q^2-21 q+9}}{3 q}$  &
$-\frac{9 \left(108 t^4+504 t^3+1005 t^2+973 t+1208\right)}{486 t^5+1701
t^4+7965 t^3+20412 t^2+23298 t+1120}$
\\ \cline{3-4} &
& $ 0<q<\frac{7}{2}$ & $-\frac{1359}{140}<q<0$ \\
\cline{2-4} &\multicolumn{3}{|c||}{ $\frac{3 \left(9 t^2+21 t+2\right) \sqrt{2916 t^6+20412 t^5+79461 t^4+185598 t^3+368577 t^2+455616 t+317536}}{(486 t^5+1701 t^4+7965 t^3+20412
t^2+23298 t+1120)(1 - q t)}
\begin{matrix} {} \cr {} \cr {} \end{matrix}$} \\ \hline
\end{tabular}
\caption{BPS and non-BPS extremal black holes in two-moduli CICY models, 2/3.}
\label{cicyt2}
\end{table}
}

\afterpage{
\clearpage
\thispagestyle{empty}
\begin{table}
  \centering
    \begin{tabular}{ || c | c |c| c|| }
\hline\multirow{3}{*} {$\begin{matrix}  \cr {\rm CICY\ label,} \cr
{\rm Configuration\ Matrix}
\end{matrix}$}
& $\begin{pmatrix}
c&d\cr b&a
\end{pmatrix}$ & BPS solution  & Non-BPS solution  \\
\cline{3-4} & & Range of validity & Range of validity \\
\cline{2-4} &\multicolumn{3}{|c||}{ Recombination Factor $
\begin{matrix} {} \cr {} \cr {} \end{matrix}$} \\
\hline\hline
\multirow{3}{*} {$\begin{matrix}  \cr 7844 \cr
\begin{pmatrix}
2&1\cr 2&2
\end{pmatrix}^{2,62}_{-120}
\end{matrix}$} & $\frac{1}{3}\begin{pmatrix}
3&1\cr 2&0
\end{pmatrix}$ &
$\frac{2-3q+\sqrt{7 q^2-6 q+4}}{2 q}$ &
$-\frac{96 t^4+288 t^3+362 t^2+219 t+142}{48 t^5+108 t^4+282 t^3+444 t^2+319 t+18}$
\\ \cline{3-4} & & $0<q<3$ & $-\frac{71}{9}<q<0$ \\
\cline{2-4} &\multicolumn{3}{|c||}{ $\frac{\left(6 t^2+9 t+1\right) \sqrt{576 t^6+2592 t^5+6228 t^4+8964 t^3+10329 t^2+7596 t+3068}}{(48 t^5+108 t^4+282 t^3+444 t^2+319 t+18)(1 - q t)}
\begin{matrix} {} \cr {} \cr {} \end{matrix}$} \\ \hline
\multirow{3}{*} {$\begin{matrix}  \cr 7853 \cr
\begin{pmatrix}
0&2&1\cr 2&2&1
\end{pmatrix}^{2,64}_{-124}
\end{matrix}$} &$\frac{2}{3}\begin{pmatrix}
2&1\cr 1&0
\end{pmatrix}$ &
$\frac{1-2q+\sqrt{3 q^2-2 q+1}}{q}$  &
$-\frac{6 t^4+24 t^3+40 t^2+32 t+26}{3 t^5+9 t^4+30 t^3+62 t^2+59 t+5}$
\\ \cline{3-4} & & $0<q<2$ & $-\frac{26}{5}<q<0$ \\
\cline{2-4} &\multicolumn{3}{|c||}{ $\frac{\left(3 t^2+6 t+1\right) \sqrt{9 t^6+54 t^5+171 t^4+324 t^3+483 t^2+462 t+241}}{(3 t^5+9 t^4+30 t^3+62 t^2+59 t+5)(1 - q t)}
\begin{matrix} {} \cr {} \cr {} \end{matrix}$} \\ \hline
\multirow{3}{*} {$\begin{matrix}  \cr 7863 \cr
\begin{pmatrix}
2  & 1 & 1 \cr  2  & 1 & 1
\end{pmatrix}^{2,66}_{-128}\end{matrix}$}
& $ \frac{1}{3}\begin{pmatrix}
3&1\cr 3&1
\end{pmatrix}$ &
$\frac{3(1-q)+\sqrt{6q^2-8q+6}}{3q-1}$
&
$-\frac{7t^5+123t^4+258t^3+238t^2+123t+51}{51t^5+123t^4+258t^3+238t^2+123t+7}$
\\ \cline{3-4} & & $\frac{1}{3}<q<3$ &
$-\frac{51}{7}<q<-\frac{7}{51}$ \\
\cline{2-4} &\multicolumn{3}{|c||}{ $\frac{\left(t^3+9 t^2+9 t+1\right) \sqrt{481 t^6+1854 t^5+3555 t^4+4220 t^3+3555 t^2+1854 t+481}}{(51 t^5+123 t^4+238 t^3+258 t^2+123 t+7)(1 - q t)}
\begin{matrix} {} \cr {} \cr {} \end{matrix}$} \\ \hline
\multirow{3}{*} {$\begin{matrix}  \cr 7868 \cr
\begin{pmatrix}
1&1&1\cr 3&1&1
\end{pmatrix}^{2,68}_{-132}
\end{matrix}$} & $\frac{1}{6}\begin{pmatrix}
7&5\cr 3&0
\end{pmatrix}$  &
$\frac{3-7q+\sqrt{34 q^2-21 q+9}}{3 q}$  &
$-\frac{9 \left(108 t^4+504 t^3+969 t^2+889 t+739\right)}{486 t^5+1701 t^4+6021
t^3+13986 t^2+15297 t+1855}$
\\ \cline{3-4} & & $0<q<\frac{7}{5}$ & $-\frac{6651}{1855}<q<0$ \\
\cline{2-4} &\multicolumn{3}{|c||}{ $\frac{3 \left(9 t^2+21 t+5\right) \sqrt{2916 t^6+20412 t^5+73629 t^4+158382 t^3+258579 t^2+273042 t+155041}}{(486 t^5+1701 t^4+6021 t^3+13986
t^2+15297 t+1855)(1 - q t)}
\begin{matrix} {} \cr {} \cr {} \end{matrix}$} \\ \hline
\multirow{3}{*} {$\begin{matrix}  \cr 7883 \cr
\begin{pmatrix}
2&1\cr 3&1
\end{pmatrix}^{2,77}_{-150}
\end{matrix}$} & $\frac{1}{6} \begin{pmatrix}
5&2\cr 3&0
\end{pmatrix}$ &
$\frac{3-5q+\sqrt{19 q^2-15 q+9}}{3 q}$  &
$-\frac{9 \left(108 t^4+360 t^3+501 t^2+335 t+232\right)}{486 t^5+1215 t^4+3429
t^3+5940 t^2+4722 t+320}$
\\ \cline{3-4} &
& $0<q<\frac{5}{2}$ & $-\frac{261}{40}<q<0$ \\
\cline{2-4} &\multicolumn{3}{|c||}{ $\frac{3 \left(9 t^2+15 t+2\right) \sqrt{2916 t^6+14580 t^5+38637 t^4+61290 t^3+76977 t^2+61920 t+27232}}{(486 t^5+1215 t^4+3429 t^3+5940
t^2+4722 t+320)(1 - q t)}
\begin{matrix} {} \cr {} \cr {} \end{matrix}$} \\ \hline
\multirow{3}{*} {$\begin{matrix}  \cr 7884 \cr
\begin{pmatrix}
3\cr 3
\end{pmatrix}^{2,83}_{-162}
\end{matrix}$} & $ \frac{1}{2} \begin{pmatrix}
1&0\cr 1&0
\end{pmatrix}$ &
$\frac{1-q+\sqrt{q^2-q+1}}{q}$  & $-\frac{4 t^4+8 t^3+7 t^2+3 t+2}{2 t^5+3 t^4+7 t^3+8 t^2+4 t}$
\\ \cline{3-4} & & $q>0$ & $q<0$
 \\ \cline{2-4} &\multicolumn{3}{|c||}{ $\frac{3 (t+1) \sqrt{4 t^6+12 t^5+21 t^4+22 t^3+21 t^2+12 t+4}}{(2 t^4+3 t^3+7 t^2+8 t+4)(1 - q t)}
\begin{matrix} {} \cr {} \cr {} \end{matrix}$} \\ \hline
\end{tabular}
\caption{BPS and non-BPS extremal black holes in two-moduli CICY models, 3/3.}
\label{cicyt3}
\end{table}
}

\section{THCY Black Holes\label{thcys}}

In this Appendix we will report the results of the study of extremal black
hole attractors in the 48 two-moduli toric hypersurface Calabi-Yau (THCY),
recently discussed in \cite{Ruehle} (cfr. App. B therein), and whose cubic
forms (and thus intersection numbers) can be obtained e.g. from the database
\cite{CY-database}.

In Tables \ref{thcyt1}-\ref{thcyt5}, we report the extremal black hole
attractor solutions for 37 two-moduli THCY models, and we do not include the
model\footnote{%
Such a two-moduli THCY model corresponds to the polytope label $%
(3,1)_{-144}^{2,74}$, and the corresponding charge matrix is given by (\ref%
{thcym1}); note that this model has the same set of intersection numbers as
in the models $(4,1)_{-144}^{2,74}$ and $(4,2)_{-144}^{2,74}$, i.e., the 3rd
and 4th models of Table \ref{thcyt1}.} already treated in detail in\ Sec. %
\ref{thcy}. The first column of Tables \ref{thcyt1}-\ref{thcyt5} indicates
the polytope label as well as the charge matrix of the ambient toric
variety, as from \cite{Ruehle}. The superscripts in the configuration matrix
are the Hodge numbers $h_{1,1}=2$ and $h_{2,1}$, respectively, whereas the
subscript is the Euler number $\chi $ of the THCY model. On the other hand,
as indicated in the first row of Tables \ref{thcyt1}-\ref{thcyt5} and as
reported in Tables \ref{cicyt1}-\ref{cicyt3} for two-moduli CICY models, in
the various partitions of their second column we respectively specify : the
intersection numbers of the THCY model, the critical values of $t$ as a
function of the charge ratio $q$ for the BPS black hole attractors, along
with the range of $q$ for which the BPS moduli lie inside the K\"{a}hler
cone, the expression for $q$ as a rational polynomial function of $t$ for
the non-BPS black hole attractors, along with the range of $q$ for which the
non-BPS moduli lie within the K\"{a}hler cone, and finally the expression
for the recombination factor. Again, in all such Tables $t$ takes its
critical value.

Again, as for two-moduli CICY models, our findings for two-moduli THCY
models show that \textit{there are no multiple extremal BPS black holes} for
a given value of (supporting electric charge ratio) $q$. Analogously, for
extremal non-BPS attractors, for any given $q$ within the specified range
there is a \textit{unique} $t$ inside the K\"{a}hler cone; thus, all 37
two-moduli THCY models under consideration give \textit{unique} non-BPS
black hole attractors. Moreover, we should remark that all BPS and non-BPS
solutions are \textit{mutually exclusive}.

We have numerically evaluated the recombination factor for the non-BPS black
hole attractors in the entire moduli space for all allowed values of $q$; we
have found that the recombination factor is always greater than $1$ for all
the 37 two-moduli THCY models under consideration : therefore, all non-BPS
black holes in the 37 two-moduli THCY models listed in Tables \ref{thcyt1}-%
\ref{thcyt5} are \textit{unstable}.

Since the Calabi-Yau threefolds with $h_{1,1}=2$ constructed as
hypersurfaces in toric varieties (THCYs) associated with the 36 reflexive
four-dimensional polytopes with six rays, and their various triangulations,
have been classified in a set of 48 models (cfr. App. B of \cite{Ruehle}), a
natural question arises : What about the remaining 10 two-moduli THCY
models, not reported in Tables \ref{thcyt1}-\ref{thcyt5}? Some of such
unlisted models have either $c=d=0$ or $a=b=0$, and thus the uniqueness of
their BPS and non-BPS black hole attractors has been discussed in Secs. \ref%
{cdzero} and \ref{cdzero2}, respectively. Apart from these, we have also not
included a few models for which the triple intersection numbers are
identical to some of the models already discussed here; they are related to
each other by a flop (for more detail, see e.g. the discussion in \cite%
{Ruehle}).\medskip

Some remarks are in order :

\begin{itemize}
\item The model $\left( 2,1\right) _{-72}^{2,38}$, i.e. the second model of
Table \ref{thcyt1} is marked red, because for such a model one of the two
eigenvalues of $G_{IJ}$ is always negative in the ranges of $q$ supporting
either the BPS or the non-BPS black hole attractors. Thus, this model does
not give rise to physically consistent black hole attractors.

\item There are models (like the models $\left( 1,1\right) _{-54}^{2,29}$
and $\left( 5,1\right) _{-162}^{2,83}$, i.e. the first and the fifth of
Table \ref{thcyt1}) with the same entries in the second column and
intersection numbers {related to one another by an overall rescaling}; they
have different polytope labels, $h_{2,1}$, $\chi $\ and charge matrices. {%
This can be explained by observing that the equations of motion (\ref{5deom}%
) is invariant under the rescaling $C_{IJK}\rightarrow \alpha
C_{IJN},t^{I}\rightarrow \alpha ^{-1/3}t^{I}$, }$\alpha \in \mathbb{R}$; t{%
hus, the attractor moduli will differ by an overall factor, but the ratio $%
t=t^{1}/t^{2}$ will remain invariant. Thus, the solution listed in Table \ref%
{thcyt1} which relate the charge ratio with $t$ remain the same for both the
models. The recombination factor too remains the same upon an overall
rescaling of the intersection numbers.}

\item There are models (like the models $\left( 4,1\right) _{-144}^{2,74}$
and $\left( 4,2\right) _{-144}^{2,74}$, i.e. the third and the fourth of
Table \ref{thcyt1}) which\texttt{\ }share the intersection numbers of the
model explicitly treated in Sec. \ref{thcy}, labelled as $%
(3,1)_{-144}^{2,74} $, but with different charge matrices. Such models share
the same entries in the second column as well as the same intersection
numbers and polytope labels; they seem to differ only for the charge
matrices. {This can be explained by noticing that the polytope corresponding
to such models admits more than one triangulation, such that different
triangulations correspond to different charge matrices. However all such
models are related to one another by a flop.}

\item There are models (like the models $\left( 20,1\right) _{-208}^{2,106}$
and $\left( 21,1\right) _{-208}^{2,106}$, i.e. the first and the second of
Table \ref{thcyt3}, or like the models $\left( 25,1\right) _{-240}^{2,122}$
and $\left( 26,1\right) _{-240}^{2,122}$, i.e. the eighth of Table \ref%
{thcyt3} and the first of Table \ref{thcyt4}) which share the same entries
in the second column and the same intersection numbers; they seem to differ
only for the charge matrices and the polytope labels. {This is somewhat a
surprising result, because the corresponding two-dimensional ambient spaces
are not related by any symmetry. As far as we know, it looks like a mere
coincidence that such pairs exist.}

\item There are models (like the models $\left( 26,1\right) _{-240}^{2,122}$
and $\left( 27,1\right) _{-240}^{2,122}$, i.e. the 1first and the second of
Table \ref{thcyt4}) with the same $h_{2,1}$, $\chi $, and intersecting
numbers related by $\left( a,b,c,d\right) \leftrightarrow \left(
d,c,b,a\right) $. They have different entries in the second column and
different charge matrices, as well. These models correspond to different
polytopes; though the Hodge numbers coincide, they give rise to different
Calabi-Yau threefolds. As far as we know, there is not any deeper reason why
the intersection numbers are related by $\left( a,b,c,d\right)
\leftrightarrow \left( d,c,b,a\right) $; e.g., no explanation has been given
for such pairs in \cite{Ruehle}.
\end{itemize}

\afterpage{
\clearpage
\thispagestyle{empty}
\begin{table}
\centering
\begin{tabular}{|c||c|c|c|}
\hline
\multirow{3}{*}{$\begin{matrix}
{\rm Polytope\ Label,}\cr {\rm Charge\ Matrix}
\end{matrix} $} &
$
\begin{pmatrix}
c & d \cr b & a
\end{pmatrix}
 $ & BPS Solution & non-BPS Solution \\
\cline{3-4}&  & Range of validity & Range of validity \\
\cline{2-4} & \multicolumn{3}{|c|}{Recombination Factor $ \begin{matrix}
{}\cr{}
\end{matrix}$}
\\ \hline\hline
\multirow{3}{*}{$\begin{matrix}
(1,1)^{2,29}_{-54}\cr {} \cr \begin{pmatrix}
1&0&0&0&1&1\cr 0&1&1&1&0&0
\end{pmatrix}
\end{matrix} $} &
$ \frac{1}{6}
\begin{pmatrix}
1 & 0 \cr 1 & 0
\end{pmatrix}
 $ & $\frac{\sqrt{q^2-q+1}-q+1}{q}$ & $-\frac{4 t^4+8 t^3+7 t^2+3 t+2}{2 t^5+3 t^4+7 t^3+8 t^2+4 t}$ \\
\cline{3-4}&  & $q>0$ & $q<0$ \\
\cline{2-4} & \multicolumn{3}{|c|}{$\frac{3  \sqrt{(t+1) \left(4 t^7+16 t^6+33 t^5+43 t^4+43 t^3+33 t^2+16 t+4\right)}}{6 t^4+11 t^3+14 t^2+11 t +6 }
\begin{matrix}
{}\cr{}
\end{matrix} $}
\\ \hline \color{red}
\multirow{3}{*}{$\begin{matrix}
(2,1)^{2,38}_{-72}\cr {}\cr \begin{pmatrix}
0&0&1&1&0&1\cr 1&1&0&0&1&-3
\end{pmatrix}\end{matrix}$} &
$\frac{1}{6}
\begin{pmatrix}
1 & 9 \cr 3 & 9
\end{pmatrix}
 $ & $\frac{3-q- \sqrt{26 q(3-q) }}{3 (q-3)}$ & $\frac{(3 t+1)^2 \left(9 t^3+9 t^2+3 t+61\right)}{27 t^5+45 t^4-516
t^3-354 t^2-59 t-75}$ \\
\cline{3-4}&  & $\frac{1}{9}<q<3$ & $q<-\frac{61}{75} \ \& \ q>3$ \\
\cline{2-4} & \multicolumn{3}{|c|}{$\frac{3 \left(3 t^3+3 t^2+t+3\right) \sqrt{81 t^6+162 t^5+135 t^4-2592 t^3-2637 t^2-882 t-23}}{\left(-27 t^5-45 t^4+516 t^3+354 t^2+59
t+75\right)(1+|q|t)} \begin{matrix}
{}\cr{}
\end{matrix}$}
\\ \hline
\multirow{3}{*}{$\begin{matrix}
(4,1)^{2,74}_{-144}\cr {}\cr
\begin{pmatrix}
-1&2&1&1&3&0\cr 1&-1&0&0&-1&1
\end{pmatrix}
\end{matrix} $} &
$ \frac{1}{6}
\begin{pmatrix}
3 & 3 \cr 3 & 2
\end{pmatrix}
 $ & $ \frac{3(1-q)+\sqrt{3(1-q)}}{3q-2}$ &
 $-\frac{8 t^5+66 t^4+132 t^3+111 t^2+45 t+9}{3 (t+1)^2 \left(4 t^3-9 t^2-9 t-3\right)}$ \\
\cline{3-4}&  & $\frac{2}{3}<q<1$ & $q<-\frac{2}{3} \ \& \ q>1$ \\
\cline{2-4} & \multicolumn{3}{|c|}{$
\frac{2^{1/6} \left(2 t^3+9 t^2+9 t+3\right)^{5/6} \sqrt{112 t^6+360 t^5+441 t^4+282 t^3+135 t^2+54 t+9}}{3^{5/6} (t+1)^2 \left(4
t^3-9 t^2-9 t-3\right)(|q|t +1)} \begin{matrix}
{}\cr{}
\end{matrix}$}
\\ \hline
\multirow{3}{*}{$\begin{matrix}
(4,2)^{2,74}_{-144}\cr {}\cr
\begin{pmatrix}
2&-1&1&1&0&3\cr -1&1&0&0&1&-1
\end{pmatrix}
\end{matrix} $} &
$ \frac{1}{6}
\begin{pmatrix}
3 & 3 \cr 3 & 2
\end{pmatrix}
 $ & $ \frac{3(1-q)+\sqrt{3(1-q)}}{3q-2}$ &
 $-\frac{8 t^5+66 t^4+132 t^3+111 t^2+45 t+9}{3 (t+1)^2 \left(4 t^3-9 t^2-9 t-3\right)}$ \\
\cline{3-4}&  & $\frac{2}{3}<q<1$ & $q<-\frac{2}{3} \ \& \ q>1$ \\
\cline{2-4} & \multicolumn{3}{|c|}{$
\frac{2^{1/6} \left(2 t^3+9 t^2+9 t+3\right)^{5/6} \sqrt{112 t^6+360 t^5+441 t^4+282 t^3+135 t^2+54 t+9}}{3^{5/6} (t+1)^2 \left(4
t^3-9 t^2-9 t-3\right)(|q|t +1)} \begin{matrix}
{}\cr{}
\end{matrix}$}
\\ \hline
\multirow{3}{*}{$\begin{matrix}
(5,1)^{2,83}_{-162}\cr {} \cr \begin{pmatrix}
1&0&1&0&0&1\cr 0&1&0&1&1&0
\end{pmatrix}
\end{matrix} $} &
$ \frac{1}{2}
\begin{pmatrix}
1 & 0 \cr 1 & 0
\end{pmatrix}
 $ & $\frac{\sqrt{q^2-q+1}-q+1}{q}$ & $-\frac{4 t^4+8 t^3+7 t^2+3 t+2}{2 t^5+3 t^4+7 t^3+8 t^2+4 t}$ \\
\cline{3-4}&  & $q>0$ & $q<0$ \\
\cline{2-4} & \multicolumn{3}{|c|}{$\frac{3  \sqrt{(t+1) \left(4 t^7+16 t^6+33 t^5+43 t^4+43 t^3+33 t^2+16 t+4\right)}}{6 t^4+11 t^3+14 t^2+11 t +6 }
\begin{matrix}
{}\cr{}
\end{matrix} $}
\\ \hline
\multirow{3}{*}{$\begin{matrix}
(6,1)^{2,84}_{-164}\cr {}\cr \begin{pmatrix}
1&-1&1&1&2&0\cr 0&1&0&0&-1&1
\end{pmatrix}
\end{matrix} $} &
$\frac{1}{6}
\begin{pmatrix}
5&5\cr 5&3
\end{pmatrix}
 $ & $\frac{5(1-q)+ \sqrt{10(1-q)}}{5 q-3}$ & $-\frac{21 t^5+195 t^4+390 t^3+320 t^2+125 t+25}{5 (t+1)^2 \left(9 t^3-15
t^2-15 t-5\right)}$ \\
\cline{3-4}&  & $\frac{3}{5}<q<1$ & $q<-\frac{7}{15}\ \& \ q > 1$ \\
\cline{2-4} & \multicolumn{3}{|c|}{$\frac{\left(3 t^3+15 t^2+15 t+5\right) \sqrt{369 t^6+1170 t^5+1395 t^4+840 t^3+375 t^2+150 t+25}}{5 (t+1)^2 \left(9 t^3-15 t^2-15
t-5\right)(1+|q|t)} \begin{matrix}
{}\cr{}
\end{matrix}$}
\\ \hline
\multirow{3}{*}{$\begin{matrix}
(10,1)^{2,86}_{-168}\cr {}\cr \begin{pmatrix}
1&0&0&1&0&1\cr 0&1&1&0&1&-1
\end{pmatrix}
\end{matrix} $} &
$\frac{1}{6}
\begin{pmatrix}
3&0\cr 5&5
\end{pmatrix}
 $ & $\frac{5-3 q+\sqrt{9 q^2-15 q+10}}{5 (q-1)}$ & $-\frac{125 t^5+825 t^4+1350 t^3+1485 t^2+1215 t+486}{9 t \left(75 t^4+275 t^3+465 t^2+360 t+108\right)}$ \\
\cline{3-4}&  & $q>1$ & $q<-\frac{5}{27}$ \\
\cline{2-4} & \multicolumn{3}{|c|}{$\frac{\left(5 t^2+15 t+9\right) \sqrt{3625 t^6+15750 t^5+33075 t^4+41850 t^3+32805 t^2+14580 t+2916}}{3 \left(75 t^4+275 t^3+465
t^2+360 t+108\right)(1-\text{qt})} \begin{matrix}
{}\cr{}
\end{matrix}$}
\\ \hline
\multirow{3}{*}{$\begin{matrix}
(11,1)^{2,86}_{-168}\cr {}\cr
 \begin{pmatrix}
  1&0&0&1&1&0\cr 0&1&1&-1&-1&1
  \end{pmatrix}
\end{matrix} $} &
$\frac{1}{6}
\begin{pmatrix}
3&0\cr 7&11
\end{pmatrix}
 $ & $ \frac{7-3 q+\sqrt{9 q^2-21 q+16}}{7 q-11}$  & $ \frac{77 t^5-267 t^4-1134 t^3-2133 t^2-1701 t-486}{9
t \left(233 t^4+721 t^3+897 t^2+504 t+108\right)}$ \\
\cline{3-4}&  & $q>\frac{11}{7}$ & $q<\frac{77}{2097}$ \\
\cline{2-4} & \multicolumn{3}{|c|}{$\frac{\left(11 t^2+21 t+9\right) \sqrt{14569 t^6+57582 t^5+102735 t^4+103950 t^3+61965 t^2+20412 t+2916}}{3 \left(233 t^4+721 t^3+897
t^2+504 t+108\right)(1+|q|t)} \begin{matrix}
{}\cr{}
\end{matrix}$}
\\ \hline
\end{tabular}
\caption{BPS and non-BPS extremal black holes in two-moduli THCY models, 1/5.}
\label{thcyt1}
\end{table}
}

\afterpage{
\clearpage
\thispagestyle{empty}
\begin{table}
\centering
\begin{tabular}{|c|c|c|c|}
\hline
\multirow{3}{*}{$\begin{matrix}
{\rm Polytope\ Label,}\cr {\rm Charge\ Matrix}
\end{matrix} $} &
$
\begin{pmatrix}
c & d \cr b & a
\end{pmatrix}
 $ & BPS Solution & non-BPS Solution \\
\cline{3-4}&  & Range of validity & Range of validity \\
\cline{2-4} & \multicolumn{3}{|c|}{Recombination Factor $ \begin{matrix}
{}\cr{}
\end{matrix}$}
\\ \hline
\hline
\multirow{3}{*}{$\begin{matrix}
(12,1)^{2,86}_{-168}\cr {}\cr
\begin{pmatrix}
1&-1&-1&1&2&0\cr 0&1&1&0&-1&1
\end{pmatrix}
\end{matrix} $} &
$\frac{1}{6}
\begin{pmatrix}
10&14\cr 6&3
\end{pmatrix}
 $ & $ \frac{6 -10 q+\sqrt{2(8 q^2-9 q+3)}}{6 q-3}$ & $-\frac{3 \left(9 t^5+102 t^4+372 t^3+642 t^2+572 t+228\right)}{2
\left(45 t^5+288 t^4+843 t^3+1308 t^2+978 t+238\right)}$ \\
\cline{3-4}& & $\frac{1}{2}<q<\frac{5}{7}$  & $-\frac{171}{119}<q<-\frac{3}{10}$ \\
\cline{2-4} & \multicolumn{3}{|c|}{$\frac{ \left(3 t^3+18 t^2+30 t+14\right) \sqrt{3\left(603 t^6+5292 t^5+20160 t^4+42780 t^3+53460 t^2+37272 t+11308\right)}}{2 \left(45
t^5+288 t^4+843 t^3+1308 t^2+978 t+238\right)(1-q t)}\begin{matrix}
{}\cr{}
\end{matrix}$}
\\ \hline
\multirow{3}{*}{$\begin{matrix}
(14,1)^{2,86}_{-168}\cr {}\cr \begin{pmatrix}
1&-1&-1&1&3&0\cr 0&1&1&0&-2&1
\end{pmatrix}
\end{matrix} $} &
$\frac{1}{6}
\begin{pmatrix}
13&17\cr 9&6
\end{pmatrix}
 $ & $ \frac{9 -13 q+\sqrt{16 q^2-15 q+3}}{9 q-6}$ &

$\frac{3 \left(72 t^5+534 t^4+1500 t^3+1881 t^2+895 t+15\right)}{396
t^5+3195 t^4+10137 t^3+15450 t^2+10977 t+2737}$ \\
\cline{3-4}&  & $\frac{2}{3}<q<\frac{13}{17}$ & $\frac{45}{2737}<q<\frac{6}{11} $ \\
\cline{2-4} & \multicolumn{3}{|c|}{$\frac{ \left(6 t^3+27 t^2+39 t+17\right) \sqrt{3\left(720 t^6+8424 t^5+42651 t^4+117750 t^3+184077 t^2+152826 t+52331\right)}}{(396
t^5+3195 t^4+10137 t^3+15450 t^2+10977 t+2737) (1+q t)} \begin{matrix}
{}\cr{}
\end{matrix}$}
\\ \hline
\multirow{3}{*}{$\begin{matrix}
(15,1)^{2,86}_{-168}\cr {}\cr \begin{pmatrix}
1&0&0&0&0&1\cr 0&1&1&2&1&-1
\end{pmatrix}
\end{matrix} $} &
$\frac{1}{6}
\begin{pmatrix}
3&1\cr 3&3
\end{pmatrix}
 $ & $\frac{3 (1- q) + \sqrt{6q(q-1)}}{3 (q-1)}$ &
$\frac{3 (t+1)^2 \left(3 t^3+9 t^2+9 t-11\right)}{9 t^5+45 t^4+132 t^3+174
t^2+87 t+5}$  \\
\cline{3-4}&   & $1<q<3$ & $-\frac{33}{5}<q<1$ \\
\cline{2-4} & \multicolumn{3}{|c|}{$\frac{\left(3 t^3+9 t^2+9 t+1\right) \sqrt{9 t^6+54 t^5+135 t^4+384 t^3+747 t^2+666 t+217}}{\left(9 t^5+45 t^4+132 t^3+174 t^2+87
t+5\right)(1+|q|t)} \begin{matrix}
{}\cr{}
\end{matrix}$}
\\ \hline
\multirow{3}{*}{$\begin{matrix}
(16,1)^{2,90}_{-176}\cr {}\cr \begin{pmatrix}
1&0&0&0&0&1\cr 0&1&1&1&1&-1
\end{pmatrix}
\end{matrix} $} &
$ \frac{1}{6}
\begin{pmatrix}
5&2\cr 5&5
\end{pmatrix}
 $ & $
\frac{5(1- q) + \sqrt{15 q(q-1)}}{5 (q-1)}$ &
$\frac{5 (t+1)^2 \left(5 t^3+15 t^2+15 t-16\right)}{25 t^5+125 t^4+355
t^3+460 t^2+230 t+16}$   \\
\cline{3-4}&  & $1<q<\frac{5}{2}$ & $-5<q<1$ \\
\cline{2-4} & \multicolumn{3}{|c|}{$\frac{\left(5 t^3+15 t^2+15 t+2\right) \sqrt{25 t^6+150 t^5+375 t^4+1010 t^3+1905 t^2+1680 t+544}}{\left(25 t^5+125 t^4+355 t^3+460
t^2+230 t+16\right)(1+|q|t)} \begin{matrix}
{}\cr{}
\end{matrix}$}
\\ \hline
\multirow{3}{*}{$\begin{matrix}
(17,1)^{2,92}_{-180} \cr {} \cr \begin{pmatrix}
1&-1&-1&-1&2&0\cr 0&1&1&1&-1&1
\end{pmatrix}
\end{matrix} $} &
$\frac{1}{2}
\begin{pmatrix}
4&7\cr 2&1
\end{pmatrix}
 $ & $\frac{2 -4 q+\sqrt{q (2 q-1)}}{2 q-1}$ &
$\frac{(t+2)^2 \left(t^3+6 t^2+12 t+1\right)}{2 t^5+20 t^4+87 t^3+188
t^2+188 t+63}$ \\
\cline{3-4}&  & $\frac{1}{2}<q<\frac{4}{7}$ & $\frac{4}{63}<q<\frac{1}{2}$  \\
\cline{2-4} & \multicolumn{3}{|c|}{$\frac{\left(t^3+6 t^2+12 t+7\right) \sqrt{t^6+12 t^5+60 t^4+194 t^3+444 t^2+600 t+337}}{\left(2 t^5+20 t^4+87 t^3+188 t^2+188 t+63\right)(1+q
t)} \begin{matrix}
{}\cr{}
\end{matrix}$}
\\ \hline
\multirow{3}{*}{$\begin{matrix}
(17,2)^{2,92}_{-180}\cr {} \cr \begin{pmatrix}
1&0&0&0&1&1\cr -1&1&1&1&-2&0
\end{pmatrix}
\end{matrix} $} &
$\frac{1}{2}
\begin{pmatrix}
1&0\cr 3&7
\end{pmatrix}
 $ & $\frac{3-q+\sqrt{q^2-3 q+2}}{3 q-7}$ &
$\frac{63 t^5+127 t^4+66 t^3-21 t^2-27 t-6}{t \left(59 t^4+159
t^3+161 t^2+72 t+12\right)}$  \\
\cline{3-4}&  &  $ q>\frac{7}{3}$ & $q<\frac{63}{59}$ \\
\cline{2-4} & \multicolumn{3}{|c|}{$\frac{\left(7 t^2+9 t+3\right) \sqrt{337 t^6+1422 t^5+2499 t^4+2322 t^3+1197 t^2+324 t+36}}{\left(59 t^4+159 t^3+161 t^2+72 t+12\right)(1+|q|t)} \begin{matrix}
{}\cr{}
\end{matrix}$}
\\ \hline
\multirow{3}{*}{$\begin{matrix}
(18,1)^{2,95}_{-186} \cr {} \cr \begin{pmatrix}
1&0&0&1&1&0\cr 0&1&1&0&-2&1
\end{pmatrix}
\end{matrix} $} &
$ \frac{1}{6}
\begin{pmatrix}
3&0\cr 7&14
\end{pmatrix}
 $ & $\frac{7-3 q+\sqrt{9 q^2-21 q+7}}{7 (q-2)}$ &
$\frac{2744 t^5+6762 t^4+5292 t^3-189 t^2-1701 t-486}{9 t \left(196
t^4+637 t^3+861 t^2+504 t+108\right)}$\\
\cline{3-4}&  & $ q>2$ & $q<\frac{14}{9} $ \\
\cline{2-4} & \multicolumn{3}{|c|}{$\frac{\left(14 t^2+21 t+9\right) \sqrt{5488 t^6+24696 t^5+56889 t^4+76734 t^3+56133 t^2+20412 t+2916}}{3 \left(196 t^4+637 t^3+861
t^2+504 t+108\right)(1+|q|t)} \begin{matrix}
{}\cr{}
\end{matrix}$}
\\ \hline
\multirow{3}{*}{$\begin{matrix}
(19,1)^{2,102}_{-200}\cr {} \cr \begin{pmatrix}
1&0&0&0&1&0\cr 0&1&1&1&-2&1
\end{pmatrix}
\end{matrix} $} &
$ \frac{1}{3}
\begin{pmatrix}
3&1\cr 6&12
\end{pmatrix}
 $ & $\frac{3(2-q) + \sqrt{3 q(q-2)}}{6 (q-2)}$ &
$\frac{3 (2 t+1)^2 \left(12 t^3+18 t^2+9 t-2\right)}{72 t^5+180 t^4+222 t^3+132
t^2+33 t+2}$  \\
\cline{3-4}&  & $2<q<3$ & $-3<q<2$ \\
\cline{2-4} & \multicolumn{3}{|c|}{$\frac{\left(12 t^3+18 t^2+9 t+1\right) \sqrt{144 t^6+432 t^5+540 t^4+564 t^3+441 t^2+180 t+28}}{\left(72 t^5+180 t^4+222 t^3+132
t^2+33 t+2\right)(1+|q|t)} \begin{matrix}
{}\cr{}
\end{matrix}$}
\\ \hline
\end{tabular}
\caption{BPS and non-BPS extremal black holes in two-moduli THCY models, 2/5.}
\label{thcyt2}
\end{table}
}

\afterpage{
\clearpage
\thispagestyle{empty}
\begin{table}
\centering
\begin{tabular}{|c|c|c|c|}
\hline
\multirow{3}{*}{$\begin{matrix}
{\rm Polytope\ Label,}\cr {\rm Charge\ Matrix}
\end{matrix} $} &
$
\begin{pmatrix}
c & d \cr b & a
\end{pmatrix}
 $ & BPS Solution & non-BPS Solution \\
\cline{3-4}&  & Range of validity & Range of validity \\
\cline{2-4} & \multicolumn{3}{|c|}{Recombination Factor $ \begin{matrix}
{}\cr{}
\end{matrix}$}
\\ \hline
\hline
\multirow{3}{*}{$\begin{matrix}
(20,1)^{2,106}_{-208} \cr {} \cr \begin{pmatrix}
  1&1&1&0&-3&8\cr 0&0&0&1&1&-2
  \end{pmatrix}
\end{matrix} $} &
$ \frac{1}{6}
\begin{pmatrix}
12&36\cr 4&1
\end{pmatrix}
 $ & $\frac{2 \left(2-6 q+\sqrt{1-3 q}\right)}{4 q-1}$
  & $-\frac{5 t^5+108 t^4+648 t^3+1692 t^2+2160 t+1296}{12
(t+3)^2 \left(t^3-12 t^2-36 t-36\right)}$
 \\
\cline{3-4}&  & $\frac{1}{4}<q<\frac{1}{3}$ & $q< -\frac{5}{12}\ \& \ q > \frac{1}{3}$ \\
\cline{2-4} & \multicolumn{3}{|c|}{$\frac{\left(t^3+12 t^2+36 t+36\right) \sqrt{17 t^6+168 t^5+648 t^4+1368 t^3+2160 t^2+2592 t+1296}}{4 (t+3)^2 \left(t^3-12 t^2-36
t-36\right)(1+|q|t)} \begin{matrix}
{}\cr{}
\end{matrix}$}
\\ \hline
\multirow{3}{*}{$\begin{matrix}
(21,1)^{2,106}_{-208} \cr {} \cr \begin{pmatrix}
1&-3&1&1&5&0\cr 0&1&0&0&-1&1
\end{pmatrix}
\end{matrix} $} &
$ \frac{1}{6}
\begin{pmatrix}
12&36\cr 4&1
\end{pmatrix}
 $ & $\frac{2 \left(2-6 q+\sqrt{1-3 q}\right)}{4 q-1}$
  & $-\frac{5 t^5+108 t^4+648 t^3+1692 t^2+2160 t+1296}{12
(t+3)^2 \left(t^3-12 t^2-36 t-36\right)}$
 \\
\cline{3-4}&  & $\frac{1}{4}<q<\frac{1}{3}$ & $q< -\frac{5}{12}\ \& \ q > \frac{1}{3}$ \\
\cline{2-4} & \multicolumn{3}{|c|}{$\frac{\left(t^3+12 t^2+36 t+36\right) \sqrt{17 t^6+168 t^5+648 t^4+1368 t^3+2160 t^2+2592 t+1296}}{4 (t+3)^2 \left(t^3-12 t^2-36
t-36\right)(1+|q|t)} \begin{matrix}
{}\cr{}
\end{matrix}$}
\\ \hline
\multirow{3}{*}{$\begin{matrix}
(22,1)^{2,106}_{-208}\cr \begin{pmatrix}
1&0&0&0&1&0\cr 0&1&1&2&-3&1
\end{pmatrix}
\end{matrix} $} &
$\frac{1}{6}
\begin{pmatrix}
4&1\cr 12&36
\end{pmatrix}
 $ & $ \frac{2 (3 - q)+\sqrt{q(q-3)}}{6 (q-3)}$ &
$\frac{12 (3 t+1)^2 \left(36 t^3+36 t^2+12 t-1\right)}{1296 t^5+2160 t^4+1692 t^3+648
t^2+108 t+5}$  \\
\cline{3-4}&  & $3<q<4$ & $-\frac{12}{5}<q<3 $  \\
\cline{2-4} & \multicolumn{3}{|c|}{$\frac{3 \left(36 t^3+36 t^2+12 t+1\right) \sqrt{1296 t^6+2592 t^5+2160 t^4+1368 t^3+648 t^2+168 t+17}}{\left(1296 t^5+2160 t^4+1692
t^3+648 t^2+108 t+5\right)(1+|q|t)} \begin{matrix}
{}\cr{}
\end{matrix}$}
\\ \hline
\multirow{3}{*}{$\begin{matrix}
(23,1)^{2,116}_{-228} \cr {} \cr \begin{pmatrix}
-2&3&-2&-4&5&0\cr 1&0&1&2&-1&3
\end{pmatrix}
\end{matrix} $} &
$\frac{1}{6}
\begin{pmatrix}
25&98\cr 5&1
\end{pmatrix}
 $ & $ \frac{5(1 - 5 q) +3  \sqrt{3q (5 q-1)}}{5 q-1}$
& $\frac{(t+5)^2 \left(t^3+15 t^2+75 t-64\right)}{5
t^5+125 t^4+1439 t^3+8140 t^2+20350 t+14896}$ \\
\cline{3-4}&  & $\frac{1}{5}<q<\frac{25}{98}$ & $-\frac{100}{931}<q<\frac{1}{5}$ \\
\cline{2-4} & \multicolumn{3}{|c|}{$\frac{\left(t^3+15 t^2+75 t+98\right) \sqrt{t^6+30 t^5+375 t^4+3418 t^3+23145 t^2+87600 t+131104}}{\left(5 t^5+125 t^4+1439 t^3+8140
t^2+20350 t+14896\right)(1+|q|t)} \begin{matrix}
{}\cr{}
\end{matrix}$}
\\ \hline
\multirow{3}{*}{$\begin{matrix}
(23,2)^{2,116}_{-228} \cr {} \cr \begin{pmatrix}
0&1&0&0&1&2\cr 2&-3&2&4&-5&0
\end{pmatrix}
\end{matrix} $} &
$\frac{1}{3}
\begin{pmatrix}
1&0\cr 8&49
\end{pmatrix}
 $ & $ \frac{8-q+\sqrt{q^2-8 q+15}}{8 q-49}$ &
$\frac{3724 t^5+2815 t^4+520 t^3-84 t^2-36 t-3}{2 t \left(754
t^4+760 t^3+287 t^2+48 t+3\right)}$  \\
\cline{3-4}&  & $q>\frac{49}{8}$ & $q<\frac{931}{377} $  \\
\cline{2-4} & \multicolumn{3}{|c|}{$\frac{\left(49 t^2+24 t+3\right) \sqrt{32776 t^6+50952 t^5+32865 t^4+11232 t^3+2142 t^2+216 t+9}}{2 \left(754 t^4+760 t^3+287 t^2+48
t+3\right)(1+|q|t)} \begin{matrix}
{}\cr{}
\end{matrix}$}
\\ \hline
\multirow{3}{*}{$\begin{matrix}
(24,1)^{2,120}_{-236}\cr {} \cr \begin{pmatrix}
1&-1&1&2&0&3\cr -1&4&-1&-5&3&0
\end{pmatrix}
\end{matrix} $} &
$\frac{1}{6}
\begin{pmatrix}
8&2\cr 32&101
\end{pmatrix}
 $ & $\frac{8(4-q)+3 \sqrt{6(4-q)}}{32 q-101}$
& $-\frac{15655 t^5+26528 t^4+13264 t^3+2938 t^2+320 t+16}{2 (4
t+1)^2 \left(61 t^3-96 t^2-24 t-2\right)}$
 \\
\cline{3-4}&  & $\frac{101}{32}<q<4$ & $q< -\frac{15655}{1952}\ \& \ q > 4$ \\
\cline{2-4} & \multicolumn{3}{|c|}{$\frac{\left(101 t^3+96 t^2+24 t+2\right) \sqrt{134617 t^6+112704 t^5+37392 t^4+6956 t^3+960 t^2+96 t+4}}{2 (4 t+1)^2 \left(61 t^3-96
t^2-24 t-2\right)(1+|q|t)}\begin{matrix}
{}\cr{}
\end{matrix}$}
\\ \hline
\multirow{3}{*}{$\begin{matrix}
(24,2)^{2,120}_{-236}\cr {} \cr \begin{pmatrix}
1&-4&1&5&-3&0\cr 0&1&0&-1&1&1
\end{pmatrix}
\end{matrix} $} &
$\frac{1}{6}
\begin{pmatrix}
23&101\cr 5&1
\end{pmatrix}
 $ & $\frac{5-23 q+\sqrt{24 q^2-14 q+2}}{5 q-1}$ &
$\frac{3 t^5+79 t^4+746 t^3+3312 t^2+7067 t+5869}{-t^5+79
t^4+1268 t^3+7078 t^2+17249 t+15655}$  \\
\cline{3-4}&  & $\frac{1}{5}<q<\frac{23}{101}$ & $q < -3\ \& \ q > \frac{5869}{15655}$\\
\cline{2-4} & \multicolumn{3}{|c|}{$\frac{\left(t^3+15 t^2+69 t+101\right) \sqrt{73 t^6+1542 t^5+13539 t^4+63256 t^3+165903 t^2+231666 t+134617}}{\left(t^5-79 t^4-1268
t^3-7078 t^2-17249 t-15655\right)(1+|q|t)} \begin{matrix}
{}\cr{}
\end{matrix}$}
\\ \hline
\multirow{3}{*}{$\begin{matrix}
(25,1)^{2,122}_{-240}\cr {} \cr \begin{pmatrix}
-3&1&1&1&0&7\cr 1&0&0&0&1&-2
\end{pmatrix}
\end{matrix} $} &
$\frac{1}{6}
\begin{pmatrix}
21&63\cr 7&2
\end{pmatrix}
 $ & $\frac{7(1 - 3 q) +\sqrt{7(1 -3 q)}}{7 q-2}$ &
$\frac{16 t^5+294 t^4+1764 t^3+4851 t^2+6615 t+3969}{441 (t+3)^2 \left(t^2+3
t+3\right)}$ \\
\cline{3-4}&  & $\frac{2}{7}<q<\frac{1}{3}$ & $ q>\frac{1}{3}$  \\
\cline{2-4} & \multicolumn{3}{|c|}{$\frac{\left(2 t^3+21 t^2+63 t+63\right) \sqrt{32 t^6+336 t^5+1449 t^4+3654 t^3+6615 t^2+7938 t+3969}}{147 (t+3)^2 \left(t^2+3 t+3\right)(1+ q t)}\begin{matrix}
{}\cr{}
\end{matrix}$}
\\ \hline
\end{tabular}
\caption{BPS and non-BPS extremal black holes in two-moduli THCY models, 3/5.}
\label{thcyt3}
\end{table}
}

\afterpage{
\clearpage
\thispagestyle{empty}
\begin{table}
\centering
\begin{tabular}{|c|c|c|c|}
\hline
\multirow{3}{*}{$\begin{matrix}
{\rm Polytope\ Label,}\cr {\rm Charge\ Matrix}
\end{matrix} $} &
$
\begin{pmatrix}
c & d \cr b & a
\end{pmatrix}
 $ & BPS Solution & non-BPS Solution \\
\cline{3-4}&  & Range of validity & Range of validity \\
\cline{2-4} & \multicolumn{3}{|c|}{Recombination Factor $ \begin{matrix}
{}\cr{}
\end{matrix}$}
\\ \hline
\hline
\multirow{3}{*}{$\begin{matrix}
(26,1)^{2,122}_{-240}\cr {} \cr \begin{pmatrix}
1&-3&1&1&4&0\cr 0&1&0&0&-1&1
\end{pmatrix}
\end{matrix} $} &
$\frac{1}{6}
\begin{pmatrix}
21&63\cr 7&2
\end{pmatrix}
 $ & $\frac{7(1 - 3 q) +\sqrt{7(1 -3 q)}}{7 q-2}$ &
$\frac{16 t^5+294 t^4+1764 t^3+4851 t^2+6615 t+3969}{441 (t+3)^2 \left(t^2+3
t+3\right)}$ \\
\cline{3-4}&  & $\frac{2}{7}<q<\frac{1}{3}$ & $ q>\frac{1}{3}$  \\
\cline{2-4} & \multicolumn{3}{|c|}{$\frac{\left(2 t^3+21 t^2+63 t+63\right) \sqrt{32 t^6+336 t^5+1449 t^4+3654 t^3+6615 t^2+7938 t+3969}}{147 (t+3)^2 \left(t^2+3 t+3\right)(1+ q t)}\begin{matrix}
{}\cr{}
\end{matrix}$}
\\ \hline
\multirow{3}{*}{$\begin{matrix}
(27,1)^{2,122}_{-240}\cr {} \cr \begin{pmatrix}
1&0&0&0&1&0\cr 0&1&1&1&-3&1
\end{pmatrix}
\end{matrix} $} &
$\frac{1}{6}
\begin{pmatrix}
7&2\cr 21&63
\end{pmatrix}
 $ & $\frac{7(3 - q) +\sqrt{7 q (q-3)}}{21 (q-3)}$ &
$\frac{441 t (3 t+1)^2 \left(3 t^2+3 t+1\right)}{3969 t^5+6615
t^4+4851 t^3+1764 t^2+294 t+16}$   \\
\cline{3-4}&  & $3<q<\frac{7}{2}$ & $0<q<3$ \\
\cline{2-4} & \multicolumn{3}{|c|}{$\frac{3 \left(63 t^3+63 t^2+21 t+2\right) \sqrt{3969 t^6+7938 t^5+6615 t^4+3654 t^3+1449 t^2+336 t+32}}{\left(3969 t^5+6615 t^4+4851
t^3+1764 t^2+294 t+16\right)(1+q t)} \begin{matrix}
{}\cr{}
\end{matrix}$}
\\ \hline
\multirow{3}{*}{$\begin{matrix}
(30,2)^{2,128}_{-252}\cr {} \cr \begin{pmatrix}
  0&1&0&0&1&2\cr 2&-3&2&2&-3&0
  \end{pmatrix}
\end{matrix} $} &
$\frac{2}{3}
\begin{pmatrix}
1&0\cr 6&27
\end{pmatrix}
 $ & $\frac{1}{2 q-9}$  & $\frac{3t - 1}{2t}$  \\
\cline{3-4}&  & $ q>\frac{9}{2}$ & $q<\frac{3}{2}$ \\
\cline{2-4} & \multicolumn{3}{|c|}{$\frac{3(3-q)}{3-2q+|q|} \begin{matrix}
{}\cr{}
\end{matrix}$}
\\ \hline
\multirow{3}{*}{$\begin{matrix}
(31,1)^{2,128}_{-252}\cr {}\cr \begin{pmatrix}
  1&-1&-1&2&0&1\cr -2&3&3&-5&1&0
\end{pmatrix}
\end{matrix} $} &
$\frac{1}{6}
\begin{pmatrix}
16&6\cr 42&109
\end{pmatrix}
 $ & $\frac{42-16 q+\sqrt{4 q^2-18 q+20}}{42 q-109}$ &

  $\frac{15369 t^5+29210 t^4+21840 t^3+7986 t^2+1416 t+96}{7522
t^5+14700 t^4+11374 t^3+4344 t^2+816 t+60}$  \\
\cline{3-4}&  & $\frac{109}{42}<q<\frac{8}{3}$ & $\frac{8}{5}<q<\frac{15369}{7522}$ \\
\cline{2-4} & \multicolumn{3}{|c|}{$\frac{\left(109 t^3+126 t^2+48 t+6\right) \sqrt{83881 t^6+221868 t^5+244500 t^4+143652 t^3+47448 t^2+8352 t+612}}{2 \left(3761 t^5+7350
t^4+5687 t^3+2172 t^2+408 t+30\right)(1+q t)} \begin{matrix}
{}\cr{}
\end{matrix}$}
\\ \hline
\multirow{3}{*}{$\begin{matrix}
(31,2)^{2,128}_{-252}\cr {} \cr \begin{pmatrix}
2&-3&-3&5&-1&0\cr 0&1&1&-1&1&2
\end{pmatrix}
\end{matrix} $} &
$\frac{1}{6}
\begin{pmatrix}
25&109\cr 5&1
\end{pmatrix}
 $ & $\frac{5(1 -5 q) +4 \sqrt{q (5 q-1)}}{5 q-1}$
& $\frac{(t+5)^2 \left(t^3+15 t^2+75 t+13\right)}{5 t^5+125
t^4+1362 t^3+7370 t^2+18425 t+15369}$  \\
\cline{3-4}&  & $\frac{1}{5}<q<\frac{25}{109}$ & $\frac{325}{15369}<q<\frac{1}{5}$ \\
\cline{2-4} & \multicolumn{3}{|c|}{$\frac{\left(t^3+15 t^2+75 t+109\right) \sqrt{t^6+30 t^5+375 t^4+3044 t^3+17535 t^2+59550 t+83881}}{\left(5 t^5+125 t^4+1362 t^3+7370
t^2+18425 t+15369\right)(1+q t)} \begin{matrix}
{}\cr{}
\end{matrix}$}
\\ \hline
\multirow{3}{*}{$\begin{matrix}
(32,1)^{2,128}_{-252}\cr {} \cr \begin{pmatrix}
1&-3&-3&1&4&0\cr 0&1&1&0&-1&1
\end{pmatrix}
\end{matrix} $} &
$\frac{1}{3}
\begin{pmatrix}
15&54\cr 4&1
\end{pmatrix}
 $ & $ \frac{5 - 18 q}{4 q-1}$ &
$\frac{t+9}{18}$   \\
\cline{3-4}&  & $\frac{1}{4}<q<\frac{5}{18}$ & $q>\frac{1}{2} $ \\
\cline{2-4} & \multicolumn{3}{|c|}{ $ 3 \begin{matrix}
{}\cr{}
\end{matrix}$}
\\ \hline
\multirow{3}{*}{$\begin{matrix}
(33,1)^{2,132}_{-260}\cr {} \cr \begin{pmatrix}
1&-1&-1&-1&2&0\cr 0&1&1&1&-1&2
\end{pmatrix}
\end{matrix} $} &
$ \frac{1}{3}
\begin{pmatrix}
4&7\cr 2&1
\end{pmatrix}
 $ & $\frac{2(1 -2 q) +\sqrt{q (2 q-1)}}{2 q-1}$ &
$\frac{(t+2)^2 \left(t^3+6 t^2+12 t+1\right)}{2 t^5+20 t^4+87 t^3+188
t^2+188 t+63}$  \\
\cline{3-4}&  & $\frac{1}{2}<q<\frac{4}{7}$ & $\frac{4}{63}<q<\frac{1}{2}$  \\
\cline{2-4} & \multicolumn{3}{|c|}{$\frac{\left(t^3+6 t^2+12 t+7\right) \sqrt{t^6+12 t^5+60 t^4+194 t^3+444 t^2+600 t+337}}{\left(2 t^5+20 t^4+87 t^3+188 t^2+188 t+63\right)(1+q
t)} \begin{matrix}
{}\cr{}
\end{matrix}$}
\\ \hline
\multirow{3}{*}{$\begin{matrix}
(33,2)^{2,132}_{-260}\cr {}\cr \begin{pmatrix}
1&0&0&0&1&2\cr -1&1&1&1&-2&0
\end{pmatrix}
\end{matrix} $} &
$\frac{1}{3}
\begin{pmatrix}
1&0\cr 3&7
\end{pmatrix}
 $ & $\frac{3-q+\sqrt{q^2-3 q+2}}{3 q-7}$  &
$\frac{63 t^5+127 t^4+66 t^3-21 t^2-27 t-6}{t \left(59 t^4+159
t^3+161 t^2+72 t+12\right)}$   \\
\cline{3-4}&  & $ q>\frac{7}{3}$ & $q<\frac{63}{59}$ \\
\cline{2-4} & \multicolumn{3}{|c|}{$\frac{\left(7 t^2+9 t+3\right) \sqrt{337 t^6+1422 t^5+2499 t^4+2322 t^3+1197 t^2+324 t+36}}{\left(59 t^4+159 t^3+161 t^2+72 t+12\right)(1+|q|t)} \begin{matrix}
{}\cr{}
\end{matrix}$}
\\ \hline
\end{tabular}
\caption{BPS and non-BPS extremal black holes in two-moduli THCY models, 4/5.}
\label{thcyt4}
\end{table}
}

\afterpage{
\clearpage
\thispagestyle{empty}
\begin{table}
\centering
\begin{tabular}{|c|c|c|c|}
\hline
\multirow{3}{*}{$\begin{matrix}
{\rm Polytope\ Label,}\cr {\rm Charge\ Matrix}
\end{matrix} $} &
$
\begin{pmatrix}
c & d \cr b & a
\end{pmatrix}
 $ & BPS Solution & non-BPS Solution \\
\cline{3-4}&  & Range of validity & Range of validity \\
\cline{2-4} & \multicolumn{3}{|c|}{Recombination Factor $ \begin{matrix}
{}\cr{}
\end{matrix}$}
\\ \hline
\hline
\multirow{3}{*}{$\begin{matrix}
(34,1)^{2,132}_{-260}\cr {}\cr \begin{pmatrix}
1&-2&-2&-4&7&0\cr 0&1&1&2&-3&1
\end{pmatrix}
\end{matrix} $} &
$\frac{1}{6}
\begin{pmatrix}
49&144 \cr 21&9
\end{pmatrix}
 $ & $ \frac{7( 3 -7 q) +\sqrt{q (7 q-3)}}{3(7q-3)}$ &
$\frac{(3 t+7)^2 \left(9 t^3+63 t^2+147 t+112\right)}{189
t^5+2205 t^4+10311 t^3+24108 t^2+28126 t+13072}$
 \\
\cline{3-4}&  & $\frac{3}{7}<q<\frac{49}{114}$ & $\frac{343}{817}<q<\frac{3}{7}$ \\
\cline{2-4} & \multicolumn{3}{|c|}{$\frac{3 \left(3 t^3+21 t^2+49 t+38\right) \sqrt{81 t^6+1134 t^5+6615 t^4+20682 t^3+36729 t^2+35280 t+14368}}{\left(189 t^5+2205 t^4+10311
t^3+24108 t^2+28126 t+13072\right)(1+ q t)} \begin{matrix}
{}\cr{}
\end{matrix}$}
\\ \hline
\multirow{3}{*}{$\begin{matrix}
(34,2)^{2,132}_{-260}\cr {}\cr \begin{pmatrix}
1&0&0&0&1&2\cr -1&2&2&4&-7&0
\end{pmatrix}
\end{matrix} $} &
$\frac{1}{3}
\begin{pmatrix}
1&0\cr 8&57
\end{pmatrix}
 $ & $\frac{8-q+\sqrt{q^2-8 q+7}}{8 q-57}$ &
$\frac{3268 t^5+2277 t^4+536 t^3+4 t^2-12 t-1}{2 t \left(262
t^4+232 t^3+93 t^2+16 t+1\right)}$
 \\
\cline{3-4}&  & $q>\frac{57}{8}$ & $q<\frac{817}{131}$ \\
\cline{2-4} & \multicolumn{3}{|c|}{$\frac{3 \left(19 t^2+8 t+1\right) \sqrt{3592 t^6+3912 t^5+2409 t^4+992 t^3+222 t^2+24 t+1}}{\left(524 t^4+464 t^3+186 t^2+32 t+2\right)(1+|q|t)} \begin{matrix}
{}\cr{}
\end{matrix}$}
\\ \hline
\multirow{3}{*}{$\begin{matrix}
(35,1)^{2,144}_{-284}\cr {} \cr \begin{pmatrix}
1&-2&-2&-2&5&0\cr 0&1&1&1&-2&1
\end{pmatrix}
\end{matrix} $} &
$\frac{1}{3}
\begin{pmatrix}
25&62\cr 10&4
\end{pmatrix}
 $ & $ \frac{5 (2 - 5 q) +\sqrt{q (5 q-2)}}{2 (5q - 2)}$ &
$\frac{(2 t+5)^2 \left(4 t^3+30 t^2+75 t+59\right)}{40 t^5+500
t^4+2514 t^3+6320 t^2+7900 t+3906}$  \\
\cline{3-4}&  & $\frac{2}{5}<q<\frac{25}{62}$ &
$\frac{1475}{3906}<q<\frac{2}{5}$ \\
\cline{2-4} & \multicolumn{3}{|c|}{$\frac{\left(4 t^3+30 t^2+75 t+62\right) \sqrt{16 t^6+240 t^5+1500 t^4+5068 t^3+9885 t^2+10650 t+4969}}{\left(40 t^5+500 t^4+2514
t^3+6320 t^2+7900 t+3906\right)(1+q t)} \begin{matrix}
{}\cr{}
\end{matrix}$}
\\ \hline
\multirow{3}{*}{$\begin{matrix}
(35,2)^{2,144}_{-284}\cr {}\cr \begin{pmatrix}
1&0&0&0&1&2\cr -1&2&2&2&-5&0
\end{pmatrix}
\end{matrix} $} &
$\frac{2}{3}
\begin{pmatrix}
1&0\cr 6&31
\end{pmatrix}
 $ & $ \frac{6-q+\sqrt{q^2-6 q+5}}{6 q-31}$ &
$-\frac{-1953 t^5-1865 t^4-570 t^3+6 t^2+27 t+3}{478 t^5+600
t^4+316 t^3+72 t^2+6 t}$  \\
\cline{3-4}&  & $q>\frac{31}{6}$ & $q<\frac{1953}{478}$ \\
\cline{2-4} & \multicolumn{3}{|c|}{$\frac{\left(31 t^2+18 t+3\right) \sqrt{4969 t^6+8514 t^5+7575 t^4+3996 t^3+1143 t^2+162 t+9}}{\left(478 t^4+600 t^3+316 t^2+72 t+6\right)(1+|q|
t)} \begin{matrix}
{}\cr{}
\end{matrix}$}
\\ \hline
\multirow{3}{*}{$\begin{matrix}
(36,1)^{2,272}_{-540}\cr {} \cr \begin{pmatrix}
0&0&0&2&3&1\cr 1&1&1&0&0&-3
\end{pmatrix}
\end{matrix} $} &
$\frac{1}{6}
\begin{pmatrix}
1&0\cr 3&9
\end{pmatrix}
 $ & $\frac{3-q+\sqrt{(q-3) q}}{3 (q-3)}$ &
$\frac{(3 t+1)^2 \left(9 t^3+9 t^2+3 t-2\right)}{t \left(27 t^4+45
t^3+51 t^2+24 t+4\right)}$  \\
\cline{3-4}&  & $q>3$ & $q<3$ \\
\cline{2-4} & \multicolumn{3}{|c|}{$\frac{3 \left(3 t^2+3 t+1\right) \sqrt{81 t^6+162 t^5+135 t^4+162 t^3+117 t^2+36 t+4}}{\left(27 t^4+45 t^3+51 t^2+24 t+4\right)(1+|q|t)} \begin{matrix}
{}\cr{}
\end{matrix}$}
\\ \hline
\end{tabular}
\caption{BPS and non-BPS extremal black holes in two-moduli THCY models, 5/5.}
\label{thcyt5}
\end{table}
}

\section{\label{cicys2}CICY Black Strings}

In this Appendix we will report the results of the study of extremal black
string attractors in two-moduli complete intersection Calabi-Yau (CICY)
models. In Tables \ref{cicymt1}-\ref{cicymt4}, we report the extremal black
string attractor solutions for 20 two-moduli CICY models. In particular,
again, the first model of Table \ref{cicymt1} is the one explicitly treated
in Sec. \ref{mcicy}. The first column of Tables \ref{cicymt1}-\ref{cicymt4}
indicates the CICY label as well as its configuration matrix, as from \cite%
{Ruehle}. Again, the superscripts in the configuration matrix are the Hodge
numbers $h_{1,1}=2$ and $h_{2,1}$, respectively, whereas the subscript is
the Euler number $\chi $ of the CICY model. On the other hand, as indicated
in the first row of Tables \ref{cicymt1}-\ref{cicymt4}, for black strings
only the non-BPS attractors need to be reported (the BPS ones are always
\textit{unique}) : so, the second column reports the intersection numbers of
the CICY model, the expression for $p$ as a rational polynomial function of $%
t$ for the non-BPS black string attractors, along with the range of $%
p=p^{1}/p^{2}$ for which the non-BPS moduli lie within the K\"{a}hler cone,
and finally the expression for the recombination factor. Note that in all
such Tables $t$ takes its critical value.

Our findings show that \textit{there are no multiple extremal non-BPS black
strings} for a given value of (supporting magnetic charge ratio) $p$.

We have numerically evaluated the recombination factor for the non-BPS black
string attractors in the entire moduli space for all allowed values of $p$;
we have found that the recombination factor is always greater than $1$ for
all the 20 two-moduli CICY models under consideration : therefore, all
non-BPS black strings in the 20 two-moduli CICY models listed in Tables \ref%
{cicymt1}-\ref{cicymt4} are \textit{stable }(for all the allowed values of $%
p $) against decay into their constituent BPS/anti-BPS black string pairs.

Since the two-moduli CICY models have been classified in a set of 36 models
(cfr. e.g. \cite{Ruehle}), a natural question arises : What about the
remaining 16 two-moduli CICY models, not reported in Tables \ref{cicymt1}-%
\ref{cicymt4}? Again, all such unlisted models have either $c=d=0$ or $a=b=0$%
, and thus the uniqueness of their non-BPS black string attractors has been
discussed in Secs. \ref{cdzero-2} and \ref{cdzero-3}, respectively.
Moreover, in such models the recombination factor of non-BPS black holes can
be computed exactly. For $c=d=0$, we find
\begin{equation}
R=3\left( \theta (-p)\frac{3b+ap}{9b+ap}+\theta (p)\right) .
\end{equation}%
Similarly, for $a=b=0$, we find
\begin{equation}
R=3\left( \theta (-p)\frac{d+3cp}{d+9cp}+\theta (p)\right) .
\end{equation}%
Using the K\"{a}hler cone condition, we can see that $R<1$ for all allowed
values of $p$, and hence all these non-BPS string attractors are stable.
Moreover we have numerically analysed the recombination factor for all the
models listed in tables \ref{cicymt1}-\ref{cicymt4} and we found that the
non-BPS attractors are stable for all the allowed values of $p$.

Let us observe that :

\begin{itemize}
\item As already noticed in remark 1 in App. \ref{cicymodel}, the first and
the third model of Table \ref{cicymt1} have the matrices of intersection
numbers reciprocally proportional, and they also share the same entries in
all partitions of the second column, such as the same expressions of $p$ as
a rational polynomial function of $t$ for non-BPS string attractors.
However, the fact that their intersection numbers are different
(notwithstanding being proportional) implies that the cubic constraint
defining the 5D scalar manifold $\mathcal{M}$ (i.e. $%
C_{IJK}t^{I}t^{J}t^{K}=1 $) yields different expressions for the moduli, and
the similarity of the various expressions observed here is a mere
coincidence. Furthermore, such models also have different $h_{2,1}$ (and
thus different $\chi $), as well as different configuration matrices.
\end{itemize}

\afterpage{
\clearpage
\thispagestyle{empty}
\begin{table}
  \centering
    \begin{tabular}{ ||c| c|c| c|| }
\hline\multirow{2}{*} {$\begin{matrix}  \cr {\rm CICY\ label,} \cr
{\rm Configuration\ Matrix}
\end{matrix}$} & $\begin{pmatrix}
c&d\cr b&a
\end{pmatrix}$
  & Non-BPS solution & Range of validity \\
  \cline{2-4} &\multicolumn{3}{|c||}{ Recombination factor $
\begin{matrix} {} \cr {} \cr {} \end{matrix}$} \\
  \hline\hline
  \multirow{2}{*} {$\begin{matrix}  \cr 7643 \cr {}
\begin{pmatrix}
0&0&2&1\cr 2&2&1&1
\end{pmatrix}^{2,46}_{-88}
\end{matrix}$} & $\frac{2}{3}\begin{pmatrix}
3&2\cr 1&0
\end{pmatrix}$
& $- \frac{6 t^4 + 45 t^3 + 93 t^2 + 156 t + 56}{6 t^3+27 t^2+107 t+120}$ &
$p<-\frac{7}{15}$
\\
\cline{2-4} &\multicolumn{3}{|c||}{ $\frac{\sqrt{p^2 \left(6 t^2+18 t+23\right)+2 p \left(9 t^2+8 t+6\right)+3 t^4+18 t^3+54 t^2+24 t+4}}{-p (2 t+3)+t^2+6 t+2}
\begin{matrix} {} \cr {} \cr {} \end{matrix}$} \\
\hline
\multirow{2}{*} {$\begin{matrix}  \cr 7644 \cr
\begin{pmatrix}
2&0&1&1&1\cr 0&2&1&1&1
\end{pmatrix}^{2,46}_{-88}
\end{matrix}$} & $\frac{2}{3}\begin{pmatrix}
3&1\cr 3&1
\end{pmatrix}$
& $-\frac{25 t^4+69 t^3+63 t^2+37 t+6}{6 t^4+37 t^3+63 t^2+69 t+25}$ & $-\frac{25}{6}<p<-\frac{6}{25}$
\\
\cline{2-4} &\multicolumn{3}{|c||}{ $\frac{\sqrt{p^2 \left(t^4+12 t^3+54 t^2+52 t+21\right)+6 p \left(t^4+4 t^3+10 t^2+4 t+1\right)+21 t^4+52 t^3+54 t^2+12 t+1}}{p \left(t^2+6
t+3\right)-3 t^2-6 t-1}
\begin{matrix} {} \cr {} \cr {} \end{matrix}$} \\
\hline
\multirow{2}{*} {$\begin{matrix}  \cr 7668 \cr
\begin{pmatrix}
0&2&1\cr 3&1&1
\end{pmatrix}^{2,47}_{-90}
\end{matrix}$} &$\frac{1}{2} \begin{pmatrix}
3&2\cr 1&0
\end{pmatrix}$
 &
$ -\frac{6 t^4+45 t^3+93 t^2+156 t+56}{6 t^3+27 t^2+107 t+120}$ & $p<-\frac{7}{15}$
\\
\cline{2-4} &\multicolumn{3}{|c||}{ $\frac{\sqrt{p^2 \left(6 t^2+18 t+23\right)+2 p \left(9 t^2+8 t+6\right)+3 t^4+18 t^3+54 t^2+24 t+4}}{-p (2 t+3)+t^2+6 t+2}
\begin{matrix} {} \cr {} \cr {} \end{matrix}$} \\
 \hline
\multirow{2}{*} {$\begin{matrix}  \cr 7725 \cr
\begin{pmatrix}
0&0&1&1&1\cr 2&2&1&1&1
\end{pmatrix}^{2,50}_{-96} \end{matrix}$} & $\frac{2}{3}\begin{pmatrix}
3&3\cr 1&0
\end{pmatrix}$
& $-\frac{2 t^4+15 t^3+33 t^2+51 t+24}{2 t^3+9 t^2+31 t+33}$ &
$p<-\frac{8}{11}$ \\
\cline{2-4} &\multicolumn{3}{|c||}{ $\frac{\sqrt{3} \sqrt{p^2 \left(2 t^2+6 t+7\right)+p \left(6 t^2+8 t+6\right)+t^4+6 t^3+18 t^2+12 t+3}}{-p (2 t+3)+t^2+6 t+3}
\begin{matrix} {} \cr {} \cr {} \end{matrix}$} \\ \hline
\multirow{2}{*} {$\begin{matrix}  \cr 7726 \cr
\begin{pmatrix}
0&1&1&1&1\cr 2&1&1&1&1
\end{pmatrix}^{2,50}_{-96}\end{matrix}$} &$\frac{1}{3}\begin{pmatrix}
6&4\cr 4&1
\end{pmatrix}$ &
$-\frac{11 t^4+46 t^3+66 t^2+56 t+16}{2 t^4+17 t^3+42 t^2+62 t+32}$ &
$-\frac{11}{2}<p<-\frac{1}{2}$ \\
\cline{2-4} &\multicolumn{3}{|c||}{ $\frac{\sqrt{p^2 \left(t^4+16 t^3+96 t^2+136 t+76\right)+8 p \left(t^4+6 t^3+21 t^2+16 t+6\right)+4 \left(9 t^4+34 t^3+54 t^2+24 t+4\right)}}{p
\left(t^2+8 t+6\right)-4 \left(t^2+3 t+1\right)}
\begin{matrix} {} \cr {} \cr {} \end{matrix}$} \\ \hline
\multirow{2}{*} {$\begin{matrix}  \cr 7758 \cr
\begin{pmatrix}
0&2&1\cr 2&1&2
\end{pmatrix}^{2,52}_{-100}\end{matrix}$} &$\frac{1}{3}\begin{pmatrix}
5&2\cr 2&0
\end{pmatrix}$ &
$-\frac{24 t^4+150 t^3+249 t^2+365 t+84}{24 t^3+90 t^2+319 t+305}$ & $p<-\frac{84}{305}$ \\
\cline{2-4} &\multicolumn{3}{|c||}{ $\frac{\sqrt{p^2 \left(24 t^2+60 t+67\right)+4 p \left(15 t^2+8 t+5\right)+2 \left(6 t^4+30 t^3+75 t^2+20 t+2\right)}}{p (4 t+5)-2 \left(t^2+5
t+1\right)}
\begin{matrix} {} \cr {} \cr {} \end{matrix}$} \\
\hline
\multirow{2}{*} {$\begin{matrix}  \cr 7759 \cr
\begin{pmatrix}
0&2&1&1\cr 2&1&1&1
\end{pmatrix}^{2,52}_{-100}\end{matrix}$} & $\frac{1}{3}\begin{pmatrix}
5&2\cr 4&1
\end{pmatrix}$
& $-\frac{123 t^4+421 t^3+471 t^2+345 t+68}{22 t^4+177 t^3+375 t^2+521 t+237}$
& $-\frac{123}{22}<p<-\frac{68}{237}$ \\
\cline{2-4} &\multicolumn{3}{|c||}{ $\frac{\sqrt{p^2 \left(t^4+16 t^3+96 t^2+116 t+59\right)+4 p \left(2 t^4+10 t^3+33 t^2+16 t+5\right)+2 \left(19 t^4+58 t^3+75 t^2+20 t+2\right)}}{p
\left(t^2+8 t+5\right)-2 \left(2 t^2+5 t+1\right)}
\begin{matrix} {} \cr {} \cr {} \end{matrix}$} \\ \hline
\multirow{2}{*} {$\begin{matrix}  \cr 7761 \cr
\begin{pmatrix}
1&1&1&1&1\cr 1&1&1&1&1
\end{pmatrix}^{2,52}_{-100}\end{matrix}$} & $\frac{5}{6}\begin{pmatrix}
2&1\cr 2&1
\end{pmatrix}$ &
$-\frac{21 t^4+62 t^3+66 t^2+39 t+8}{8 t^4+39 t^3+66 t^2+62 t+21}$ &
$-\frac{21}{8}<p<-\frac{8}{21}$ \\
\cline{2-4} &\multicolumn{3}{|c||}{ $\frac{\sqrt{p^2 \left(t^4+8 t^3+24 t^2+22 t+8\right)+2 p \left(2 t^4+8 t^3+15 t^2+8 t+2\right)+8 t^4+22 t^3+24 t^2+8 t+1}}{p \left(t^2+4
t+2\right)-2 t^2-4 t-1}
\begin{matrix} {} \cr {} \cr {} \end{matrix}$} \\ \hline
  \end{tabular}
  \caption{Non-BPS extremal black strings in two-moduli CICY models, 1/3.}
\label{cicymt1}
\end{table}
  }

\afterpage{
\clearpage
\thispagestyle{empty}
\begin{table}
  \centering
    \begin{tabular}{ ||c| c|c| c|| }
\hline\multirow{2}{*} {$\begin{matrix}  \cr {\rm CICY\ label,} \cr
{\rm Configuration\ Matrix}
\end{matrix}$} &  $\begin{pmatrix}
c&d\cr b&a
\end{pmatrix}$
  & Non-BPS solution & Range of validity \\
  \cline{2-4} &\multicolumn{3}{|c||}{Recombination factor $
\begin{matrix} {} \cr {} \cr {} \end{matrix}$} \\
  \hline\hline
  \multirow{2}{*} {$\begin{matrix}  \cr 7799 \cr
\begin{pmatrix}
2&1&1\cr 1&2&1
\end{pmatrix}^{2,55}_{-106}\end{matrix}$} & $\frac{1}{6}\begin{pmatrix}
7&2\cr 7&2
\end{pmatrix}$ &
$-\frac{276 t^4+749 t^3+651 t^2+384 t+56}{56 t^4+384 t^3+651 t^2+749 t+276}$ &
$-\frac{69}{14}<p<-\frac{14}{69}$ \\
\cline{2-4} &\multicolumn{3}{|c||}{ $\frac{\sqrt{p^2 \left(4 t^4+56 t^3+294 t^2+286 t+119\right)+2 p \left(14 t^4+56 t^3+159 t^2+56 t+14\right)+119 t^4+286 t^3+294 t^2+56
t+4}}{p \left(2 t^2+14 t+7\right)-7 t^2-14 t-2}
\begin{matrix} {} \cr {} \cr {} \end{matrix}$} \\ \hline
\multirow{2}{*} {$\begin{matrix}  \cr 7807 \cr
\begin{pmatrix}
0&1&1&1\cr 2&2&1&1
\end{pmatrix}^{2,56}_{-108}\end{matrix}$} & $\frac{1}{3}\begin{pmatrix}
5&4\cr 2&0
\end{pmatrix}$ &
$-\frac{24 t^4+150 t^3+273 t^2+355 t+136}{24 t^3+90 t^2+263 t+235}$
&$p<-\frac{136}{235}$\\
\cline{2-4} &\multicolumn{3}{|c||}{ $\frac{\sqrt{p^2 \left(24 t^2+60 t+59\right)+p \left(60 t^2+64 t+40\right)+2 \left(6 t^4+30 t^3+75 t^2+40 t+8\right)}}{p (4 t+5)-2 \left(t^2+5
t+2\right)}
\begin{matrix} {} \cr {} \cr {} \end{matrix}$} \\ \hline
\multirow{2}{*} {$\begin{matrix}  \cr 7808 \cr
\begin{pmatrix}
0&1&1&1\cr 3&1&1&1
\end{pmatrix}^{2,56}_{-108}\end{matrix}$} & $\frac{1}{2}\begin{pmatrix}
3&3\cr 1&0
\end{pmatrix}$ &
$-\frac{2 t^4+15 t^3+33 t^2+51 t+24}{2 t^3+9 t^2+31 t+33}$ & $p<-\frac{8}{11}$\\
\cline{2-4} &\multicolumn{3}{|c||}{ $\frac{\sqrt{3} \sqrt{p^2 \left(2 t^2+6 t+7\right)+p \left(6 t^2+8 t+6\right)+t^4+6 t^3+18 t^2+12 t+3}}{-p (2 t+3)+t^2+6 t+3}
\begin{matrix} {} \cr {} \cr {} \end{matrix}$} \\ \hline
\multirow{2}{*} {$\begin{matrix}  \cr 7809 \cr
\begin{pmatrix}
1&1&1&1\cr 2&1&1&1
\end{pmatrix}^{2,56}_{-108}\end{matrix}$} &$\frac{1}{6}\begin{pmatrix}
9&5\cr 7&2
\end{pmatrix}$ &
$-\frac{1196 t^4+4277 t^3+5223 t^2+3809 t+920}{248 t^4+1832 t^3+3891 t^2+4985 t+2219}$ & $-\frac{299}{62}<p<-\frac{920}{2219}$\\
\cline{2-4} &\multicolumn{3}{|c||}{ $\frac{\sqrt{p^2 \left(4 t^4+56 t^3+294 t^2+358 t+173\right)+2 p \left(14 t^4+72 t^3+219 t^2+140 t+45\right)+111 t^4+358 t^3+486 t^2+180
t+25}}{p \left(2 t^2+14 t+9\right)-7 t^2-18 t-5}
\begin{matrix} {} \cr {} \cr {} \end{matrix}$} \\ \hline
\multirow{2}{*} {$\begin{matrix}  \cr 7821 \cr
 \begin{pmatrix}
1&1&1\cr 2&2&1
\end{pmatrix}^{2,58}_{-112}\end{matrix}$} &$\frac{1}{6}\begin{pmatrix}
8&5\cr 4&0
\end{pmatrix}$ &
$ -\frac{24 t^4+120 t^3+174 t^2+182 t+55}{24 t^3+72 t^2+170 t+122}$ & $p<-\frac{55}{122}$
\\
\cline{2-4} &\multicolumn{3}{|c||}{ $\frac{\sqrt{8 p^2 \left(12 t^2+24 t+19\right)+16 p \left(12 t^2+10 t+5\right)+48 t^4+192 t^3+384 t^2+160 t+25}}{-8 p (t+1)+4 t^2+16 t+5}
\begin{matrix} {} \cr {} \cr {} \end{matrix}$} \\ \hline
\multirow{2}{*} {$\begin{matrix}  \cr 7833 \cr{}
 \begin{pmatrix}
2&1\cr 1&3
\end{pmatrix}^{2,59}_{-114}\end{matrix}$} &$\frac{1}{6}\begin{pmatrix}
7&2\cr 3&0
\end{pmatrix}$ &
$-\frac{162 t^4+945 t^3+1431 t^2+2016 t+344}{9 \left(18 t^3+63 t^2+217 t+196\right)}$
& $p<-\frac{86}{441}$ \\ \cline{2-4} &\multicolumn{3}{|c||}{ $\frac{\sqrt{9 p^2 \left(6 t^2+14 t+15\right)+2 p \left(63 t^2+24 t+14\right)+27 t^4+126 t^3+294 t^2+56 t+4}}{-p (6 t+7)+3 t^2+14 t+2}
\begin{matrix} {} \cr {} \cr {} \end{matrix}$} \\ \hline
\multirow{2}{*} {$\begin{matrix}  \cr 7844 \cr {}
 \begin{pmatrix}
2&1\cr 2&2
\end{pmatrix}^{2,62}_{-120}\end{matrix}$} &$\frac{1}{3}\begin{pmatrix}
3&1\cr 2&0
\end{pmatrix}$ &
$ -\frac{24 t^4+90 t^3+93 t^2+78 t+14}{24 t^3+54 t^2+107 t+60}$ & $p<-\frac{7}{30}$ \\ \cline{2-4} &\multicolumn{3}{|c||}{ $\frac{\sqrt{p^2 \left(24 t^2+36 t+23\right)+2 p \left(18 t^2+8 t+3\right)+12 t^4+36 t^3+54 t^2+12 t+1}}{-p (4 t+3)+2 t^2+6 t+1}
\begin{matrix} {} \cr {} \cr {} \end{matrix}$} \\ \hline
\multirow{2}{*} {$\begin{matrix}  \cr 7853 \cr {}
\begin{pmatrix}
0&2&1\cr 2&2&1
\end{pmatrix}^{2,64}_{-124}\end{matrix}$} &$\frac{2}{3}\begin{pmatrix}
2&1\cr 1&0
\end{pmatrix}$ &
$ -\frac{3 t^4+15 t^3+21 t^2+23 t+6}{3 t^3+9 t^2+23 t+17}$ & $p<-\frac{6}{17}$ \\ \cline{2-4} &\multicolumn{3}{|c||}{ $\frac{\sqrt{2 p^2 \left(3 t^2+6 t+5\right)+4 p \left(3 t^2+2 t+1\right)+3 t^4+12 t^3+24 t^2+8 t+1}}{-2 p (t+1)+t^2+4 t+1}
\begin{matrix} {} \cr {} \cr {} \end{matrix}$}\\ \hline
 \end{tabular}
   \caption{Non-BPS extremal black strings in two-moduli CICY models, 2/3.}
\label{cicymt3}
\end{table}
  }

\afterpage{
\clearpage
\thispagestyle{empty}
\begin{table}
  \centering
    \begin{tabular}{ ||c| c|c| c|| }
\hline\multirow{2}{*} {$\begin{matrix}  \cr {\rm CICY\ label,} \cr
{\rm Configuration\ Matrix}
\end{matrix}$} &  $\begin{pmatrix}
c&d\cr b&a
\end{pmatrix}$
  & Non-BPS solution & Range of validity \\
  \cline{2-4} &\multicolumn{3}{|c||}{ Recombination factor $
\begin{matrix} {} \cr {} \cr {} \end{matrix}$} \\
  \hline\hline
\multirow{2}{*} {$\begin{matrix}  \cr 7863 \cr {} \begin{pmatrix}
2&1&1\cr 2&1&1
\end{pmatrix}^{2,66}_{-128}\end{matrix}$} &$\frac{1}{3}\begin{pmatrix}
3&1\cr 3&1
\end{pmatrix}$ & $
-\frac{25 t^4+69 t^3+63 t^2+37 t+6}{6 t^4+37 t^3+63 t^2+69 t+25}$ &
$-\frac{25}{6}<p<-\frac{6}{25}$\\ \cline{2-4} &\multicolumn{3}{|c||}{ $\frac{\sqrt{p^2 \left(t^4+12 t^3+54 t^2+52 t+21\right)+6 p \left(t^4+4 t^3+10 t^2+4 t+1\right)+21 t^4+52 t^3+54 t^2+12 t+1}}{p \left(t^2+6
t+3\right)-3 t^2-6 t-1}
\begin{matrix} {} \cr {} \cr {} \end{matrix}$} \\ \hline
\multirow{2}{*} {$\begin{matrix}  \cr 7868 \cr {} \begin{pmatrix}
1&1&1\cr 3&1&1
\end{pmatrix}^{2,68}_{-132}\end{matrix}$} &$\frac{1}{6}\begin{pmatrix}
7&5\cr 3&0
\end{pmatrix}$ & $
-\frac{162 t^4+945 t^3+1593 t^2+1953 t+680}{9 \left(18 t^3+63 t^2+175 t+147\right)}$
& $p<-\frac{680}{1323}$ \\ \cline{2-4} &\multicolumn{3}{|c||}{ $\frac{\sqrt{9 p^2 \left(6 t^2+14 t+13\right)+2 p \left(63 t^2+60 t+35\right)+27 t^4+126 t^3+294 t^2+140 t+25}}{-p (6 t+7)+3 t^2+14 t+5}
\begin{matrix} {} \cr {} \cr {} \end{matrix}$} \\ \hline
\multirow{2}{*} {$\begin{matrix}  \cr 7883 \cr {} \begin{pmatrix}
2&1\cr 3&1
\end{pmatrix}^{2,77}_{-150}\end{matrix}$} &$\frac{1}{6}\begin{pmatrix}
5&2\cr 3&0
\end{pmatrix}$ & $
-\frac{162 t^4+675 t^3+783 t^2+720 t+152}{9 \left(18 t^3+45 t^2+97 t+60\right)}$
& $p<-\frac{38}{135}$\\ \cline{2-4} &\multicolumn{3}{|c||}{ $\frac{\sqrt{9 p^2 \left(6 t^2+10 t+7\right)+p \left(90 t^2+48 t+20\right)+27 t^4+90 t^3+150 t^2+40 t+4}}{-p (6 t+5)+3 t^2+10 t+2}
\begin{matrix} {} \cr {} \cr {} \end{matrix}$} \\ \hline
\multirow{2}{*} {$\begin{matrix}  \cr 7884 \cr {} \begin{pmatrix}
3\cr 3
\end{pmatrix}^{2,83}_{-162}\end{matrix}$} &$\frac{1}{2}\begin{pmatrix}
1\cr 1
\end{pmatrix}$ & $
-\frac{t \left(2 t^3+5 t^2+3 t+2\right)}{2 t^3+3 t^2+5 t+2}$ & $p<0$\\ \cline{2-4} &\multicolumn{3}{|c||}{ $\frac{\sqrt{3} \sqrt{p^2 \left(2 t^2+2 t+1\right)+2 p t^2+t^2 \left(t^2+2 t+2\right)}}{2 p t+p-t (t+2)}
\begin{matrix} {} \cr {} \cr {} \end{matrix}$}
\\ \hline
\end{tabular}
   \caption{Non-BPS extremal black strings in two-moduli CICY models, 3/3.}
\label{cicymt4}
\end{table}
  }

\section{\label{thcy2}THCY Black Strings}

In this Appendix we will report the most interesting results of our paper,
namely the study of non-BPS extremal black string attractors in two-moduli
toric hypersurface Calabi-Yau (THCY) models. Our findings show that \textit{%
there are multiple extremal non-BPS black strings} in most of the two-moduli
THCY models under consideration, and that they are also \textit{stable},
\textit{at least} in some subsector of the allowed range for the supporting
magnetic charge ratio $p$. We should remark that the existence of \textit{%
multiple} non-BPS black string attractors was not observed in the analysis
of \cite{LSVY}, and the fact that we also obtain that they can be stable
adds interest and physical relevance to such a finding.

In Tables \ref{thcymt1}-\ref{thcymt8}, we report the non-BPS extremal black
string attractor solutions for 37 two-moduli THCY models, and we do not
include the model $(3,1)_{-144}^{2,74}$ already treated in detail in\ Sec. %
\ref{thcy}. The first column of Tables \ref{thcymt1}-\ref{thcymt8} indicates
the polytope label as well as the charge matrix of the ambient toric
variety, as from \cite{Ruehle}. Again, the superscripts in the configuration
matrix are the Hodge numbers $h_{1,1}=2$ and $h_{2,1}$, respectively,
whereas the subscript is the Euler number $\chi $ of the THCY model. On the
other hand, in the various partitions of their second column we respectively
specify : the triple intersection numbers, the non-BPS black string
attractor solution, along with the constraints on $p$ for the existence of
the single solution, and of the \textit{multiple} solution (if any), as
well. Also, we report the analytical form of the recombination factor. As in
previous Tables, $t$ takes the critical value in all these formulae. A
numerical evaluation of the recombination factor allows us to conclude that
most of the non-BPS black string attractors, both in their single and
\textit{multiple} solution regimes, remain stable for certain range of $p$
and then become unstable; in particular, the range of values of $p$ which
supports stable non-BPS black string attractors is specified in all models.

Again, since the Calabi-Yau threefolds with $h_{1,1}=2$ constructed as
hypersurfaces in toric varieties (THCYs) associated with the 36 reflexive
four-dimensional polytopes with six rays, and their various triangulations,
have been classified in a set of 48 models (cfr. App. B of \cite{Ruehle}),
one might ask : What about the remaining 10 two-moduli THCY models, not
reported in Tables \ref{thcymt1}-\ref{thcymt8}? Some of such unlisted models
have either $c=d=0$ or $a=b=0$, and thus the uniqueness of their non-BPS
black string attractors has been discussed in Secs. \ref{cdzero-2} and \ref%
{cdzero-3}. Apart from these, we have again not included a few models for
which the triple intersection numbers are identical to some of the models
already discussed here; they are related to each other by a flop (for more
detail, see e.g. the discussion in \cite{Ruehle}).\medskip

Again, we remark that

\begin{itemize}
\item The second model of Table \ref{thcymt1} is marked red, because for
such a model one of the two eigenvalues of $G_{IJ}$ is always negative in
the ranges of $p$ supporting either the BPS or the non-BPS black string
attractors. Thus, this model does not give rise to physically consistent
black string attractors.
\end{itemize}

\afterpage{
\clearpage
\thispagestyle{empty}
\begin{sidewaystable}
  \centering
    \begin{tabular}{ ||c| c|c|c|c|| }
\hline\multirow{3}{*} {$\begin{matrix}  \cr {\rm Polytope\ label,} \cr {}
{\rm Charge\ Matrix}
\end{matrix}$}
 & $\begin{pmatrix}
c&d\cr b&a
\end{pmatrix}$  & Non-BPS solution &
$\begin{matrix}\rm
 \rm Range\ of\ validity \cr \rm for\ Single\ solution
\end{matrix}$ & $\begin{matrix}\rm
 \rm Range\ of\ validity \cr \rm for\ Multiple\ solutions
\end{matrix}$
\\ \cline{2-5} &\multicolumn{4}{|c||}{ Recombination factor $
\begin{matrix} {} \cr {} \cr {} \end{matrix}$} \\
\cline{2-5} & \multicolumn{4}{|c||}{Stability of the non-BPS attractor} \\
  \hline\hline
  \multirow{3}{*} {$\begin{matrix}   (1,1)^{2,29}_{-54}\cr {} \cr {}
\begin{pmatrix}
1&0&0&0&1&1\cr 0&1&1&1&0&0
\end{pmatrix}
\end{matrix}$}
&$\frac{1}{6}\begin{pmatrix}
1&0\cr 1&0
\end{pmatrix}$ & $-\frac{t \left(2 t^3+5 t^2+3 t+2\right)}{2 t^3+3 t^2+5 t+2}$
 &  $p<0$ & NA \\ \cline{2-5} &\multicolumn{4}{|c||}{ $\frac{\sqrt{3} \sqrt{2 p t^2+t^2 \left(2+2 t+t^2\right)+p^2 \left(1+2 t+2 t^2\right)}}{t (2+t)+\text{$|$p$|$} (1+2 t)}
\begin{matrix} {} \cr {} \cr {} \end{matrix}$}
\\ \cline{2-5} & \multicolumn{4}{|c||}{All non-BPS attractors are stable}
\\ \hline
  \multirow{3}{*} {\color{red}
  $\begin{matrix}   (2,1)^{2,38}_{-72}\cr {} \cr {}
\begin{pmatrix}
0&0&1&1&0&1\cr 1&1&0&0&1&-3
\end{pmatrix}
\end{matrix}$}
&$\frac{1}{6}\begin{pmatrix}
1&9\cr 3&9
\end{pmatrix}$ &
$\frac{9 t^4-3 t^3-9 t^2-47 t-12}{45 t^3+45 t^2+15 t-7}$
 & $p<\frac{12}{7}$ & $p>\frac{12}{7}$  \\ \cline{2-5} &\multicolumn{4}{|c||}{ $\frac{\sqrt{3} \sqrt{27+12 t+2 t^2-48 t^3+3 t^4+6 p \left(1+12 t+28 t^2+4 t^3+3 t^4\right)+p^2 \left(-17-48 t+18 t^2+36 t^3+27 t^4\right)}}{9+2
t+3 t^2+\text{$|$p$|$} (1+3 t)^2}
\begin{matrix} {} \cr {} \cr {} \end{matrix}$}
\\ \cline{2-5} & \multicolumn{4}{|c||}{ }
\\ \hline
  \multirow{3}{*} {$\begin{matrix}   (4,1)^{2,74}_{-144}\cr {} \cr {}
\begin{pmatrix}
-1&2&1&1&3&0\cr 1&-1&0&0&-1&1
\end{pmatrix}
\end{matrix}$}
&$\frac{1}{6}\begin{pmatrix}
3&3\cr 3&2
\end{pmatrix}$ &
$-\frac{t \left(16 t^3+45 t^2+45 t+15\right)}{8 t^4+28 t^3+27 t^2+3 t-3}$
 & $p>0 \ \& \ p<-2$  & $-2<p<-1.78$ \\ \cline{2-5} &\multicolumn{4}{|c||}{ $\frac{\sqrt{9+36 t+54 t^2+42 t^3+15 t^4+6 p \left(3+12 t+15 t^2+8 t^3+2 t^4\right)+p^2 \left(9+42 t+54 t^2+24 t^3+4 t^4\right)}}{3 (1+t)^2+\text{$|$p$|$}
\left(3+6 t+2 t^2\right)}
\begin{matrix} {} \cr {} \cr {} \end{matrix}$}
\\ \cline{2-5} & \multicolumn{4}{|c||}{ Stable solution for $-44.856<p<-1.78$ }
\\ \hline
  \multirow{3}{*} {$\begin{matrix}   (4,2)^{2,74}_{-144}\cr {} \cr {}
\begin{pmatrix}
2&-1&1&1&0&3\cr -1&1&0&0&1&-1
\end{pmatrix}
\end{matrix}$}
&$\frac{1}{6}\begin{pmatrix}
3&3\cr 3&2
\end{pmatrix}$ &
$-\frac{t \left(16 t^3+45 t^2+45 t+15\right)}{8 t^4+28 t^3+27 t^2+3 t-3}$
 & $p>0 \ \& \ p<-2$  & $-2<p<-1.78$ \\ \cline{2-5} &\multicolumn{4}{|c||}{ $\frac{\sqrt{9+36 t+54 t^2+42 t^3+15 t^4+6 p \left(3+12 t+15 t^2+8 t^3+2 t^4\right)+p^2 \left(9+42 t+54 t^2+24 t^3+4 t^4\right)}}{3 (1+t)^2+\text{$|$p$|$}
\left(3+6 t+2 t^2\right)}
\begin{matrix} {} \cr {} \cr {} \end{matrix}$}
\\ \cline{2-5} & \multicolumn{4}{|c||}{Stable solution for $-44.856<p<-1.78$ }
\\ \hline
  \multirow{3}{*} {$\begin{matrix}   (5,1)^{2,83}_{-162}\cr {} \cr {}
\begin{pmatrix}
1&0&1&0&0&1\cr 0&1&0&1&1&0
\end{pmatrix}
\end{matrix}$}
&$\frac{1}{2}\begin{pmatrix}
1&0\cr 1&0
\end{pmatrix}$ &
$-\frac{t \left(2 t^3+5 t^2+3 t+2\right)}{2 t^3+3 t^2+5 t+2}$
 & $p<0$ & NA  \\ \cline{2-5} &\multicolumn{4}{|c||}{ $\frac{\sqrt{3} \sqrt{2 p t^2+t^2 \left(2+2 t+t^2\right)+p^2 \left(1+2 t+2 t^2\right)}}{t (2+t)+\text{$|$p$|$} (1+2 t)}
\begin{matrix} {} \cr {} \cr {} \end{matrix}$}
\\ \cline{2-5} & \multicolumn{4}{|c||}{All non-BPS attractors are stable }
\\ \hline
    \end{tabular}
\caption{Non-BPS extremal (multiple) black strings in two-moduli THCY models, 1/8.}
\label{thcymt1}
\end{sidewaystable}
\clearpage
}

\afterpage{
\clearpage
\thispagestyle{empty}
\begin{sidewaystable}
  \centering
    \begin{tabular}{ ||c| c|c|c|c|| }
\hline\multirow{3}{*} {$\begin{matrix}{\rm Polytope\ label,}  \cr {} \cr {}
{\rm Charge\ Matrix}
\end{matrix}$}
 & $\begin{pmatrix}
c&d\cr b&a
\end{pmatrix}$  & Non-BPS solution &
$\begin{matrix}\rm
 \rm Range\ of\ validity \cr \rm for\ Single\ solution
\end{matrix}$ & $\begin{matrix}\rm
 \rm Range\ of\ validity \cr \rm for\ Multiple\ solutions
\end{matrix}$
\\ \cline{2-5} &\multicolumn{4}{|c||}{ Recombination factor $
\begin{matrix} {} \cr {} \cr {} \end{matrix}$} \\ \cline{2-5} & \multicolumn{4}{|c||}{Stability of non-BPS attractors }
\\ \hline \hline
\multirow{3}{*} {$\begin{matrix}   (6,1)^{2,84}_{-164}\cr {} \cr {}
\begin{pmatrix}
1&-1&1&1&2&0\cr 0&1&0&0&-1&1
\end{pmatrix}
\end{matrix}$}
&$\frac{1}{6}\begin{pmatrix}
5&5\cr 5&3
\end{pmatrix}$ &
$-\frac{t \left(27 t^3+75 t^2+75 t+25\right)}{12 t^4+45 t^3+45 t^2+5 t-5}$
 & $p>0\ \& \ p<-\frac{9}{4}$  & $-\frac{9}{4}<p<-1.87$  \\ \cline{2-5} &\multicolumn{4}{|c||}{ $\frac{\sqrt{10 p \left(5+20 t+24 t^2+12 t^3+3 t^4\right)+5 \left(5+20 t+30 t^2+24 t^3+9 t^4\right)+p^2 \left(25+120 t+150 t^2+60 t^3+9
t^4\right)}}{5 (1+t)^2+\text{$|$p$|$} \left(5+10 t+3 t^2\right)}
\begin{matrix} {} \cr {} \cr {} \end{matrix}$}
\\ \cline{2-5} & \multicolumn{4}{|c||}{Stable solution for $-37.02<p<-1.87$ }
\\ \hline
  \multirow{3}{*} {$\begin{matrix}   (10,1)^{2,86}_{-168} \cr {} \cr {}
\begin{pmatrix}
1&0&0&1&0&1\cr 0&1&1&0&1&-1
\end{pmatrix}
\end{matrix}$}
&$\frac{1}{6}\begin{pmatrix}
3&0\cr 5&5
\end{pmatrix}$ &
$-\frac{9 t (25 t^3+55 t^2+45 t+18)}{200 t^4+675 t^3+945 t^2+675 t+162}$
 & $-\frac{9}{8}<p<0$ & NA \\ \cline{2-5} &\multicolumn{4}{|c||}{ $\frac{\sqrt{9 t^2 \left(6+10 t+5 t^2\right)+10 p t^2 \left(9+12 t+5 t^2\right)+p^2 \left(27+90 t+150 t^2+100 t^3+25 t^4\right)}}{t (6+5
t)+\text{$|$p$|$} \left(3+10 t+5 t^2\right)}
\begin{matrix} {} \cr {} \cr {} \end{matrix}$}
 \\ \cline{2-5} & \multicolumn{4}{|c||}{All non-BPS attractors are stable }
\\ \hline
  \multirow{3}{*} {$\begin{matrix} (11,1)^{2,86}_{-168} \cr {} \cr {}
 \begin{pmatrix}
  1&0&0&1&1&0\cr 0&1&1&-1&-1&1
  \end{pmatrix} \end{matrix}$} &$\frac{1}{6}\begin{pmatrix}
  3&0\cr 7&11
  \end{pmatrix}$
  & $ - \frac{9 t (49 t^3+91 t^2+63 t+18)}{704 t^4+1827 t^3+1917 t^2+945 t+162}$
  & $- \frac{441}{704}<p<0$ & NA\\ \cline{2-5} &\multicolumn{4}{|c||}{ $\frac{\sqrt{9 t^2 \left(6+14 t+9 t^2\right)+2 p t^2 \left(63+132 t+77 t^2\right)+p^2 \left(27+126 t+294 t^2+308 t^3+121 t^4\right)}}{t
(6+7 t)+\text{$|$p$|$} \left(3+14 t+11 t^2\right)}
\begin{matrix} {} \cr {} \cr {} \end{matrix}$}
\\ \cline{2-5} & \multicolumn{4}{|c||}{All non-BPS attractors are stable }
\\ \hline
  \multirow{3}{*} {$\begin{matrix} (12,1)^{2,86}_{-168}  \cr {} \cr {}
\begin{pmatrix}
1&-1&-1&1&2&0\cr 0&1&1&0&-1&1
\end{pmatrix}\end{matrix}$} &$\frac{1}{6}\begin{pmatrix}
10&14\cr 6&3
\end{pmatrix}$ &
$ - \frac{81 t^4+474 t^3+1062 t^2+1110 t+448}{36 t^4+243 t^3+630 t^2+762 t+354}$
& $-\frac{9}{4}<p<-\frac{224}{177} $ & NA\\ \cline{2-5} &\multicolumn{4}{|c||}{ $\frac{\sqrt{3 p^2 \left(44+92 t+72 t^2+24 t^3+3 t^4\right)+4 p \left(70+168 t+153 t^2+60 t^3+9 t^4\right)+4 \left(49+140 t+150 t^2+69
t^3+12 t^4\right)}}{14+20 t+6 t^2+\text{$|$p$|$} \left(10+12 t+3 t^2\right)}
\begin{matrix} {} \cr {} \cr {} \end{matrix}$}
\\ \cline{2-5} & \multicolumn{4}{|c||}{All non-BPS attractors are stable }
\\ \hline
  \multirow{3}{*} {$\begin{matrix} (14,1)^{2,86}_{-168}  \cr {} \cr {}
 \begin{pmatrix}
1&-1&-1&1&3&0\cr 0&1&1&0&-2&1
\end{pmatrix}\end{matrix}$} &$\frac{1}{6}\begin{pmatrix}
13&17\cr 9&6
\end{pmatrix}$ &
$ - \frac{72 t^4+681 t^3+2061 t^2+2523 t+1088}{72 t^4+612 t^3+1791 t^2+2211 t+993}$
& $ - \frac{1088}{993}<p<-1 $ & NA\\ \cline{2-5} &\multicolumn{4}{|c||}{ $\frac{\sqrt{289+884 t+1014 t^2+498 t^3+87 t^4+3 p^2 \left(67+166 t+162 t^2+72 t^3+12 t^4\right)+2 p \left(221+612 t+657 t^2+312 t^3+54
t^4\right)}}{17+26 t+9 t^2+\text{$|$p$|$} \left(13+18 t+6 t^2\right)}
\begin{matrix} {} \cr {} \cr {} \end{matrix}$}
\\ \cline{2-5} & \multicolumn{4}{|c||}{All non-BPS attractors are stable }
\\ \hline
\end{tabular}
\caption{Non-BPS extremal (multiple) black strings in two-moduli THCY models, 2/8.}
\label{thcymt2}
\end{sidewaystable}
\clearpage
}

\afterpage{
\clearpage
\thispagestyle{empty}
\begin{sidewaystable}
  \centering
    \begin{tabular}{ ||c| c|c|c|c|| }
\hline\multirow{3}{*} {$\begin{matrix} {\rm Polytope\ label,} \cr {} \cr {}
{\rm Charge\ Matrix}
\end{matrix}$}
 & $\begin{pmatrix}
c&d\cr b&a
\end{pmatrix}$  & Non-BPS solution &
$\begin{matrix}\rm
 \rm Range\ of\ validity \cr \rm for\ Single\ solution
\end{matrix}$ & $\begin{matrix}\rm
 \rm Range\ of\ validity \cr \rm for\ Multiple\ solutions
\end{matrix}$
\\ \cline{2-5} &\multicolumn{4}{|c||}{ Recombination factor $
\begin{matrix} {} \cr {} \cr {} \end{matrix}$} \\ \cline{2-5} & \multicolumn{4}{|c||}{Stability of non-BPS attractors }
\\ \hline \hline
  \multirow{3}{*} {$\begin{matrix}(15,1)^{2,86}_{-168}  \cr {} \cr {}
\begin{pmatrix}
1&0&0&0&0&1\cr 0&1&1&2&1&-1
\end{pmatrix}\end{matrix}$} &$\frac{1}{6}\begin{pmatrix}
3&1\cr 3&3
\end{pmatrix}$ &
$\frac{3 t^4-3 t^3-27 t^2-23 t-4}{15 t^3+45 t^2+45 t+17}$
& $p>-\frac{4}{17}$ & $ -0.45<p<-\frac{4}{17}$\\ \cline{2-5} &\multicolumn{4}{|c||}{ $\frac{\sqrt{1+12 t+54 t^2+48 t^3+9 t^4+6 p \left(1+4 t+12 t^2+12 t^3+3 t^4\right)+3 p^2 \left(7+16 t+18 t^2+12 t^3+3 t^4\right)}}{1+6
t+3 t^2+3 \text{$|$p$|$} (1+t)^2}
\begin{matrix} {} \cr {} \cr {} \end{matrix}$}
\\ \cline{2-5} & \multicolumn{4}{|c||}{Stable attractors for $-0.45<p<-0.05$ }
\\ \hline
  \multirow{3}{*} {$\begin{matrix}(16,1)^{2,90}_{-176}  \cr {} \cr {}
\begin{pmatrix}
1&0&0&0&0&1\cr 0&1&1&1&1&-1
\end{pmatrix}
\end{matrix}$} &$\frac{1}{6}\begin{pmatrix}
5&2\cr 5&5
\end{pmatrix}$ & $ \frac{5t^4-5 t^3-45 t^2-40 t-8}{25 t^3 + 75 t^2 + 75 t + 28}$
 & $p>-\frac{2}{7}$ & $-0.47<p<-\frac{2}{7}$\\ \cline{2-5} &\multicolumn{4}{|c||}{ $\frac{\sqrt{4+40 t+150 t^2+130 t^3+25 t^4+10 p \left(2+8 t+21 t^2+20 t^3+5 t^4\right)+5 p^2 \left(11+26 t+30 t^2+20 t^3+5 t^4\right)}}{2+10
t+5 t^2+5 \text{$|$p$|$} (1+t)^2}
\begin{matrix} {} \cr {} \cr {} \end{matrix}$} \\ \cline{2-5} & \multicolumn{4}{|c||}{Stable attractors for $-0.47<p<-0.04$ }
\\ \hline
  \multirow{3}{*} {$\begin{matrix} (17,1)^{2,92}_{-180} \cr {} \cr {}
\begin{pmatrix}
1&-1&-1&-1&2&0\cr 0&1&1&1&-1&1
\end{pmatrix} \end{matrix}$} &$\frac{1}{2}\begin{pmatrix}
4&7\cr 2&1
\end{pmatrix}$ &
 $ \frac{t^4 - 2 t^3 - 36 t^2 - 83 t - 56}{5 t^3 + 30 t^2 + 60 t + 41}$
&$p>-\frac{56}{41}$ &  $ -1.367<p<-\frac{56}{41}$\\ \cline{2-5} &\multicolumn{4}{|c||}{ $\frac{\sqrt{49+112 t+96 t^2+34 t^3+4 t^4+p^2 \left(20+34 t+24 t^2+8 t^3+t^4\right)+2 p \left(28+56 t+45 t^2+16 t^3+2 t^4\right)}}{7+8
t+2 t^2+\text{$|$p$|$} (2+t)^2}
\begin{matrix} {} \cr {} \cr {} \end{matrix}$} \\ \cline{2-5} & \multicolumn{4}{|c||}{Stable attractors for $-1.367<p<-0.016$ }
\\ \hline
  \multirow{3}{*} {$\begin{matrix}(17,2)^{2,92}_{-180}  \cr {} \cr {}
 \begin{pmatrix}
1&0&0&0&1&1\cr -1&1&1&1&-2&0
\end{pmatrix} \end{matrix}$} &$\frac{1}{2}\begin{pmatrix}
1&0\cr 3&7
\end{pmatrix}$ &
$ -\frac{t (15 t^3+37 t^2+27 t+6)}{56 t^4+141 t^3+123 t^2+45 t+6}$
& $- \frac{15}{56}<p<0$ & NA\\ \cline{2-5} &\multicolumn{4}{|c||}{ $\frac{\sqrt{t^2 \left(6+18 t+13 t^2\right)+2 p t^2 \left(9+28 t+21 t^2\right)+p^2 \left(3+18 t+54 t^2+84 t^3+49 t^4\right)}}{t (2+3 t)+\text{$|$p$|$}
\left(1+6 t+7 t^2\right)}
\begin{matrix} {} \cr {} \cr {} \end{matrix}$}\\ \cline{2-5} & \multicolumn{4}{|c||}{All non-BPS attractors are stable }
\\ \hline
  \multirow{3}{*} {$\begin{matrix} (18,1)^{2,95}_{-186}  \cr {} \cr {}
 \begin{pmatrix}
1&0&0&1&1&0\cr 0&1&1&0&-2&1
\end{pmatrix}\end{matrix}$} &$\frac{1}{6}\begin{pmatrix}
3&0\cr 7&14
\end{pmatrix}$ &
$ - \frac{9 t (49 t^2+63 t+18)}{392 t^4+1764 t^3+2079 t^2+945 t+162}$
& NA & $-0.22<p<0$\\ \cline{2-5} &\multicolumn{4}{|c||}{ $\frac{\sqrt{9 t^2 \left(6+14 t+7 t^2\right)+14 p t^2 \left(9+24 t+14 t^2\right)+p^2 \left(27+126 t+294 t^2+392 t^3+196 t^4\right)}}{t
(6+7 t)+\text{$|$p$|$} \left(3+14 t+14 t^2\right)}
\begin{matrix} {} \cr {} \cr {} \end{matrix}$}
\\ \cline{2-5} & \multicolumn{4}{|c||}{$\begin{matrix}
{\rm All\ solutions\ for}\ -0.22<p<-0.037\
{\rm and\ one\ branch\ of\ solutions}\cr {\rm  with\ smaller}\
t_c\ {\rm for\ the\ remaining\ allowed\ range} -0.037<p<0\
 {\rm are\ stable}\end{matrix}$ }
\\ \hline
\end{tabular}
\caption{Non-BPS extremal (multiple) black strings in two-moduli THCY models, 3/8.}
\label{thcymt3}
\end{sidewaystable}
\clearpage
}

\afterpage{
\clearpage
\thispagestyle{empty}
\begin{sidewaystable}
  \centering
    \begin{tabular}{ ||c| c|c|c|c|| }
\hline\multirow{3}{*} {$\begin{matrix} {\rm Polytope\ label,} \cr {} \cr {}
{\rm Charge\ Matrix}
\end{matrix}$}
 & $\begin{pmatrix}
c&d\cr b&a
\end{pmatrix}$  & Non-BPS solution &
$\begin{matrix}\rm
 \rm Range\ of\ validity \cr \rm for\ Single\ solution
\end{matrix}$ & $\begin{matrix}\rm
 \rm Range\ of\ validity \cr \rm for\ Multiple\ solutions
\end{matrix}$
\\ \cline{2-5} &\multicolumn{4}{|c||}{ Recombination factor $
\begin{matrix} {} \cr {} \cr {} \end{matrix}$} \\ \cline{2-5} & \multicolumn{4}{|c||}{Stability of non-BPS attractors }
\\ \hline \hline
  \multirow{3}{*} {$\begin{matrix}(19,1)^{2,102}_{-200}  \cr {} \cr {}
 \begin{pmatrix}
1&0&0&0&1&0\cr 0&1&1&1&-2&1
\end{pmatrix} \end{matrix}$} &$\frac{1}{3}\begin{pmatrix}
3&1\cr 6&12
\end{pmatrix}$ &
$ \frac{12 t^4 -6 t^3-27 t^2-14 t-2}{60 t^3 + 90 t^2 + 45 t + 8}$
& $p>-\frac{1}{4}$ & $-0.28<p<-\frac{1}{4}$\\ \cline{2-5} &\multicolumn{4}{|c||}{ $\frac{\sqrt{1+12 t+54 t^2+84 t^3+36 t^4+6 p \left(1+8 t+30 t^2+48 t^3+24 t^4\right)+3 p^2 \left(5+28 t+72 t^2+96 t^3+48 t^4\right)}}{1+6
t+6 t^2+3 \text{$|$p$|$} (1+2 t)^2}
\begin{matrix} {} \cr {} \cr {} \end{matrix}$}
\\ \cline{2-5} & \multicolumn{4}{|c||}{All non-BPS attractors for $-0.28<p<-0.01$
are stable }
\\ \hline
  \multirow{3}{*} {$\begin{matrix} (20,1)^{2,106}_{-208}  \cr {} \cr {}
\begin{pmatrix}
  1&1&1&0&-3&8\cr 0&0&0&1&1&-2
  \end{pmatrix}\end{matrix}$} & $\frac{1}{6}\begin{pmatrix}
12&36\cr 4&1
\end{pmatrix}$ &
  $ - \frac{3 t (7 t^3+60 t^2+180 t+180)}{4 t^4+39 t^3+108 t^2+36 t-108}$
  & $p>0\ \& \ p<-\frac{21}{4}$ & $-\frac{21}{4}<p<-4.99$\\ \cline{2-5} &\multicolumn{4}{|c||}{ $\frac{\sqrt{24 \left(54+72 t+36 t^2+9 t^3+t^4\right)+8 p \left(108+144 t+63 t^2+12 t^3+t^4\right)+p^2 \left(144+216 t+96 t^2+16 t^3+t^4\right)}}{4
(3+t)^2+\text{$|$p$|$} \left(12+8 t+t^2\right)}
\begin{matrix} {} \cr {} \cr {} \end{matrix}$}
\\ \cline{2-5} & \multicolumn{4}{|c||}{All non-BPS attractors
for $-181.52<p<-4.99$ are stable }
\\ \hline
  \multirow{3}{*} {$\begin{matrix} (21,1)^{2,106}_{-208}  \cr {} \cr {}
\begin{pmatrix}
  1&-3&1&1&5&0\cr 0&1&0&0&-1&1
  \end{pmatrix}\end{matrix}$} & $\frac{1}{6}\begin{pmatrix}
12&36\cr 4&1
\end{pmatrix}$ &
  $ - \frac{3 t (7 t^3+60 t^2+180 t+180)}{4 t^4+39 t^3+108 t^2+36 t-108}$
  & $p>0\ \& \ p<-\frac{21}{4}$ & $-\frac{21}{4}<p<-4.99$\\ \cline{2-5} &\multicolumn{4}{|c||}{ $\frac{\sqrt{24 \left(54+72 t+36 t^2+9 t^3+t^4\right)+8 p \left(108+144 t+63 t^2+12 t^3+t^4\right)+p^2 \left(144+216 t+96 t^2+16 t^3+t^4\right)}}{4
(3+t)^2+\text{$|$p$|$} \left(12+8 t+t^2\right)}
\begin{matrix} {} \cr {} \cr {} \end{matrix}$}
\\ \cline{2-5} & \multicolumn{4}{|c||}{All non-BPS attractors
for $-181.52<p<-4.99$ are stable }
\\ \hline
  \multirow{3}{*} {$\begin{matrix}(22,1)^{2,106}_{-208}  \cr {} \cr {}
\begin{pmatrix}
1&0&0&0&1&0\cr 0&1&1&2&-3&1
\end{pmatrix} \end{matrix}$} &$\frac{1}{6}\begin{pmatrix}
4&1\cr 12&36
\end{pmatrix}$ &
$ \frac{108 t^4-36 t^3-108 t^2-39 t-4}{540 t^3 + 540 t^2 + 180 t + 21}$
& $ p> - \frac{4}{21} $ & $ - 0.2<p<-\frac{4}{21}$\\ \cline{2-5} &\multicolumn{4}{|c||}{ $\frac{\sqrt{1+16 t+96 t^2+216 t^3+144 t^4+24 p^2 \left(1+9 t+36 t^2+72 t^3+54 t^4\right)+8 p \left(1+12 t+63 t^2+144 t^3+108 t^4\right)}}{1+8
t+12 t^2+4 \text{$|$p$|$} (1+3 t)^2}
\begin{matrix} {} \cr {} \cr {} \end{matrix}$}
\\ \cline{2-5} & \multicolumn{4}{|c||}{All non-BPS attractors for
$-0.2<p<-0.005$ are stable }
\\ \hline
  \multirow{3}{*} {$\begin{matrix} (23,1)^{2,116}_{-228}  \cr {} \cr {}
\begin{pmatrix}
-2&3&-2&-4&5&0\cr 1&0&1&2&-1&3
\end{pmatrix} \end{matrix}$} &$\frac{1}{6}\begin{pmatrix}
25&98\cr 5&1
\end{pmatrix}$ &
$ \frac{t^4 - 5 t^3 - 225 t^2 - 1240 t - 1960}{5 t^3 + 75 t^2 + 375 t + 652}$
& $p>- \frac{490}{163}$ & $-3.1<p<-\frac{490}{163}$\\ \cline{2-5} &\multicolumn{4}{|c||}{ $\frac{\sqrt{9604+9800 t+3750 t^2+554 t^3+25 t^4+p^2 \left(895+554 t+150 t^2+20 t^3+t^4\right)+2 p \left(2450+1960 t+669 t^2+100 t^3+5
t^4\right)}}{98+50 t+5 t^2+\text{$|$p$|$} (5+t)^2}
\begin{matrix} {} \cr {} \cr {} \end{matrix}$}
\\ \cline{2-5} & \multicolumn{4}{|c||}{All non-BPS attractors for
$-3.1<p<-0.07$ are stable }
\\ \hline
  \end{tabular}
\caption{Non-BPS extremal (multiple) black strings in two-moduli THCY models, 4/8.}
\label{thcymt4}
\end{sidewaystable}
\clearpage
}

\afterpage{
\clearpage
\thispagestyle{empty}
\begin{sidewaystable}
  \centering
    \begin{tabular}{ ||c| c|c|c|c|| }
\hline\multirow{3}{*} {$\begin{matrix}{\rm Polytope\ label,}  \cr {} \cr {}
{\rm Charge\ Matrix}
\end{matrix}$}
 & $\begin{pmatrix}
c&d\cr b&a
\end{pmatrix}$  & Non-BPS solution &
$\begin{matrix}\rm
 \rm Range\ of\ validity \cr \rm for\ Single\ solution
\end{matrix}$ & $\begin{matrix}\rm
 \rm Range\ of\ validity \cr \rm for\ Multiple\ solutions
\end{matrix}$
\\ \cline{2-5} &\multicolumn{4}{|c||}{ Recombination factor $
\begin{matrix} {} \cr {} \cr {} \end{matrix}$} \\ \cline{2-5} & \multicolumn{4}{|c||}{Stability of non-BPS attractors }
\\ \hline \hline
  \multirow{3}{*} {$\begin{matrix} (23,2)^{2,116}_{-228} \cr {} \cr {}
\begin{pmatrix}
0&1&0&0&1&2\cr 2&-3&2&4&-5&0
\end{pmatrix} \end{matrix}$} & $\frac{1}{3}\begin{pmatrix}
1&0\cr 8&49
\end{pmatrix}$ &
$ -\frac{t (164 t^3+137 t^2+36 t+3)}{1470 t^4+1340 t^3+435 t^2+60 t+3}$
& $- \frac{82}{735}<p<0$ & NA\\ \cline{2-5} &\multicolumn{4}{|c||}{ $\frac{\sqrt{2 t^2 \left(3+24 t+47 t^2\right)+8 p t^2 \left(6+49 t+98 t^2\right)+p^2 \left(3+48 t+384 t^2+1568 t^3+2401 t^4\right)}}{2
t (1+4 t)+\text{$|$p$|$} \left(1+16 t+49 t^2\right)}
\begin{matrix} {} \cr {} \cr {} \end{matrix}$}
\\ \cline{2-5} & \multicolumn{4}{|c||}{ All non-BPS attractors are stable }
\\ \hline
  \multirow{3}{*} {$\begin{matrix} (24,1)^{2,120}_{-236}  \cr {} \cr {}
\begin{pmatrix}
1&-1&1&2&0&3\cr -1&4&-1&-5&3&0
\end{pmatrix}\end{matrix}$} &$\frac{1}{6}\begin{pmatrix}
8&2\cr 32&101
\end{pmatrix}$ &
$ - \frac{t (667 t^3+480 t^2+120 t+10)}{1616 t^4+1273 t^3+288 t^2+8 t-2}$
& $p>0\ \& \ p< - \frac{667}{1616}$ & $- \frac{667}{1616}<p<-0.4$\\ \cline{2-5} &\multicolumn{4}{|c||}{ $\frac{\sqrt{4 \left(1+16 t+96 t^2+283 t^3+364 t^4\right)+4 p \left(8+128 t+687 t^2+1616 t^3+1616 t^4\right)+p^2 \left(64+1132 t+6144 t^2+12928
t^3+10201 t^4\right)}}{2 (1+4 t)^2+\text{$|$p$|$} \left(8+64 t+101 t^2\right)}
\begin{matrix} {} \cr {} \cr {} \end{matrix}$}
\\ \cline{2-5} & \multicolumn{4}{|c||}{All non-BPS attractors for $-18.02<p<-0.4$
are stable }
\\ \hline
  \multirow{3}{*} {$\begin{matrix} (24,2)^{2,120}_{-236} \cr {} \cr {}
 \begin{pmatrix}
1&-4&1&5&-3&0\cr 0&1&0&-1&1&1
\end{pmatrix}\end{matrix}$} & $\frac{1}{6}\begin{pmatrix}
23&101\cr 5&1
\end{pmatrix}$ &
$-\frac{23 t^4+357 t^3+2055 t^2+5191 t+4848}{4 t^4+65 t^3+387 t^2+1001 t+949}$
& $- \frac{23}{4}<p<- \frac{4848}{949}$ & NA\\ \cline{2-5} &\multicolumn{4}{|c||}{ $\frac{\sqrt{10201+9292 t+3174 t^2+488 t^3+29 t^4+p^2 \left(577+488 t+150 t^2+20 t^3+t^4\right)+2 p \left(2323+2020 t+648 t^2+92 t^3+5
t^4\right)}}{101+46 t+5 t^2+\text{$|$p$|$} \left(23+10 t+t^2\right)}
\begin{matrix} {} \cr {} \cr {} \end{matrix}$}
\\ \cline{2-5} & \multicolumn{4}{|c||}{All non-BPS attractors are stable }
\\ \hline
  \multirow{3}{*} {$\begin{matrix} (25,1)^{2,122}_{-240}  \cr {} \cr {}
\begin{pmatrix}
-3&1&1&1&0&7\cr 1&0&0&0&1&-2
\end{pmatrix} \end{matrix}$} &$\frac{1}{6}\begin{pmatrix}
21&63\cr 7&2
\end{pmatrix}$ &
$ - \frac{9 t (4 t^3 + 35 t^2 + 105 t + 105)}{8 t^4 + 72 t^3 + 189 t^2 + 63 t - 189}$
& $p>0 \ \& \ p<-\frac{9}{2}$ & $-\frac{9}{2}<p< -4.48$\\ \cline{2-5} &\multicolumn{4}{|c||}{ $\frac{\sqrt{63 \left(63+84 t+42 t^2+10 t^3+t^4\right)+14 p \left(189+252 t+117 t^2+24 t^3+2 t^4\right)+p^2 \left(441+630 t+294 t^2+56
t^3+4 t^4\right)}}{7 (3+t)^2+\text{$|$p$|$} \left(21+14 t+2 t^2\right)}
\begin{matrix} {} \cr {} \cr {} \end{matrix}$}\\ \cline{2-5} & \multicolumn{4}{|c||}{All non-BPS attractors for $-322.24<p<-4.48$ are stable }
\\ \hline
  \multirow{3}{*} {$\begin{matrix} (26,1)^{2,122}_{-240}  \cr {} \cr {}
\begin{pmatrix}
1&-3&1&1&4&0\cr 0&1&0&0&-1&1
\end{pmatrix} \end{matrix}$} &$\frac{1}{6}\begin{pmatrix}
21&63\cr 7&2
\end{pmatrix}$ &
$ - \frac{9 t (4 t^3 + 35 t^2 + 105 t + 105)}{8 t^4 + 72 t^3 + 189 t^2 + 63 t - 189}$
& $p>0 \ \& \ p<-\frac{9}{2}$ & $-\frac{9}{2}<p< -4.48$\\ \cline{2-5} &\multicolumn{4}{|c||}{ $\frac{\sqrt{63 \left(63+84 t+42 t^2+10 t^3+t^4\right)+14 p \left(189+252 t+117 t^2+24 t^3+2 t^4\right)+p^2 \left(441+630 t+294 t^2+56
t^3+4 t^4\right)}}{7 (3+t)^2+\text{$|$p$|$} \left(21+14 t+2 t^2\right)}
\begin{matrix} {} \cr {} \cr {} \end{matrix}$}\\ \cline{2-5} & \multicolumn{4}{|c||}{All non-BPS attractors for $-322.24<p<-4.48$ are stable  }
\\ \hline
  \end{tabular}
\caption{Non-BPS extremal (multiple) black strings in two-moduli THCY models, 5/8.}
\label{thcymt5}
\end{sidewaystable}
\clearpage
}

\afterpage{
\clearpage
\thispagestyle{empty}
\begin{sidewaystable}
  \centering
    \begin{tabular}{ ||c| c|c|c|c|| }
\hline\multirow{3}{*} {$\begin{matrix} {\rm Polytope\ label,} \cr {} \cr {}
{\rm Charge\ Matrix}
\end{matrix}$}
 & $\begin{pmatrix}
c&d\cr b&a
\end{pmatrix}$  & Non-BPS solution &
$\begin{matrix}\rm
 \rm Range\ of\ validity \cr \rm for\ Single\ solution
\end{matrix}$ & $\begin{matrix}\rm
 \rm Range\ of\ validity \cr \rm for\ Multiple\ solutions
\end{matrix}$
\\ \cline{2-5} &\multicolumn{4}{|c||}{ Recombination factor $
\begin{matrix} {} \cr {} \cr {} \end{matrix}$} \\ \cline{2-5} & \multicolumn{4}{|c||}{Stability of non-BPS attractors }
\\ \hline \hline
  \multirow{3}{*} {$\begin{matrix} (27,1)^{2,122}_{-240} \cr {} \cr {}
\begin{pmatrix}
1&0&0&0&1&0\cr 0&1&1&1&-3&1
\end{pmatrix} \end{matrix}$} & $\frac{1}{6}\begin{pmatrix}
7&2\cr 21&63
\end{pmatrix}$ &
$ \frac{189 t^4-63 t^3-189 t^2-72 t-8}{945 t^3 + 945 t^2 + 315 t + 36}$
& $p>-\frac{2}{9}$ & $-0.223<p<-\frac{2}{9}$\\ \cline{2-5} &\multicolumn{4}{|c||}{ $\frac{\sqrt{4+56 t+294 t^2+630 t^3+441 t^4+63 p^2 \left(1+10 t+42 t^2+84 t^3+63 t^4\right)+14 p \left(2+24 t+117 t^2+252 t^3+189 t^4\right)}}{2+14
t+21 t^2+7 \text{$|$p$|$} (1+3 t)^2}
\begin{matrix} {} \cr {} \cr {} \end{matrix}$}
\\ \cline{2-5} & \multicolumn{4}{|c||}{All non-BPS attractors for
$-0.223<p<-0.003$ are stable }
\\ \hline
  \multirow{3}{*} {$\begin{matrix} (30,2)^{2,128}_{-252}  \cr {} \cr {}
 \begin{pmatrix}
  0&1&0&0&1&2\cr 2&-3&2&2&-3&0
  \end{pmatrix}\end{matrix}$} &$\frac{2}{3}\begin{pmatrix}
  1&0\cr 6&27
\end{pmatrix}$   & $-\frac{t}{6t+1}$
  & $-\frac{1}{6}< p < 0$ & NA\\ \cline{2-5} &\multicolumn{4}{|c||}{ $
  \frac{1+3 p}{1-p}
  \begin{matrix} {} \cr {} \cr {} \end{matrix}$}
\\ \cline{2-5} & \multicolumn{4}{|c||}{All non-BPS attractors are stable }
\\ \hline
  \multirow{3}{*} {$\begin{matrix} (31,1)^{2,128}_{-252} \cr {} \cr {}
\begin{pmatrix}
  1&-1&-1&2&0&1\cr -2&3&3&-5&1&0
\end{pmatrix}   \end{matrix}$}
 &$\frac{1}{6}\begin{pmatrix}
 16&6\cr 42&109
\end{pmatrix}$  & $ - \frac{1539 t^4+2618 t^3+1656 t^2+462 t+48}{4360 t^4+7425 t^3+4710 t^2+1320 t+138}$
  & $ - \frac{1539}{4360}<p<-\frac{8}{23}$ & NA\\ \cline{2-5} &\multicolumn{4}{|c||}{ $\frac{\sqrt{4 \left(9+96 t+384 t^2+681 t^3+451 t^4\right)+4 p \left(48+504 t+1989 t^2+3488 t^3+2289 t^4\right)+p^2 \left(264+2724 t+10584
t^2+18312 t^3+11881 t^4\right)}}{6+32 t+42 t^2+\text{$|$p$|$} \left(16+84 t+109 t^2\right)}
\begin{matrix} {} \cr {} \cr {} \end{matrix}$}
\\ \cline{2-5} & \multicolumn{4}{|c||}{All non-BPS attractors are stable }
\\ \hline
  \multirow{3}{*} {$\begin{matrix}(31,2)^{2,128}_{-252}   \cr {} \cr {}
\begin{pmatrix}
2&-3&-3&5&-1&0\cr 0&1&1&-1&1&2
\end{pmatrix} \end{matrix}$} & $\frac{1}{6}\begin{pmatrix}
25&109\cr 5&1
\end{pmatrix}$ &
$ \frac{t^4-5 t^3-225 t^2-1295 t-2180}{5 t^3+75 t^2+375 t+641}$
& $ p>- \frac{2180}{641}$ & $-3.405<p<- \frac{2180}{641}$\\ \cline{2-5} &\multicolumn{4}{|c||}{ $\frac{\sqrt{11881+10900 t+3750 t^2+532 t^3+25 t^4+p^2 \left(785+532 t+150 t^2+20 t^3+t^4\right)+2 p \left(2725+2180 t+702 t^2+100 t^3+5
t^4\right)}}{109+50 t+5 t^2+\text{$|$p$|$} (5+t)^2}
\begin{matrix} {} \cr {} \cr {} \end{matrix}$}
\\ \cline{2-5} & \multicolumn{4}{|c||}{All non-BPS attractors for
$-3.405<p<-0.042$ are stable }
\\ \hline
  \multirow{3}{*} {$\begin{matrix} (32,1)^{2,128}_{-252} \cr {} \cr {}
\begin{pmatrix}
1&-3&-3&1&4&0\cr 0&1&1&0&-1&1
\end{pmatrix} \end{matrix}$} &$\frac{1}{3}\begin{pmatrix}
15&54\cr
4&1
\end{pmatrix}$ &
$-\frac{9(t+4)}{2 t + 9}$
& $-\frac{9}{2}<p<-4$ & NA\\ \cline{2-5} &\multicolumn{4}{|c||}{ $
3 \left(\frac{p^2+9 p+18}{p^2+9p-18}\right)
\begin{matrix} {} \cr {} \cr {} \end{matrix}$} \\ \cline{2-5} & \multicolumn{4}{|c||}{All non-BPS attractors are stable }
\\ \hline
\end{tabular}
\caption{Non-BPS extremal (multiple) black strings in two-moduli THCY models, 6/8.}
\label{thcymt6}
\end{sidewaystable}
\clearpage
}

\afterpage{
\clearpage
\thispagestyle{empty}
\begin{sidewaystable}
  \centering
    \begin{tabular}{ ||c| c|c|c|c|| }
\hline\multirow{3}{*} {$\begin{matrix} {\rm Polytope\ label,} \cr {}
{\rm Charge\ Matrix}
\end{matrix}$}
 & $\begin{pmatrix}
c&d\cr b&a
\end{pmatrix}$  & Non-BPS solution &
$\begin{matrix}\rm
 \rm Range\ of\ validity \cr \rm for\ Single\ solution
\end{matrix}$ & $\begin{matrix}\rm
 \rm Range\ of\ validity \cr \rm for\ Multiple\ solutions
\end{matrix}$
\\ \cline{2-5} &\multicolumn{4}{|c||}{ Recombination factor $
\begin{matrix} {} \cr {} \end{matrix}$} \\ \cline{2-5} & \multicolumn{4}{|c||}{Stability of non-BPS attractors }
\\ \hline \hline
  \multirow{3}{*} {$\begin{matrix} (33,1)^{2,132}_{-260} \cr {} \cr {}
\begin{pmatrix}
1&-1&-1&-1&2&0\cr 0&1&1&1&-1&2
\end{pmatrix} \end{matrix}$} &$\frac{1}{3}\begin{pmatrix}
4&7\cr 2&1
\end{pmatrix}$ &
$\frac{t^4-2 t^3-36 t^2-83 t-56}{5 t^3+30 t^2+60 t+41}$
& $p>- \frac{56}{41}$ & $-1.37<p< - \frac{56}{41}$\\ \cline{2-5} &\multicolumn{4}{|c||}{ $\frac{\sqrt{49+112 t+96 t^2+34 t^3+4 t^4+p^2 \left(20+34 t+24 t^2+8 t^3+t^4\right)+2 p \left(28+56 t+45 t^2+16 t^3+2 t^4\right)}}{7+8
t+2 t^2+\text{$|$p$|$} (2+t)^2}
\begin{matrix} {} \cr {} \cr {} \end{matrix}$}
\\ \cline{2-5} & \multicolumn{4}{|c||}{All non-BPS attractors for $<-1.37<p<-0.02$
are stable }
\\ \hline
  \multirow{3}{*} {$\begin{matrix}(33,2)^{2,132}_{-260}  \cr {} \cr {}
\begin{pmatrix}
1&0&0&0&1&2\cr -1&1&1&1&-2&0
\end{pmatrix} \end{matrix}$} &$\frac{1}{3}\begin{pmatrix}
1&0\cr 3&7
\end{pmatrix}$ &
$ -\frac{t (15 t^3+37 t^2+27 t+6)}{56 t^4+141 t^3+123 t^2+45 t+6}$
& $- \frac{15}{56}<p<0$ & NA\\ \cline{2-5} &\multicolumn{4}{|c||}{ $\frac{\sqrt{t^2 \left(6+18 t+13 t^2\right)+2 p t^2 \left(9+28 t+21 t^2\right)+p^2 \left(3+18 t+54 t^2+84 t^3+49 t^4\right)}}{t (2+3 t)+\text{$|$p$|$}
\left(1+6 t+7 t^2\right)}
\begin{matrix} {} \cr {} \cr {} \end{matrix}$}
\\ \cline{2-5} & \multicolumn{4}{|c||}{All non-BPS attractors are stable }
\\ \hline
  \multirow{3}{*} {$\begin{matrix} (34,1)^{2,132}_{-260} \cr {} \cr {}
\begin{pmatrix}
1&-2&-2&-4&7&0\cr 0&1&1&2&-3&1
\end{pmatrix} \end{matrix}$} & $\frac{1}{6}\begin{pmatrix}
49&114\cr 21&9
\end{pmatrix}$ &
$ \frac{9 t^4-21 t^3-441 t^2-1256 t-1064}{45 t^3+315 t^2+735 t+572}$
& $p>- \frac{266}{143}$ & NA\\ \cline{2-5} &\multicolumn{4}{|c||}{ $\frac{\sqrt{3} \sqrt{4332+7448 t+4802 t^2+1374 t^3+147 t^4+p^2 \left(805+1374 t+882 t^2+252 t^3+27 t^4\right)+2 p \left(1862+3192 t+2055
t^2+588 t^3+63 t^4\right)}}{114+98 t+21 t^2+\text{$|$p$|$} (7+3 t)^2}
\begin{matrix} {} \cr {} \cr {} \end{matrix}$}
\\ \cline{2-5} & \multicolumn{4}{|c||}{All non-BPS attractors for $- \frac{266}{143}<p<0$ are stable }
\\ \hline
  \multirow{3}{*} {$\begin{matrix}(34,2)^{2,132}_{-260}  \cr {} \cr {}
\begin{pmatrix}
1&0&0&0&1&2\cr -1&2&2&4&-7&0
\end{pmatrix} \end{matrix}$} &$\frac{1}{3}\begin{pmatrix}
1&0\cr 8&57
\end{pmatrix}$ &
 $\frac{t (20 t^3-27 t^2-12 t-1)}{266 t^4+436 t^3+153 t^2+20 t+1}$
& $0<p<\frac{10}{133}$ & $-0.06<p<0$\\ \cline{2-5} &\multicolumn{4}{|c||}{ $\frac{\sqrt{3} \sqrt{2 t^2 \left(1+8 t+13 t^2\right)+8 p t^2 \left(2+19 t+38 t^2\right)+p^2 \left(1+16 t+128 t^2+608 t^3+1083 t^4\right)}}{2
t (1+4 t)+\text{$|$p$|$} \left(1+16 t+57 t^2\right)}
\begin{matrix} {} \cr {} \cr {} \end{matrix}$}
\\ \cline{2-5} & \multicolumn{4}{|c||}{All non-BPS attractors for $-0.06<p<-0.01$
are stable }
\\ \hline
  \multirow{3}{*} {$\begin{matrix} (35,1)^{2,144}_{-284} \cr {} \cr {}
\begin{pmatrix}
1&-2&-2&-2&5&0\cr 0&1&1&1&-2&1
\end{pmatrix} \end{matrix}$} &$\frac{1}{3}\begin{pmatrix}
25&62\cr 10&4
\end{pmatrix}$ &
$ \frac{4 t^4-10 t^3-225 t^2-685 t-620}{20 t^3+150 t^2+375 t+313}$
&$ p>- \frac{620}{313}$ & NA\\ \cline{2-5} &\multicolumn{4}{|c||}{ $\frac{\sqrt{p^2 \left(635+1004 t+600 t^2+160 t^3+16 t^4\right)+4 p \left(775+1240 t+747 t^2+200 t^3+20 t^4\right)+2 \left(1922+3100 t+1875
t^2+502 t^3+50 t^4\right)}}{62+50 t+10 t^2+\text{$|$p$|$} (5+2 t)^2}
\begin{matrix} {} \cr {} \cr {} \end{matrix}$}
\\ \cline{2-5} & \multicolumn{4}{|c||}{All non-BPS attractors for $- \frac{620}{313}<p<0$ are stable }
\\ \hline
  \end{tabular}
\caption{Non-BPS extremal (multiple) black strings in two-moduli THCY models, 7/8.}
\label{thcymt7}
\end{sidewaystable}
\clearpage
}

\afterpage{
\clearpage
\thispagestyle{empty}
\begin{sidewaystable}
  \centering
    \begin{tabular}{ ||c| c|c|c|c|| }
\hline\multirow{3}{*} {$\begin{matrix} {\rm Polytope\ label,} \cr {} \cr {}
{\rm Charge\ Matrix}
\end{matrix}$}
 & $\begin{pmatrix}
c&d\cr b&a
\end{pmatrix}$  & Non-BPS solution &
$\begin{matrix}\rm
 \rm Range\ of\ validity \cr \rm for\ Single\ solution
\end{matrix}$ & $\begin{matrix}\rm
 \rm Range\ of\ validity \cr \rm for\ Multiple\ solutions
\end{matrix}$
\\ \cline{2-5} &\multicolumn{4}{|c||}{ Recombination factor $
\begin{matrix} {} \cr {} \cr {} \end{matrix}$} \\ \cline{2-5} & \multicolumn{4}{|c||}{Stability of non-BPS attractors }
\\ \hline \hline
  \multirow{3}{*} {$\begin{matrix} (35,2)^{2,144}_{-284} \cr {} \cr {}
\begin{pmatrix}
1&0&0&0&1&2\cr -1&2&2&2&-5&0
\end{pmatrix} \end{matrix}$} &$\frac{2}{3}\begin{pmatrix}
1&0\cr 6&31
\end{pmatrix}$ &
$ \frac{t (3 t^3-53 t^2-27 t-3)}{310 t^4+555 t^3+255 t^2+45 t+3}$
& $0<p<\frac{3}{310}$ & $-0.08<p<0$\\ \cline{2-5} &\multicolumn{4}{|c||}{ $\frac{\sqrt{2 t^2 \left(3+18 t+23 t^2\right)+4 p t^2 \left(9+62 t+93 t^2\right)+p^2 \left(3+36 t+216 t^2+744 t^3+961 t^4\right)}}{2 t
(1+3 t)+\text{$|$p$|$} \left(1+12 t+31 t^2\right)}
\begin{matrix} {} \cr {} \cr {} \end{matrix}$}
\\ \cline{2-5} & \multicolumn{4}{|c||}{All non-BPS attractors for $-0.08<p<-0.01$
are stable }
\\ \hline
  \multirow{3}{*} {$\begin{matrix} (36,1)^{2,272}_{-540}  \cr {}
\begin{pmatrix}
0&0&0&2&3&1\cr 1&1&1&0&0&-3
\end{pmatrix} \end{matrix}$} & $\frac{1}{6}\begin{pmatrix}
1&0\cr 3&9
\end{pmatrix}$ &
$ \frac{t (9 t^3 - 3 t^2 - 9 t - 2)}{45 t^3 + 45 t^2 + 15 t + 2}$
& $p>0$ & $-0.12<p<0$\\ \cline{2-5} &\multicolumn{4}{|c||}{ $\frac{\sqrt{3} \sqrt{6 p t^2 \left(1+4 t+3 t^2\right)+t^2 \left(2+6 t+3 t^2\right)+p^2 \left(1+6 t+18 t^2+36 t^3+27 t^4\right)}}{\text{$|$p$|$}
(1+3 t)^2+t (2+3 t)}
\begin{matrix} {} \cr {}  \end{matrix}$} \\ \cline{2-5} & \multicolumn{4}{|c||}{All non-BPS attractors for $-0.12<p<-0.025$ are stable }
\\ \hline
  \end{tabular}
\caption{Non-BPS extremal (multiple) black strings in two-moduli THCY models, 8/8.}
\label{thcymt8}
\end{sidewaystable}
\clearpage
}\pagebreak

\end{document}